\documentclass[pra,amsmath,amssymb,twocolumn]{revtex4}
\usepackage{hyperref,amssymb,amsbsy,bbold,mathtools,relsize,mathrsfs,multirow}
\pdfoutput=1

\begin{document}
\title{Quantum description of a rotating and vibrating molecule}
\author{Sylvain D. \surname{Brechet}}
\email{sylvain.brechet@epfl.ch}
\author{Francois A. \surname{Reuse}}
\author{Klaus \surname{Maschke}}
\author{Jean-Philippe \surname{Ansermet}}
\affiliation{Institute of Condensed Matter Physics, Station 3, Ecole Polytechnique F\'ed\'erale de Lausanne - EPFL, CH-1015 Lausanne,
Switzerland}

\begin{abstract}

A rigorous quantum description of molecular dynamics with a particular emphasis on internal observables is developed accounting explicitly for kinetic couplings between nuclei and electrons. Rotational modes are treated in a genuinely quantum framework by defining a molecular orientation operator. Canonical rotational commutation relations are established explicitly. Moreover, physical constraints are imposed on the observables in order to define the state of a molecular system located in the neighborhood of the ground state defined by the equilibrium condition. 

\end{abstract}

\maketitle


\section{Introduction}
\label{Introduction}

The dynamics of quantum molecular systems has been studied analytically and numerically for decades. Molecular rotations are usually characterised by Euler angles defined with respect to a molecular reference frame~\cite{Wilson:1936,Brown:2003,Bunker:2005,Yamanouchi:2013} and kinetic couplings between nuclei and electrons are neglected. However, in order to make a precise quantum description of molecular dynamics, these couplings have to be taken explicitly into account, which leads to quantum deviations in the commutation relations. It is also important to recognise that the notion of a molecular reference frame is inconsistent with quantum physics, due to nonlocality. For the same reason, the orientation of a molecule cannot be described simply using Euler angles as in a classical framework. In order to treat rotational states of molecular systems in a genuine quantum framework, the orientation and rotation of a molecule has to be described using operators. This is done in this article.

Rotational states of molecular systems are currently of great interest. For example, in small molecular systems at low temperature, the rotational degrees of freedom play an important role since they can be distinguished experimentally from the vibrational degrees of freedom. Due to technological improvement, the distinction between these degrees of freedom became increasingly important in the last decade. Rotating atoms~\cite{Natterer:2013}, rotating molecules~\cite{Daley:2013,Natterer:2014}, rotating trapped Bose-Einstein condensates~\cite{Fetter:2009} and even rotating microgyroscopes~\cite{Arita:2013} are currently studied experimentally and are attracting much attention. For example, physisorbed H$_2$, HD and D$_2$ on a substrate at low temperature form a honeycomb lattice and rotational spectroscopy revealed a resonance width of H$_2$ twice as large as the resonance widths of HD and D$_2$~\cite{Natterer:2013}. The theoretical explanation requires a rigorous quantum formalism with a genuine quantum treatment of molecular rotations.

A semi-classical approach is commonly used for the description of molecular dynamics~\cite{Thoss:2004,Gorshkov:2013,Buchholtz:2012}. An important shortcoming of such an approach is that it requires the rotational states to be in an eigenstate, thus imposing severe restrictions on the dynamics. Rotational states play an important role for low temperature spectroscopy~\cite{Brown:2003}, for THz spectroscopy~\cite{Bartalini:2014}, for molecular magnetism~\cite{Yin:2007} and for molecular superrotors~\cite{Korobenko:2014}. In order to establish a genuine quantum description of rotational molecular states, a molecular rotation operator has to be introduced.

Here, we develop a rigorous quantum description of molecule with a particular emphasis on internal observables. In this description, the vibrational and rotational modes are described by operators associated to the deformation and orientation of the molecule. The set of internal observables is related explicitly to the set of observables associated to nuclei and electrons through a rotation operator. The internal observables are chosen in order to satisfy canonical commutation relations in translation and rotation. Moreover, in order to define the state of a molecular system, physical constraints need to be imposed on the observables. In fact, the amplitudes of the vibrational modes of a molecular system have to be sufficiently small.

The quantum molecular description presented in this publication is expected to be relevant for extremely fast rotating molecular systems exhibiting a large orbital angular momentum. Such systems, called ``superrotors'', have been observed for the molecules listed in Table~\ref{type}. The molecular orbital angular momentum depends in part on internal vibrations, as we will show below. Thus, this description is of importance for molecules with large vibration amplitudes. This can be realised  for molecules with weak bonds, such as van der Waals bonds. The present quantum molecular formalism is expected also to be important for molecular systems, listed in Table~\ref{type}, for which the ``Coriolis interaction'' leads to a large rotational-vibrational coupling.

\begin{table}
\caption{\label{type} Molecular systems}
\begin{center}
\begin{tabular}{|c|c|c|}
\hline
Type & Molecules & References \\
\hline
\multirow{4}{*}{Superrotors} & $O_2$, $N_2$ & \cite{Korobenko:2014}  \\
 & $CO_2$ & \cite{Yuan:2011} \\
 & $NO_2$ & \cite{Yuan:2011b} \\
 & $Cl_2$ & \cite{Villeneuve:2000}  \\
 & $PDT$ & \cite{Kurban:2013}  \\
\hline
\multirow{7}{*}{Coriolis} & $CH_3CCl_3$ & \cite{Kisiel:2008}  \\
 & $Ne-D_2O$ & \cite{Li:2011}  \\
 & $CH_2CNH$ & \cite{Bane:2011}  \\
 & $C_2H_2D_2$ & \cite{Gabona:2014}  \\
 & $C_2H_3F$ & \cite{Tasinato:2012}  \\
 & $FHF$, $FDF$ & \cite{Sebald:2013}  \\
 & $C_5H_8$, $C_5H_7D$ & \cite{Perry:2012}  \\
\hline
\end{tabular}
\end{center}
\end{table}

The structure of this publication is the following. In Sec.~\ref{Quantum description of a system of $N$ nuclei and $n$ electrons}, we formally describe the dynamics of a system of $N$ nuclei and $n$ electrons. In Sec.~\ref{Internal observables of the molecular system}, we define the internal observables in order to obtain canonical commutations relations in translation and rotation. Sec.~\ref{Dynamical description of a rotating and vibrating molecule} is devoted to the description of the dynamics of the molecular system in terms of the internal observables. Finally, in Sec.~\ref{Molecular ground state and vibrational modes}, we determine the equilibrium conditions that define the molecular ground state and we define explicitly the angular frequency of the vibrational modes of the molecular system.

\section{Quantum description of a system of $N$ nuclei and $n$ electrons}
\label{Quantum description of a system of $N$ nuclei and $n$ electrons}
The quantum dynamics of a molecular system consisting of $N$ nuclei and $n$ electrons is obtained from the classical dynamics by applying the ``correspondence principle''. The Hilbert subspaces describing the nuclei and the electrons are denoted $\mathcal{H}_{\mathcal{N}}$ and $\mathcal{H}_{e}$ respectively. The Hilbert space describing the whole system is expressed as,
\begin{equation}\label{2.1}
\mathcal{H}=\mathcal{H}_{\mathcal{N}}\otimes\mathcal{H}_{e}\,.
\end{equation}
To investigate the molecular kinematics, it is not restrictive to assume that the nuclei are $N$ discernible particles denoted by an index $\mu=1,..,N$. These particles have a mass $M_{\mu}$, an electric charges $Z_{\mu}(-\,e)$ and a spin $S_{\mu}$. The symbol $e$ represents the electronic electric charge including its sign and $Z_{\mu}$ denotes the atomic number of the nucleus $\mu$.

The Hilbert subspace $\mathcal{H}_{e}$ associated to the electrons is assumed to be isomorphic to the tensor product of $n$ one-electron Hilbert spaces, i.e.
\begin{equation*}
\begin{split}
&\mathcal{H}_{e}\sim\big(L^{2}(\mathbb{R}^{3}, d^{3}\boldsymbol{x})\otimes\mathbb{C}^{2}\big)^{\otimes n}\sim L^{2}(\mathbb{R}^{3n}, d^{3n}\boldsymbol{x})\otimes\mathbb{C}^{2n}\,.
\end{split}
\end{equation*}
In fact, the Hilbert spaces describing the electrons are totally antisymmetric subspaces of $\mathcal{H}_{e}$. For a molecular system, the contribution of the overlap integrals between the nuclei is negligible. Thus, we do not need to take into account explicitly the fermionic nature of the nuclei.

The position, momentum and spin observables of the nucleus $\mu$ are characterised respectively by the self-adjoint operators $\boldsymbol{R}_{\mu}\,\otimes\,\mathbb{1}_{e}$, $\boldsymbol{P}_{\mu}\,\otimes\,\mathbb{1}_{e}$ and $\boldsymbol{S}_{\mu}\,\otimes\,\mathbb{1}_{e}$, where $\mu=1,..,N$, acting trivially on the Hilbert subspace $\mathcal{H}_{e}$ associated to the electrons. The components of the operators
$\boldsymbol{R}_{\mu}$ and $\boldsymbol{P}_{\mu}$ acting on the Hilbert subspace $\mathcal{H}_{\mathcal{N}}$ satisfy the canonical commutation relations, i.e.
\begin{equation}\label{2.3}
\left[\ \boldsymbol{e}_{j}\cdot\boldsymbol{P}_{\mu},\ \boldsymbol{e}^{k}\cdot\boldsymbol{R}_{\nu}\ \right] = -\,i\hbar\, \delta_{\mu\nu}\left(\boldsymbol{e}_{j}\cdot\boldsymbol{e}^{k}\right)\mathbb{1}_{\mathcal{N}}\,,
\end{equation}
and the components of the operator $\boldsymbol{S}_{\mu}$ satisfy the canonical commutation relation, i.e.
\begin{equation}\label{2.3 bis}
\left[\ \boldsymbol{e}_{j}\cdot\boldsymbol{S}_{\mu},\ \boldsymbol{e}_{k}\cdot\boldsymbol{S}_{\nu}\ \right] = i\hbar\,\delta_{\mu\nu}\left(\boldsymbol{e}_{j}\times\boldsymbol{e}_{k}\right)\cdot\boldsymbol{S}_{\mu}\,,
\end{equation}
where $\boldsymbol{e}_{j}$ are the units vectors of an orthonormal basis and $\boldsymbol{e}^{k}$ are the units vectors of the dual orthonormal basis. The other commutation relations are trivial, i.e.
\begin{equation}\label{2.4}
\begin{split}
&\left[\ \boldsymbol{e}^{j}\cdot\boldsymbol{R}_{\mu},\ \boldsymbol{e}^{k}\cdot\boldsymbol{R}_{\nu}\ \right] = 0\,,\\
&\left[\ \boldsymbol{e}_{j}\cdot\boldsymbol{P}_{\mu},\ \boldsymbol{e}_{k}\cdot\boldsymbol{P}_{\nu}\ \right] = 0\,,\\
&\left[\ \boldsymbol{e}^{j}\cdot\boldsymbol{R}_{\mu},\ \boldsymbol{e}_{k}\cdot\boldsymbol{S}_{\nu}\ \right] = 0\,,\\
&\left[\ \boldsymbol{e}_{j}\cdot\boldsymbol{P}_{\mu},\ \boldsymbol{e}_{k}\cdot\boldsymbol{S}_{\nu}\ \right] = 0\,.
\end{split} 
\end{equation}
Similarly, the position, momentum and spin observables of the electron $\nu$ are characterised respectively by the self-adjoint operators $\mathbb{1}_{\mathcal{N}}\,\otimes\,\boldsymbol{r}_{\nu}$, $\mathbb{1}_{\mathcal{N}}\,\otimes\,\boldsymbol{p}_{\nu}$ and $\mathbb{1}_{\mathcal{N}}\,\otimes\,\boldsymbol{s}_{\nu}$, where $\nu=1,..,n$, acting trivially on the Hilbert subspace $\mathcal{H}_{\mathcal{N}}$ associated to the nuclei. The components of the operators $\boldsymbol{r}_{\nu}$ and $\boldsymbol{p}_{\nu}$ acting on the Hilbert subspace $\mathcal{H}_{e}$ satisfy the canonical commutation relations, i.e.
\begin{equation}\label{2.6}
\left[\ \boldsymbol{e}_{j}\cdot\boldsymbol{p}_{\mu},\ \boldsymbol{e}^{k}\cdot\boldsymbol{r}_{\nu}\ \right] = -\,i\hbar\, \delta_{\mu\nu}\left(\boldsymbol{e}_{j}\cdot\boldsymbol{e}^{k}\right)\mathbb{1}_{e}\,,
\end{equation}
and the components of the operator $\boldsymbol{s}_{\mu}$ satisfy the canonical commutation relation, i.e.
\begin{equation}\label{2.7}
\left[\ \boldsymbol{e}_{j}\cdot\boldsymbol{s}_{\mu},\ \boldsymbol{e}_{k}\cdot\boldsymbol{s}_{\nu}\ \right] = i\hbar\,\delta_{\mu\nu}\left(\boldsymbol{e}_{j}\times\boldsymbol{e}_{k}\right)\cdot\boldsymbol{s}_{\mu}\,.
\end{equation}
The other commutation relations are trivial, i.e.
\begin{equation}\label{2.8}
\begin{split}
&\left[\ \boldsymbol{e}^{j}\cdot\boldsymbol{r}_{\mu},\ \boldsymbol{e}^{k}\cdot\boldsymbol{r}_{\nu}\ \right] = 0\,,\\
&\left[\ \boldsymbol{e}_{j}\cdot\boldsymbol{p}_{\mu},\ \boldsymbol{e}_{k}\cdot\boldsymbol{p}_{\nu}\ \right] = 0\,,\\
&\left[\ \boldsymbol{e}^{j}\cdot\boldsymbol{r}_{\mu},\ \boldsymbol{e}_{k}\cdot\boldsymbol{s}_{\nu}\ \right] = 0\,,\\
&\left[\ \boldsymbol{e}_{j}\cdot\boldsymbol{p}_{\mu},\ \boldsymbol{e}_{k}\cdot\boldsymbol{s}_{\nu}\ \right] = 0\,.
\end{split} 
\end{equation}

In order to discuss the dynamics of an electrically neutral molecular system composed of $N$ nuclei and $n$ electrons, we implicitly assume that
\begin{equation}\label{2.10}
\sum_{\mu=1}^{N}Z_{\mu}=n\,.
\end{equation}
In a non-relativistic framework, we restrict our analysis to instantaneous electromagnetic interactions between the particles, i.e. the electrons and the nuclei. In this framework, the Hamiltonian governing the evolution reads,
\begin{equation}\label{2.11}
H = H_{\mathcal{N}}\otimes\mathbb{1}_{e} + \mathbb{1}_{\mathcal{N}}\otimes H_{e} + H_{\mathcal{N}-e}\,,
\end{equation}
where the Hamiltonians $H_{\mathcal{N}}$ and $H_{e}$ associated to the nuclei and electrons are defined respectively as,
\begin{equation}\label{2.13}
\begin{split}
&H_{\mathcal{N}}=\sum_{\mu=1}^{N}\frac{\boldsymbol{P}_{\mu}^{2}}{2M_{\mu}} + V_{\mathcal{N}-\mathcal{N}} + V_{\mathcal{N}}^{SO}\,,\\
&H_{e}=\sum_{\nu=1}^{n}\frac{\boldsymbol{p}_{\nu}^{2}}{2m} + V_{e-e} + V_{e}^{SO}\,,
\end{split}
\end{equation}
where $m$ is the mass of an electron, $V_{\mathcal{N}}^{SO}$ is the nuclear spin-orbit coupling due to the interaction between the spin and the orbital angular momentum of the nuclei and $V_{e}^{SO}$ is the electronic spin-orbit coupling due to the interaction between the spin and the orbital angular momentum of the electrons. The Coulomb potentials $V_{\mathcal{N}-\mathcal{N}}$ and $V_{e-e}$ are respectively defined as,
\begin{equation}\label{2.12}
\begin{split}
&V_{\mathcal{N}-\mathcal{N}} = 
\frac{e^{2}}{8\pi\varepsilon_{0}}\sum_{\substack{\mu,\nu=1\\\mu\neq\nu}}^{N}\frac{Z_{\mu}Z_{\nu}}{\Vert\boldsymbol{R}_{\mu}-\boldsymbol{R}_{\nu}\Vert}\,,\\
&V_{e-e} = 
\frac{e^{2}}{8\pi\varepsilon_{0}}\sum_{\substack{\mu,\nu=1\\\mu\neq\nu}}^{n}\frac{1}{\Vert\boldsymbol{r}_{\mu}-\boldsymbol{r}_{\nu}\Vert}\,,
\end{split}
\end{equation}
where $\varepsilon_{0}$ is the vacuum dielectric constant. The interaction Hamiltonian $H_{\mathcal{N}-e}$ appearing in the definition~\eqref{2.11} describes the interaction between the nuclei and the electrons. It is defined as,
\begin{equation}\label{2.19}
H_{\mathcal{N}-e} = V_{\mathcal{N}-e} + V_{\mathcal{N}-e}^{SO}\,.
\end{equation}
The Coulomb potential $V_{\mathcal{N}-e}$ between the nuclei and the electrons is defined as, 
\begin{equation}\label{2.16}
V_{\mathcal{N}-e}=-\frac{e^{2}}{4\pi\varepsilon_{0}}
\sum_{\mu=1}^{N}\sum_{\nu=1}^{n}\frac{Z_{\mu}}{\Vert\boldsymbol{R}_{\mu}\otimes\mathbb{1}_{e}-\,\mathbb{1}_{\mathcal{N}}\otimes\boldsymbol{r}_{\nu}\Vert}\,,
\end{equation}
and $V_{\mathcal{N}-e}^{SO}$ is the spin-orbit coupling due to the interaction between the spin of the electrons and the orbital angular momentum of the nuclei and to the interaction between the spin of the nuclei and the orbital angular momentum of the electrons. As usual in molecular physics, the effects of the magnetic field produced by the motion of the particles are neglected.

\section{Internal observables of the molecular system}
\label{Internal observables of the molecular system}
The description of molecular dynamics in a classical framework would be much simpler than in a quantum framework since in the former a rest frame could be attached easily to the physical system. In quantum physics, the approach is slightly different because observables are described mathematically by operators, which implies that there exists no rest frame and no centre of mass frame associated to the molecular system. However, even in the absence of a centre of mass frame, the position and momentum observables of the centre of mass can be expressed mathematically as self-adjoint operators. This enables us to define other position and momentum observables with respect to the centre of mass. We shall refer to them as ``relative'' position and momentum observables because they are the quantum equivalent of the classical relative position and momentum variables defined with respect to the center of mass frame. Then, using a rotation operator, we define the ``rest'' position and momentum observables, which are the quantum equivalent of the classical position and momentum variables defined in the molecular rest frame. Finally, the ``rest'' position and momentum observables are recast in terms of internal observables characterizing the vibrational, rotational and electronic degrees of freedom.

Applying the correspondence principle, the position, momentum and angular momentum observables associated to the center of mass are respectively given by the self-adjoint operators, 
\begin{align}
\label{3.3}
&\boldsymbol{\mathcal{Q}}=\frac{1}{\mathcal{M}}\left(\ \sum_{\mu=1}^{N} M_{\mu}\,\boldsymbol{R}_{\mu}\otimes\mathbb{1}_{e}+\sum_{\nu=1}^{n}\,\mathbb{1}_{\mathcal{N}}\otimes m\,\boldsymbol{r}_{\nu}\ \right)\,,\\
\label{3.1}
&\boldsymbol{\mathcal{P}}=\sum_{\mu=1}^{N}\boldsymbol{P}_{\mu}\otimes\mathbb{1}_{e}+\sum_{\nu=1}^{n}\,\mathbb{1}_{\mathcal{N}}\otimes\boldsymbol{p}_{\nu}\,,
\end{align}
where $\mathcal{M}$ stands for the total mass of the molecule, i.e.
\begin{equation}\label{3.4}
\mathcal{M} = M + nm\,,
\end{equation}
and $M$ represents the total mass of the nuclei, i.e. 
\begin{equation}\label{3.20pre}
M=\sum_{\mu=1}^{N}M_{\mu}\,.
\end{equation}
The commutation relations~\eqref{2.3} and~\eqref{2.6} imply that the operators $\boldsymbol{\mathcal{P}}$ and $\boldsymbol{\mathcal{Q}}$ satisfy the commutation relations,
\begin{equation}
\label{3.6}
\left[\ \boldsymbol{e}_{j}\cdot\boldsymbol{\mathcal{P}}, \boldsymbol{e}^{k}\cdot\boldsymbol{\mathcal{Q}}\ \right]= -\,i\hbar\left(\boldsymbol{e}_{j}\cdot\boldsymbol{e}^{k}\right)\mathbb{1}\,.
\end{equation}

Now we define the ``relative'' position operators $\boldsymbol{R}^{\prime}_{\mu}$ and $\boldsymbol{r}^{\prime}_{\nu}$, and the ``relative'' momentum operators $\boldsymbol{P}^{\prime}_{\mu}$ and $\boldsymbol{p}^{\prime}_{\nu}$. The ``relative'' position operators $\boldsymbol{R}^{\prime}_{\mu}$ and $\boldsymbol{r}^{\prime}_{\nu}$ are related to the position operators $\boldsymbol{R}_{\mu}$ and $\boldsymbol{r}_{\nu}$ by,
\begin{equation}\label{3.9}
\begin{split}
&\boldsymbol{R}^{\prime}_{\mu}=\boldsymbol{R}_{\mu}\otimes\mathbb{1}_{e}-\boldsymbol{\mathcal{Q}}\,,\\
&\boldsymbol{r}^{\prime}_{\nu}=\mathbb{1}_{\mathcal{N}}\otimes\boldsymbol{r}_{\nu}-\boldsymbol{\mathcal{Q}}\,.
\end{split}
\end{equation}
Similarly, the ``relative'' momentum operators $\boldsymbol{P}^{\prime}_{\mu}$ and $\boldsymbol{p}^{\prime}_{\nu}$ are related to the momentum operators $\boldsymbol{P}_{\mu}$ and $\boldsymbol{p}_{\nu}$ by,  
\begin{equation}\label{3.10}
\begin{split}
&\boldsymbol{P}^{\prime}_{\mu}=\boldsymbol{P}_{\mu}\otimes\mathbb{1}_{e}-\frac{M_{\mu}}{\mathcal{M}}\ \boldsymbol{\mathcal{P}}\,,\\
&\boldsymbol{p}^{\prime}_{\nu}=\mathbb{1}_{\mathcal{N}}\otimes\boldsymbol{p}_{\nu}-\frac{m}{\mathcal{M}}\ \boldsymbol{\mathcal{P}}\,.
\end{split}
\end{equation}
The operators $\boldsymbol{R}^{\prime}_{\mu}$, $\boldsymbol{P}^{\prime}_{\mu}$, $\boldsymbol{r}^{\prime}_{\nu}$ and $\boldsymbol{p}^{\prime}_{\nu}$ commute with the operators $\boldsymbol{\mathcal{Q}}$ and $\boldsymbol{\mathcal{P}}$. In addition, these operators satisfy the condition,
\begin{equation}\label{3.11}
\sum_{\mu=1}^{N}\,M_{\mu}\,\boldsymbol{R}^{\prime}_{\mu}+\sum_{\nu=1}^{n} m\ \boldsymbol{r}^{\prime}_{\nu} = \boldsymbol{0}\,,
\end{equation}
which is a direct consequence of the definitions~\eqref{3.3} and~\eqref{3.9}, and the condition,
\begin{equation}\label{3.12}
\sum_{\mu=1}^{N}\,\boldsymbol{P}^{\prime}_{\mu}+\sum_{\nu=1}^{n}\,\boldsymbol{p}^{\prime}_{\nu} = \boldsymbol{0}\,,
\end{equation}
which is a direct consequence of the definitions~\eqref{3.1} and~\eqref{3.10}. Now, we can determine some further commutation relations. Clearly the components of the position operator  
$\boldsymbol{R}^{\prime}_{\mu}$ commute and the components of the position operator $\boldsymbol{r}^{\prime}_{\nu}$ commute as well. The components of the momenta operators $\boldsymbol{P}^{\prime}_{\mu}$ and $\boldsymbol{p}^{\prime}_{\nu}$ commute likewise, i.e. 
\begin{equation}\label{3.12 prime bis}
\begin{split}
&\left[\ \boldsymbol{e}^{j}\cdot\boldsymbol{R}^{\prime}_{\mu},\  \boldsymbol{e}^{k}\cdot\boldsymbol{R}^{\prime}_{\nu}\ \right] =  0\,,\\
&\left[\ \boldsymbol{e}_{j}\cdot\boldsymbol{P}^{\prime}_{\mu},\  \boldsymbol{e}_{k}\cdot\boldsymbol{P}^{\prime}_{\nu}\ \right] = 0\,,\\
&\left[\ \boldsymbol{e}^{j}\cdot\boldsymbol{r}^{\prime}_{\mu},\  \boldsymbol{e}^{k}\cdot\boldsymbol{r}^{\prime}_{\nu}\ \right] = 0\,,\\
&\left[\ \boldsymbol{e}_{j}\cdot\boldsymbol{p}^{\prime}_{\mu},\  \boldsymbol{e}_{k}\cdot\boldsymbol{p}^{\prime}_{\nu}\ \right] = 0\,.
\end{split}
\end{equation}
Thus, the only non-trivial commutation relations read,  
\begin{align}\label{3.13}
&\left[\ \boldsymbol{e}_j\cdot\boldsymbol{P}^{\prime}_{\mu},\  \boldsymbol{e}^k\cdot\boldsymbol{R}^{\prime}_{\nu}\ \right] =  -\,i\hbar\left(\boldsymbol{e}_j\cdot\boldsymbol{e}^k\right)\left( \delta_{\mu\nu}-\,\frac{M_{\mu}}{\mathcal{M}} \right)\,\mathbb{1}\,,\nonumber\\
&\left[\ \boldsymbol{e}_j\cdot\boldsymbol{p}^{\prime}_{\mu},\  \boldsymbol{e}^k\cdot\boldsymbol{R}^{\prime}_{\nu}\ \right] = i\hbar\left(\boldsymbol{e}_j\cdot\boldsymbol{e}^k\right)\,\frac{m}{\mathcal{M}}\,\mathbb{1}\,,\nonumber\\
&\left[\ \boldsymbol{e}_j\cdot\boldsymbol{P}^{\prime}_{\mu},\  \boldsymbol{e}^k\cdot\boldsymbol{r}^{\prime}_{\nu}\ \right] = i\hbar\left(\boldsymbol{e}_j\cdot\boldsymbol{e}^k\right)\,\frac{M_{\mu}}{\mathcal{M}}\,\mathbb{1}\,,\\
&\left[\ \boldsymbol{e}_j\cdot\boldsymbol{p}^{\prime}_{\mu},\  \boldsymbol{e}^k\cdot\boldsymbol{r}^{\prime}_{\nu}\ \right] = -\,i\hbar\left(\boldsymbol{e}_j\cdot\boldsymbol{e}^k\right)\left( \delta_{\mu\nu}-\,\frac{m}{\mathcal{M}} \right)\,\mathbb{1}\,.\nonumber
\end{align}

Now we define the ``rest'' position operators $\boldsymbol{R}^{\prime\prime}_{\mu}$ and $\boldsymbol{r}^{\prime\prime}_{\nu}$, and the ``rest'' momentum operators $\boldsymbol{P}^{\prime\prime}_{\mu}$ and $\boldsymbol{p}^{\prime\prime}_{\nu}$. These operators are related respectively to the operators ``relative'' $\boldsymbol{R}^{\prime}_{\mu}$, $\boldsymbol{r}^{\prime}_{\nu}$, $\boldsymbol{P}^{\prime}_{\mu}$ and $\boldsymbol{p}^{\prime}_{\nu}$ by a rotation operator $\mathsf{R}\left(\boldsymbol{\omega}\right)$ that is a function of the operator $\boldsymbol{\omega}$ describing the orientation of the molecular system. The rotation operator commutes with the ``rest'' position operators $\boldsymbol{R}^{\prime\prime}_{\mu}$, $\boldsymbol{r}^{\prime\prime}_{\nu}$ and the ``rest'' momentum operator $\boldsymbol{p}^{\prime\prime}_{\nu}$ but not with the ``rest'' momentum operator $\boldsymbol{P}^{\prime\prime}_{\mu}$ as explained in Appendix~\ref{A.4}. The components of the ``rest'' position operators $\boldsymbol{R}^{\prime\prime}_{\mu}$ and $\boldsymbol{r}^{\prime\prime}_{\nu}$, are related to the components of the ``relative'' position operators $\boldsymbol{R}^{\prime}_{\mu}$ and $\boldsymbol{r}^{\prime}_{\nu}$ by,
\begin{equation}\label{3.13bis}
\begin{split}
&\boldsymbol{e}^{j}\cdot\boldsymbol{R}^{\prime\prime}_{\mu} = \left(\boldsymbol{e}^{j}\cdot\mathsf{R}\left(\boldsymbol{\omega}\right)^{-1}\cdot\boldsymbol{e}_{k}\right)\,\left(\boldsymbol{e}^{k}\cdot\boldsymbol{R}^{\prime}_{\mu}\right)\,,\\
&\boldsymbol{e}^{j}\cdot\boldsymbol{r}^{\prime\prime}_{\nu} =  \left(\boldsymbol{e}^{j}\cdot\mathsf{R}\left(\boldsymbol{\omega}\right)^{-1}\cdot\boldsymbol{e}_{k}\right)\,\left(\boldsymbol{e}^{k}\cdot\boldsymbol{r}^{\prime}_{\nu}\right)\,.
\end{split}
\end{equation}
The components of the ``rest'' momentum operators $\boldsymbol{P}^{\prime\prime}_{\mu}$ and $\boldsymbol{p}^{\prime\prime}_{\nu}$ are related to the ``relative'' position operators $\boldsymbol{P}^{\prime}_{\mu}$ and $\boldsymbol{p}^{\prime}_{\nu}$ by,
\begin{equation}\label{3.13ter}
\begin{split}
&\boldsymbol{e}_{j}\cdot\boldsymbol{P}^{\prime\prime}_{\mu} = \frac{1}{2}\,\left\{\ \boldsymbol{e}^{k}\cdot\mathsf{R}\left(\boldsymbol{\omega}\right)\cdot\boldsymbol{e}_{j},\ \boldsymbol{e}_k\cdot\boldsymbol{P}^{\prime}_{\mu}\ \right\}\,,\\
&\boldsymbol{e}_{j}\cdot\boldsymbol{p}^{\prime\prime}_{\nu} = \left(\boldsymbol{e}^{k}\cdot\mathsf{R}\left(\boldsymbol{\omega}\right)\cdot\boldsymbol{e}_{j}\right)\,\left(\boldsymbol{e}_{k}\cdot\boldsymbol{p}^{\prime}_{\nu}\right)\,.
\end{split}
\end{equation}
where the brackets $\{\ ,\ \}$ denote an anticommutator accounting for the fact that the rotation operator $\mathsf{R}\left(\boldsymbol{\omega}\right)$ does not commute with the position operator $\boldsymbol{P}^{\prime}_{\mu}$ of the nuclei.

The self-adjoint molecular orientation operator $\boldsymbol{\omega}$ is fully determined by the position operators $\boldsymbol{R}_{\mu}$ and $\boldsymbol{r}_{\nu}$. Thus, the components of $\boldsymbol{\omega}$ satisfy the trivial commutation relations,
\begin{equation}\label{A.2.0}
\left[\ \boldsymbol{e}^j\cdot\boldsymbol{\omega},\ \boldsymbol{e}^k\cdot\boldsymbol{\omega}\ \right]=0\,.
\end{equation}
The orientation operator $\boldsymbol{\omega}$ belongs to the rotation algebra and it is related to the rotation operator $\mathsf{R}\left(\boldsymbol{\omega}\right)$ that belongs to the rotation group by exponentiation, i.e.
\begin{equation}\label{3.21}
\mathsf{R}\left(\boldsymbol{\omega}\right) =\exp\left(\boldsymbol{\omega}\cdot\boldsymbol{\mathsf{G}}\right)\,,
\end{equation}
taking into account the commutation relation~\eqref{A.2.0} of the components of the orientation operator $\boldsymbol{\omega}$. The elements of the rotation group have to satisfy the orthogonality condition, i.e.
\begin{equation}\label{3.21 bis}
\mathsf{R}\left(\boldsymbol{\omega}\right)^{T}\cdot\mathsf{R}\left(\boldsymbol{\omega}\right) = \mathbb{1}\,,
\end{equation}
which implies that $\mathsf{R}\left(\boldsymbol{\omega}\right)^{T} = \mathsf{R}\left(\boldsymbol{\omega}\right)^{-1}$ and in turn that,
\begin{equation}\label{A.2.1bis}
\boldsymbol{\mathsf{G}}^{\,T} = -\,\boldsymbol{\mathsf{G}}\,.
\end{equation}
where the components of the vector $\boldsymbol{\mathsf{G}}$ are rank-$2$ tensors that are generators of the rotation group acting on $\mathbb{R}^{3}$. The action of the rotation group is locally defined as,
\begin{equation}\label{3.23bis}
\left(\boldsymbol{e}_{j}\cdot\boldsymbol{\mathsf{G}}\right)\,\boldsymbol{x}=\boldsymbol{e}_{j}\times\boldsymbol{x}\,,
\end{equation}
which implies that the generators $\boldsymbol{\mathsf{G}}$ of the rotation in $\mathbb{R}^{3}$ verify the well known commutation relations
\begin{equation}\label{3.23}
\left[\ \boldsymbol{e}_{j}\cdot\boldsymbol{\mathsf{G}},\ \boldsymbol{e}_{k}\cdot\boldsymbol{\mathsf{G}}\ \right] = \left(\boldsymbol{e}_{j}\times\boldsymbol{e}_{k}\right)\cdot\boldsymbol{\mathsf{G}}\,.
\end{equation}

The operators $\boldsymbol{n}_{(j)}\left(\boldsymbol{\omega}\right)$ are Killing vectors~\cite{Szekeres:2004} of the rotation algebra that are defined in terms of the rotation operator $\mathsf{R}\left(\boldsymbol{\omega}\right)$ and the rotation generators as,
\begin{equation}\label{3.38}
\mathsf{R}\left(\boldsymbol{\omega}\right)^{-1}\cdot\left(\boldsymbol{e}_j\cdot\partial_{\boldsymbol{\omega}}\right)\,\mathsf{R}\left(\boldsymbol{\omega}\right) = \boldsymbol{n}_{(j)}\left(\boldsymbol{\omega}\right)\cdot\boldsymbol{\mathsf{G}}\,.
\end{equation}
The dual operator $\boldsymbol{m}^{(k)}\left(\boldsymbol{\omega}\right)$ satisfies the duality condition,
\begin{equation}\label{3.39}
\boldsymbol{n}_{(j)}\left(\boldsymbol{\omega}\right)\cdot\boldsymbol{m}^{(k)}\left(\boldsymbol{\omega}\right) = \boldsymbol{e}_{j}\cdot\boldsymbol{e}^{k}\,.
\end{equation}
As shown in Appendix~\ref{A.2}, the Killing form~\cite{Fulton:1991} associated to the rotation group is given by,
\begin{equation}\label{3.39.K}
\boldsymbol{n}_{(j)}\left(\boldsymbol{\omega}\right)\cdot\boldsymbol{n}_{(\ell)}\left(\boldsymbol{\omega}\right) =
\boldsymbol{e}_j\cdot\Big(\mathsf{P}_{\boldsymbol{\omega}} + A\left(\mathbb{1} -\,\mathsf{P}_{\boldsymbol{\omega}}\right)\Big)\cdot\boldsymbol{e}_{\ell}\,,
\end{equation}
where the projector $\mathsf{P}_{\boldsymbol{\omega}}$ and the scalar $A$ are respectively defined as,
\begin{equation}\label{3.39.def}
\begin{split}
&\mathsf{P}_{\boldsymbol{\omega}} \equiv \frac{\boldsymbol{\omega}\,\boldsymbol{\omega}}{\Vert\boldsymbol{\omega}\Vert^2}\,,\\
&A = \left(\frac{2}{\Vert\boldsymbol{\omega}\Vert}\,\sin\frac{\Vert\boldsymbol{\omega}\Vert}{2}\right)^{2}\,.
\end{split}
\end{equation}
According to the definition~\eqref{3.39.def}, in the limit of an infinitesimal rotation, i.e. $\Vert\boldsymbol{\omega}\Vert \longrightarrow 0$, the scalar $A \longrightarrow 1$, which implies that,
\begin{equation}\label{3.39 bis}
\begin{split}
&\lim_{\Vert\boldsymbol{\omega}\Vert\rightarrow 0}\,\boldsymbol{n}_{(j)}\left(\boldsymbol{\omega}\right) = \boldsymbol{e}_{j}\,\mathbb{1}\,,\\
&\lim_{\Vert\boldsymbol{\omega}\Vert \rightarrow 0}\,\boldsymbol{m}^{(k)}\left(\boldsymbol{\omega}\right) = \boldsymbol{e}^{k}\,\mathbb{1}\,.
\end{split}
\end{equation}
The operators $\boldsymbol{n}_{(j)}\left(\boldsymbol{\omega}\right)$ and $\boldsymbol{m}^{(k)}\left(\boldsymbol{\omega}\right)$ determine the structure of the rotation algebra.

Using the relations~\eqref{3.13bis}, the condition~\eqref{3.11} is recast as,
\begin{equation}\label{3.11bis}
\sum_{\mu=1}^{N}\,M_{\mu}\,\boldsymbol{R}^{\prime\prime}_{\mu} + \sum_{\nu=1}^{n}\,m\,\boldsymbol{r}^{\prime\prime}_{\nu} = \boldsymbol{0}\,.
\end{equation}
Similarly, using the relations~\eqref{3.13ter}, the condition~\eqref{3.12} is recast as,
\begin{equation}\label{3.12bis}
\sum_{\mu=1}^{N}\,\boldsymbol{P}^{\prime\prime}_{\mu}+\sum_{\nu=1}^{n}\,\boldsymbol{p}^{\prime\prime}_{\nu} = \boldsymbol{0}\,.
\end{equation}

Now, we can introduce operators characterizing the internal observables of the quantum molecular system. First, we introduce the scalar operators $Q^{\alpha}$, where $\alpha = 1,..,3N-6$, characterizing the amplitude of the vibrational modes of the $N$ nuclei. Second, we introduce the vectorial operators $\boldsymbol{q}_{(\nu^{\prime})}$ related to the relative position of the electrons respectively. The ``rest'' position operator $\boldsymbol{R}^{\prime\prime}_{\mu}$ is expressed in terms of the scalar operators $Q^{\alpha}$ and the vectorial operators $\boldsymbol{q}_{(\nu^{\prime})}$ as,
\begin{equation}\label{3.14}
\boldsymbol{R}^{\prime\prime}_{\mu} = \boldsymbol{R}^{(0)}_{\mu}\,\mathbb{1}+\frac{1}{\sqrt{M_{\mu}}}\,Q^{\alpha}\,\boldsymbol{X}_{\mu\alpha} -\,\frac{m}{M}\!\sum_{\nu,\nu^{\prime} = 1}^{n}\!A_{\nu\nu^{\prime}}\,\boldsymbol{q}_{(\nu^{\prime})}\,,
\end{equation}
where we used Einstein's implicit summation convention for the vibrational modes $\alpha$. The relation~\eqref{3.14} yields a kinetic coupling between the ``rest'' position operators of the nuclei and electrons. The ``rest'' position operator $\boldsymbol{r}^{\prime\prime}_{\mu}$ is expressed in terms of the position operators $\boldsymbol{q}_{(\nu^{\prime})}$ as,
\begin{equation}\label{3.15}
\boldsymbol{r}^{\prime\prime}_{\nu} = \sum_{\nu^{\prime} = 1}^{n}\,A_{\nu\nu^{\prime}}\,\boldsymbol{q}_{(\nu^{\prime})}\,,
\end{equation}
where the matrix elements $A_{\nu\nu^{\prime}}$ and $A^{-1}_{\nu^{\prime}\nu}$ are defined as,
\begin{equation}\label{3.15.A}
\begin{split}
&A_{\nu\nu^{\prime}} \equiv \delta_{\nu\nu^{\prime}} + \frac{1}{n}\,\left(\sqrt{\frac{M}{\mathcal{M}}}-\,1\right)\,,\\
&A^{-1}_{\nu^{\prime}\nu} \equiv \delta_{\nu^{\prime}\nu} + \frac{1}{n}\,\left(\sqrt{\frac{\mathcal{M}}{M}}-\,1\right)\,.
\end{split}
\end{equation}
Similarly, the ``rest'' momentum operator $\boldsymbol{P}^{\prime\prime}_{\mu}$ is expressed in terms of the scalar operators $P_{\alpha}$, the vectorial operators $\boldsymbol{p}_{(\nu^{\prime})}$ and the angular velocity pseudo-vectorial operator $\boldsymbol{\Omega}$ as,
\begin{equation}\label{3.25}
\boldsymbol{P}^{\prime\prime}_{\mu} = \boldsymbol{\Omega}\times\left(M_{\mu}\,\boldsymbol{R}^{(0)}_{\mu}\right) + \sqrt{M_{\mu}}\,P_{\alpha}\,\boldsymbol{X}^{\alpha}_{\mu}
-\,\frac{M_{\mu}}{M}\!\sum_{\nu,\nu^{\prime} = 1}^{n}\!A_{\nu\nu^{\prime}}\,\boldsymbol{p}_{(\nu^{\prime})}
\end{equation}
The relation~\eqref{3.25} yields a kinetic coupling between the ``rest'' momentum operators of the nuclei and electrons. The ``rest'' momentum operator $\boldsymbol{p}^{\prime\prime}_{\nu}$ is expressed in terms of the momentum operators $\boldsymbol{p}_{(\nu^{\prime})}$ as,
\begin{equation}\label{3.26}
\boldsymbol{p}^{\prime\prime}_{\nu} = \sum_{\nu^{\prime} = 1}^{n}\,A_{\nu\nu^{\prime}}\,\boldsymbol{p}_{(\nu^{\prime})}\,.
\end{equation}
Note that the definition of the matrix elements $A_{\nu\nu^{\prime}}$ is not unique. However, the particular choice made in relation~\eqref{3.15.A} leads to canonical commutation relations between the electronic position operator $\boldsymbol{q}_{(\nu^{\prime})}$ and the electronic momentum operator $\boldsymbol{p}_{(\nu^{\prime})}$.

The vector set $\{\boldsymbol{X}_{\mu\alpha}\}$ is the orthonormal basis characterizing the vibrational modes and the vector set $\{\boldsymbol{X}^{\beta}_{\mu}\}$ is the dual orthonormal basis, i.e.
\begin{equation}\label{3.18}
\sum_{\mu=1}^{N}\,\boldsymbol{X}_{\mu\alpha}\cdot\boldsymbol{X}^{\beta}_{\mu} = \delta^{\beta}_{\alpha}\,.
\end{equation}
The vectors $\boldsymbol{R}^{(0)}_{\mu}$ correspond to the equilibrium configurations of the nuclei. In order for the identities~\eqref{3.14} and~\eqref{3.15} to satisfy the condition~\eqref{3.11bis} and for the identities~\eqref{3.25} and~\eqref{3.26} to satisfy the condition~\eqref{3.12bis}, we need to impose conditions on the vectors $\boldsymbol{R}^{(0)}_{\mu}$ and $\boldsymbol{X}_{\mu\alpha}$. First, we choose the origin of the coordinate system such that it coincides with the center of mass, i.e.
\begin{equation}\label{3.16}
\sum_{\mu=1}^{N} M_{\mu}\,\boldsymbol{R}^{(0)}_{\mu} = \boldsymbol{0}\,.
\end{equation}
Then, we require the deformation modes of the molecule to preserve the momentum, i.e.
\begin{equation}\label{3.17}
\sum_{\mu=1}^{N}\,\sqrt{M_{\mu}}\,\boldsymbol{X}_{\mu\alpha}=\boldsymbol{0}\,.
\end{equation}
We also require the deformation modes of the molecule to preserve the orbital angular momentum, i.e.
\begin{equation}\label{3.17bis}
\sum_{\mu=1}^{N} \sqrt{M_{\mu}}\,\left(\boldsymbol{R}^{(0)}_{\mu}\times\boldsymbol{X}_{\mu\alpha}\right)=\boldsymbol{0}\,.
\end{equation}
The constraints~\eqref{3.16}-\eqref{3.17bis} are known as the Eckart conditions~\cite{Eckart:1935}. Finally, we choose the orientation of the coordinate system such that the inertia tensor of the equilibrium position of the nuclei is diagonal, i.e.
\begin{equation}\label{3.16bis}
\begin{split}
&\sum_{\mu=1}^{N} M_{\mu}\left(\boldsymbol{e}_j\cdot\boldsymbol{R}^{(0)}_{\mu}\right)\left(\boldsymbol{e}_k\cdot\boldsymbol{R}^{(0)}_{\mu}\right) =\\
&\sum_{\mu=1}^{N} M_{\mu}\left(\boldsymbol{e}_j\cdot\boldsymbol{e}_k\right)\left(\boldsymbol{e}_k\cdot\boldsymbol{R}^{(0)}_{\mu}\right)^2\,.
\end{split}
\end{equation}
As shown in Appendix~\ref{A.1}, the first relation~\eqref{3.13bis} and the physical constraints~\eqref{3.16} and~\eqref{3.17bis} determine the rotation operator $\mathsf{R}\left(\boldsymbol{\omega}\right)$, i.e.
\begin{equation}\label{3.39 ter}
\sum_{\mu=1}^{N}\, M_{\mu}\,\boldsymbol{R}^{(0)}_{\mu}\times\left(\mathsf{R}\left(\boldsymbol{\omega}\right)^{-1}\cdot\boldsymbol{R}^{\prime}_{\mu}\right)=\boldsymbol{0}\,.
\end{equation}

To emphasize the physical motivation behind the previous formal development, we consider the classical counterpart of a quantum molecular system. In a classical framework, the classical counterpart of the operatorial relation~\eqref{3.39 ter} determines the rest frame of the molecular system. Moreover, the equilibrium configuration of a molecule is given by a vector set $\{\boldsymbol{R}^{(0)}_{\mu}\}$ describing the position of the nuclei. The condition~\eqref{3.16} implies that the centre of mass of the molecule coincides with the origin of the coordinate system and the condition~\eqref{3.16bis} requires the inertial tensor of this molecule to be diagonal with respect to the coordinate system. 

The set of orthonormal vectors $\{\boldsymbol{X}_{\mu\alpha}\}$ characterize the $3N-6$ normal deformation modes of the molecule and thus account for the vibrations. The condition~\eqref{3.17} implies that the normal deformation modes preserve the momentum of the molecule and the condition~\eqref{3.17bis} requires that these modes also preserve the orbital angular momentum of the molecule. Thus, the deformation modes of the molecule do not generate translations or rotations of the molecule.

The commutation relations~\eqref{3.12 prime bis} and~\eqref{3.13} and the transformation laws~\eqref{3.13bis} and~\eqref{3.13ter} imply that the commutation relations between the operators $\boldsymbol{R}^{\prime\prime}_{\mu}$, $\boldsymbol{r}^{\prime\prime}_{\nu}$ and $\boldsymbol{p}^{\prime\prime}_{\nu}$ are given by,
\begin{align}\label{3.17ter}
&\left[\ \boldsymbol{e}^j\cdot\boldsymbol{R}^{\prime\prime}_{\mu},\  \boldsymbol{e}^k\cdot\boldsymbol{R}^{\prime\prime}_{\nu}\ \right] = 0\,,\nonumber\\	
&\left[\ \boldsymbol{e}^j\cdot\boldsymbol{r}^{\prime\prime}_{\mu},\  \boldsymbol{e}^k\cdot\boldsymbol{r}^{\prime\prime}_{\nu}\ \right] = 0\,,\nonumber\\
&\left[\ \boldsymbol{e}_j\cdot\boldsymbol{p}^{\prime\prime}_{\mu},\  \boldsymbol{e}_k\cdot\boldsymbol{p}^{\prime\prime}_{\nu}\ \right] = 0\,,\\
&\left[\ \boldsymbol{e}^j\cdot\boldsymbol{r}^{\prime\prime}_{\mu},\  \boldsymbol{e}^k\cdot\boldsymbol{R}^{\prime\prime}_{\nu}\ \right] = 0\,,\nonumber\\
&\left[\ \boldsymbol{e}_j\cdot\boldsymbol{p}^{\prime\prime}_{\mu},\  \boldsymbol{e}^k\cdot\boldsymbol{R}^{\prime\prime}_{\nu}\ \right] = i\hbar\left(\boldsymbol{e}_j\cdot\boldsymbol{e}^k\right)\,\frac{m}{\mathcal{M}}\,\mathbb{1}\,,\nonumber\\
&\left[\ \boldsymbol{e}_j\cdot\boldsymbol{p}^{\prime\prime}_{\mu},\  \boldsymbol{e}^k\cdot\boldsymbol{r}^{\prime\prime}_{\nu}\ \right] = -\,i\hbar\left(\boldsymbol{e}_j\cdot\boldsymbol{e}^k\right)\,\left(\delta_{\mu\nu}-\,\frac{m}{\mathcal{M}}\right)\,\mathbb{1}\,.\nonumber
\end{align}
The kinetic couplings~\eqref{3.14} and~\eqref{3.25} between the ``rest'' position and momentum operators of the nuclei and electrons lead to quantum deviations in the commutation relations~\eqref{3.17ter}, characterised by the mass ratio $m/\mathcal{M}$. These deviations are larger for smaller molecules. For example, for a $\text{H}_{3}^{+}$ molecule~\cite{Furtenbacher:2013} : $m/\mathcal{M} = 2\cdot 10^{-4}$.

As shown in Appendix~\ref{A.4}, using the physical identity~\eqref{3.39 ter} defining the rotation operator, the mathematical identity~\eqref{3.38} associated to the action of the rotation group, the commutation relation~\eqref{3.12 prime bis} and~\eqref{3.13} and the transformation laws~\eqref{3.13bis} and~\eqref{3.13ter}, the commutation relations between the operators $\boldsymbol{R}^{\prime\prime}_{\mu}$, $\boldsymbol{P}^{\prime\prime}_{\mu}$, $\boldsymbol{r}^{\prime\prime}_{\nu}$ and $\boldsymbol{p}^{\prime\prime}_{\nu}$ are found to be,
\begin{align}\label{3.17quad}
&\left[\ \boldsymbol{e}_j\cdot\boldsymbol{P}^{\prime\prime}_{\mu},\  \boldsymbol{e}^k\cdot\boldsymbol{r}^{\prime\prime}_{\nu}\ \right] = i\hbar\,\left(\boldsymbol{e}_{j}\cdot\boldsymbol{e}^{k}\right)\,\frac{M_{\mu}}{\mathcal{M}}\,\mathbb{1}\nonumber\\
&-\,\left[\ \boldsymbol{e}_{j}\cdot\boldsymbol{P}^{\prime\prime}_{\mu},\ \boldsymbol{e}^{\ell}\cdot\boldsymbol{\omega}\ \right]\,\boldsymbol{e}^{k}\cdot\left(\boldsymbol{n}_{(\ell)}\left(\boldsymbol{\omega}\right)\times\boldsymbol{r}^{\prime\prime}_{\nu}\right)\,,\nonumber\\
&\nonumber\\
&\left[\ \boldsymbol{e}_j\cdot\boldsymbol{P}^{\prime\prime}_{\mu},\  \boldsymbol{e}_k\cdot\boldsymbol{p}^{\prime\prime}_{\nu}\ \right] =\nonumber\\
&\left[\ \boldsymbol{e}_{j}\cdot\boldsymbol{P}^{\prime\prime}_{\mu},\ \boldsymbol{e}^{\ell}\cdot\boldsymbol{\omega}\ \right]\,\boldsymbol{e}_{k}\cdot\left(\boldsymbol{n}_{(\ell)}\left(\boldsymbol{\omega}\right)\times\boldsymbol{p}^{\prime\prime}_{\nu}\right)\,,\nonumber\\
&\\
&\left[\ \boldsymbol{e}_j\cdot\boldsymbol{P}^{\prime\prime}_{\mu},\  \boldsymbol{e}^k\cdot\boldsymbol{R}^{\prime\prime}_{\nu}\ \right] = -\,i\hbar\left(\boldsymbol{e}_j\cdot\boldsymbol{e}^k\right)\,\left(\delta_{\mu\nu}-\,\frac{M_{\mu}}{\mathcal{M}}\right)\,\mathbb{1}\nonumber\\
&-\,\left[\ \boldsymbol{e}_{j}\cdot\boldsymbol{P}^{\prime\prime}_{\mu},\ \boldsymbol{e}^{\ell}\cdot\boldsymbol{\omega}\ \right]\,\boldsymbol{e}^{k}\cdot\left(\boldsymbol{n}_{(\ell)}\left(\boldsymbol{\omega}\right)\times\boldsymbol{R}^{\prime\prime}_{\nu}\right)\,,\nonumber\\
&\nonumber\\
&\left[\ \boldsymbol{e}_j\cdot\boldsymbol{P}^{\prime\prime}_{\mu},\  \boldsymbol{e}_k\cdot\boldsymbol{P}^{\prime\prime}_{\nu}\ \right] = \nonumber\\
&\frac{1}{2}\Big\{\ \boldsymbol{e}_{\ell}\cdot\boldsymbol{P}^{\prime\prime}_{\nu},\ 
\left[\ \boldsymbol{e}_j\cdot\boldsymbol{P}^{\prime\prime}_{\mu},\ \boldsymbol{e}^{m}\cdot\boldsymbol{\omega}\ \right]\,\left(\boldsymbol{n}_{(m)}\left(\boldsymbol{\omega}\right)\times\boldsymbol{e}_{k}\right)\cdot\boldsymbol{e}^{\ell}\ \Big\}\nonumber\\
&-\frac{1}{2}\,\Big\{\,\boldsymbol{e}_{\ell}\cdot\boldsymbol{P}^{\prime\prime}_{\mu},\  \left[\ \boldsymbol{e}_k\cdot\boldsymbol{P}^{\prime\prime}_{\nu},\ \boldsymbol{e}^{m}\cdot\boldsymbol{\omega}\ \right]\,\left(\boldsymbol{n}_{(m)}\left(\boldsymbol{\omega}\right)\times\boldsymbol{e}_{j}\right)\cdot\boldsymbol{e}^{\ell}\,\Big\}\nonumber
\end{align}
The kinetic couplings~\eqref{3.14} and~\eqref{3.25} between the ``rest'' position and momentum operators of the nuclei and electrons lead to quantum deviation in the commutation relations~\eqref{3.17quad}, characterised by the mass ratio $M_\mu/\mathcal{M}$. These deviations are larger for smaller molecules. For example, for a $\text{H}_{3}^{+}$ molecule~\cite{Furtenbacher:2013} : $M_{\mu}/\mathcal{M} = 0.33$. Moreover, the fact that molecular rotations are treated in a genuine quantum framework leads to other quantum deviations in the commutation relations~\eqref{3.17quad}. These deviations are proportional to the commutator of the ``rest'' momentum operator $\boldsymbol{P}^{\prime\prime}_{\nu}$ and the molecular orientation operator $\boldsymbol{\omega}$.

The internal observables are described by the scalar operators $Q^{\alpha}$, $P_{\alpha}$, the vectorial operators $\boldsymbol{q}_{(\nu)}$, $\boldsymbol{p}_{(\nu)}$ and the pseudo-vectorial operators $\boldsymbol{\Omega}$ and $\boldsymbol{\omega}$. As shown in Appendix~\ref{A.5}, the inversion of the definitions~\eqref{3.14},~\eqref{3.15},~\eqref{3.25},~\eqref{3.26} yields explicit expressions for the internal observables $Q^{\alpha}$, $P_{\alpha}$, $\boldsymbol{q}_{(\nu)}$ and $\boldsymbol{p}_{(\nu)}$, i.e.
\begin{equation}\label{3.27 prime}
\begin{split}
&Q^{\alpha} = \sum_{\mu=1}^{N}\,\sqrt{M_{\mu}}\, \boldsymbol{X}^{\alpha}_{\mu}\cdot\Big(\boldsymbol{R}^{\prime\prime}_{\mu}-\,\boldsymbol{R}^{(0)}_{\mu}\,\mathbb{1}\Big)\,,\\
&P_{\alpha} = \sum_{\mu=1}^{N}\frac{1}{\sqrt{M_{\mu}}}\,\left(\boldsymbol{X}_{\mu\alpha}\cdot\boldsymbol{P}^{\prime\prime}_{\mu}\right)\,,\\
&\boldsymbol{q}_{(\nu)} = \sum_{\nu^{\prime}=1}^{n}\left(\delta_{\nu\nu^{\prime}} + \frac{1}{n}\,\left(\sqrt{\frac{\mathcal{M}}{M}}-\,1\right)\right)\,\boldsymbol{r}^{\prime\prime}_{\nu^{\prime}}\,,\\
&\boldsymbol{p}_{(\nu)} = \sum_{\nu^{\prime}=1}^{n}\left(\delta_{\nu\nu^{\prime}} + \frac{1}{n}\,\left(\sqrt{\frac{\mathcal{M}}{M}}-\,1\right)\right)\,\boldsymbol{p}^{\prime\prime}_{\nu^{\prime}}\,.
\end{split}
\end{equation}

The components of the inertia tensorial operator $\mathsf{I}\left(Q^{\,\boldsymbol{.}}\right)$ are defined as,
\begin{equation}\label{3.31}
\boldsymbol{e}_{k}\cdot\mathsf{I}\left(Q^{\,\boldsymbol{.}}\right)\cdot\boldsymbol{e}_{\ell} = \left(\boldsymbol{e}_{k}\cdot\mathsf{I}_{0}\cdot\boldsymbol{e}_{\ell}\right)\,\mathbb{1} + Q^{\alpha}\,\left(\boldsymbol{e}_{k}\cdot\mathsf{I}_{\alpha}\cdot\boldsymbol{e}_{\ell}\right)\,,
\end{equation}
where the dot in the argument of the operator $\mathsf{I}\left(Q^{\,\boldsymbol{.}}\right)$ refers to all the vibrational modes. The first term on the RHS of the definition~\eqref{3.31}, i.e.
\begin{align}\label{3.32}
&\boldsymbol{e}_{k}\cdot\mathsf{I}_{0}\cdot\boldsymbol{e}_{\ell} = \sum_{\mu=1}^{N}\,M_{\mu}\, \left(\boldsymbol{e}_{k}\times\boldsymbol{R}^{(0)}_{\mu}\right)\cdot\left(\boldsymbol{e}_{\ell}\times\boldsymbol{R}^{(0)}_{\mu}\right)\\
&= \sum_{\mu=1}^{N}\,M_{\mu}\,\left( {\boldsymbol{R}^{(0)}_{\mu}}^2\,\left(\boldsymbol{e}_{k}\cdot\boldsymbol{e}_{\ell}\right)-\,\left(\boldsymbol{e}_{k}\cdot\boldsymbol{R}^{(0)}_{\mu}\right)\left(\boldsymbol{e}_{\ell}\cdot\boldsymbol{R}^{(0)}_{\mu}\right)\right)\,,\nonumber
\end{align}
is required to be diagonal with respect to the rotating molecular system according to the constraint~\eqref{3.16bis}, i.e.
\begin{equation}\label{3.32bis}
\begin{split}
&\boldsymbol{e}_{k}\cdot\mathsf{I}_{0}\cdot\boldsymbol{e}_{\ell} = \left(\boldsymbol{e}_{k}\cdot\mathsf{I}_{0}\cdot\boldsymbol{e}_{k}\right)\left(\boldsymbol{e}_{k}\cdot\boldsymbol{e}_{\ell}\right)\\
&=\sum_{\mu=1}^{N}\,M_{\mu}\,\left({\boldsymbol{R}^{(0)}_{\mu}}^{2}-\,\left(\boldsymbol{e}_{k}\cdot\boldsymbol{R}^{(0)}_{\mu}\right)^{2}\right)\left(\boldsymbol{e}_{k}\cdot\boldsymbol{e}_{\ell}\right)\,,
\end{split}
\end{equation}
and the term on the RHS, i.e.
\begin{equation}\label{3.33}
\boldsymbol{e}_{k}\cdot\mathsf{I}_{\alpha}\cdot\boldsymbol{e}_{\ell} = \sum_{\mu=1}^{N}\,\sqrt{M_{\mu}}\, \left(\boldsymbol{e}_{k}\times\boldsymbol{X}_{\mu\alpha}\right)\cdot\left(\boldsymbol{e}_{\ell}\times\boldsymbol{R}^{(0)}_{\mu}\right)\,,
\end{equation}
is shown in Appendix~\ref{A.6} to be symmetric, i.e.
\begin{equation}\label{3.30}
\boldsymbol{e}_{k}\cdot\mathsf{I}_{\alpha}\cdot\boldsymbol{e}_{\ell} = \boldsymbol{e}_{\ell}\cdot\mathsf{I}_{\alpha}\cdot\boldsymbol{e}_{k}\,.
\end{equation}

As shown in detail in Appendix~\ref{A.6}, the commutation relations between the operators $Q^{\alpha}$, $P_{\alpha}$, $\boldsymbol{q}_{(\nu)}$, $\boldsymbol{p}_{(\nu)}$, $\boldsymbol{\omega}$, $\boldsymbol{\Omega}$ accounting for internal degrees of freedom are determined using the commutation relations~\eqref{3.17ter} and~\eqref{3.17quad}, the definitions~\eqref{3.27 prime} and~\eqref{3.15.A}, and the constraints~\eqref{3.18}-\eqref{3.17bis}.

The operators $Q^{\alpha}$, $\boldsymbol{q}_{(\nu)}$ and $\boldsymbol{\omega}$ related to the molecular configurations commute, i.e.
\begin{equation}\label{3.30.1}
\begin{split}
&\left[\ Q^{\alpha},\ Q^{\beta}\ \right] = 0\,,\\
&\left[\ Q^{\alpha},\ \boldsymbol{e}^{k}\cdot\boldsymbol{q}_{(\nu^{\prime})}\ \right] = 0\,,\\
&\left[\ Q^{\alpha},\ \boldsymbol{e}^{k}\cdot\boldsymbol{\omega}\ \right] = 0\,,\\
&\left[\ \boldsymbol{e}^{j}\cdot\boldsymbol{q}_{(\nu)},\ \boldsymbol{e}^{k}\cdot\boldsymbol{q}_{(\nu^{\prime})}\ \right] = 0\,,\\
&\left[\ \boldsymbol{e}^{j}\cdot\boldsymbol{q}_{(\nu)},\ \boldsymbol{e}^{k}\cdot\boldsymbol{\omega}\ \right] = 0\,.
\end{split}
\end{equation}
The other non-canonical commutation relations vanish as well, i.e.
\begin{equation}\label{3.31.1}
\begin{split}
&\left[\ P_{\alpha},\ P_{\beta}\ \right] = 0\,,\\
&\left[\ Q^{\alpha},\ \boldsymbol{e}_{k}\cdot\boldsymbol{p}_{(\nu^{\prime})}\ \right] = 0\,,\\
&\left[\ P_{\alpha},\ \boldsymbol{e}_{k}\cdot\boldsymbol{p}_{(\nu^{\prime})}\ \right] = 0\,,\\
&\left[\ P_{\alpha},\ \boldsymbol{e}^{k}\cdot\boldsymbol{q}_{(\nu^{\prime})}\ \right] = 0\,,\\
&\left[\ P_{\alpha},\ \boldsymbol{e}^{k}\cdot\boldsymbol{\omega}\ \right] = 0\,,\\
&\left[\ \boldsymbol{e}_{j}\cdot\boldsymbol{p}_{(\nu)},\ \boldsymbol{e}_{k}\cdot\boldsymbol{p}_{(\nu^{\prime})}\ \right] = 0\,,\\
&\left[\ \boldsymbol{e}_{j}\cdot\boldsymbol{p}_{(\nu)},\ \boldsymbol{e}^{k}\cdot\boldsymbol{\omega}\ \right] = 0\,.
\end{split}
\end{equation}
The canonical commutation relations are given by,
\begin{equation}\label{3.32.1}
\begin{split}
&\left[\ P_{\alpha},\ Q^{\beta}\ \right] = -\,i\hbar\,\delta^{\beta}_{\alpha}\,\mathbb{1}\,,\\
&\left[\ \boldsymbol{e}_{j}\cdot\boldsymbol{p}_{(\nu)},\ \boldsymbol{e}^{k}\cdot\boldsymbol{q}_{(\nu^{\prime})}\ \right] = -\,i\hbar\,\left(\boldsymbol{e}_{j}\cdot\boldsymbol{e}^{k}\right)\,\delta_{\nu\nu^{\prime}}\,\mathbb{1}\,.
\end{split}
\end{equation}
The operator $\boldsymbol{\Omega}$ does not commute with the operators $Q^{\alpha}$, $P_{\alpha}$, $\boldsymbol{q}_{(\nu)}$, $\boldsymbol{p}_{(\nu)}$ and $\boldsymbol{\omega}$, i.e.
\begin{align}\label{3.33.1}
&\left[\ \boldsymbol{e}^{j}\cdot\boldsymbol{\Omega},\ Q^{\alpha}\ \right] =\nonumber\\
&-\,i\hbar\sum_{\mu=1}^{N}\,\left(\boldsymbol{e}^{j}\cdot\mathsf{I}\left(Q^{\,\boldsymbol{.}}\right)^{-1}\cdot\left(\boldsymbol{X}^{\alpha}_{\mu}\times\boldsymbol{X}_{\mu\beta}\right)\,Q^{\beta}\right)\,,\nonumber\\
&\nonumber\\
&\left[\ \boldsymbol{e}^{j}\cdot\boldsymbol{\Omega},\ P_{\alpha}\ \right] =\nonumber\\
&-\,\frac{1}{2}\,i\hbar\sum_{\mu=1}^{N}\,\left[\ \boldsymbol{e}^{j}\cdot\mathsf{I}\left(Q^{\,\boldsymbol{.}}\right)^{-1}\cdot\boldsymbol{X}_{\mu\alpha},\ P_{\beta}\,\boldsymbol{X}^{\beta}_{\mu}\ \right]_{\times}\,,\nonumber\\
&\nonumber\\
&\left[\ \boldsymbol{e}^{j}\cdot\boldsymbol{\Omega},\ \boldsymbol{e}^{k}\cdot\boldsymbol{q}_{(\nu)}\ \right] =\nonumber\\
&-\,i\hbar\left(\boldsymbol{e}^{j}\cdot\mathsf{I}\left(Q^{\,\boldsymbol{.}}\right)^{-1}\cdot\boldsymbol{e}^{\ell}\right)\left(\boldsymbol{e}_{\ell}\times\boldsymbol{e}^{k}\right)\cdot\boldsymbol{q}_{(\nu)}\,,\\
&\nonumber\\
&\left[\ \boldsymbol{e}^{j}\cdot\boldsymbol{\Omega},\ \boldsymbol{e}_{k}\cdot\boldsymbol{p}_{(\nu)}\ \right] =\nonumber\\
&-\,i\hbar\left(\boldsymbol{e}^{j}\cdot\mathsf{I}\left(Q^{\,\boldsymbol{.}}\right)^{-1}\cdot\boldsymbol{e}^{\ell}\right)\left(\boldsymbol{e}_{\ell}\times\boldsymbol{e}_{k}\right)\cdot\boldsymbol{p}_{(\nu)}\,,\nonumber\\
&\nonumber\\
&\left[\ \boldsymbol{e}^{j}\cdot\boldsymbol{\Omega},\ \boldsymbol{e}^k\cdot\boldsymbol{\omega}\ \right] = -\,i\hbar\,\left(\boldsymbol{e}^{j}\cdot\mathsf{I}\left(Q^{\,\boldsymbol{.}}\right)^{-1}\cdot\boldsymbol{m}^{(k)}\left(\boldsymbol{\omega}\right)\right)\,,\nonumber
\end{align}
where we used the notation convention,
\begin{equation}\label{3.26.N}
\left[\ \boldsymbol{A},\ \boldsymbol{B}\ \right]_{\boldsymbol{\times}} = \boldsymbol{A}\times\boldsymbol{B} -\,\boldsymbol{B}\times\boldsymbol{A}\,.
\end{equation}

According to the commutation relations~\eqref{3.33.1}, the angular velocity operator $\boldsymbol{\Omega}$ does not commute with the other internal observables. Thus, it is not a suitable observable for a quantum description of molecular rotation. Therefore, we introduce the angular momentum operator $\boldsymbol{L}^{\prime}$ defined as,
\begin{equation}\label{3.17pet}
\boldsymbol{L}^{\prime} = \sum_{\mu=1}^{N}\,\boldsymbol{R}^{\prime}_{\mu}\times\boldsymbol{P}^{\prime}_{\mu} + \sum_{\nu=1}^{n}\,\boldsymbol{r}^{\prime}_{\nu}\times\boldsymbol{p}^{\prime}_{\nu}\,,
\end{equation}
and the angular momentum operator $\boldsymbol{L}$ defined as,
\begin{equation}\label{3.26bis}
\boldsymbol{L} = \frac{1}{2}\,\sum_{\mu=1}^{N}\,\left[\ \boldsymbol{R}^{\prime\prime}_{\mu},\ \boldsymbol{P}^{\prime\prime}_{\mu}\ \right]_{\boldsymbol{\times}} + \frac{1}{2}\,\sum_{\nu=1}^{n}\,\left[\ \boldsymbol{r}^{\prime\prime}_{\nu},\ \boldsymbol{p}^{\prime\prime}_{\nu}\ \right]_{\boldsymbol{\times}}\,,
\end{equation}
where we used the notation convention~\eqref{3.26.N}.

As shown in Appendix~\ref{A.7}, using the definitions~\eqref{3.14}-\eqref{3.26},~\eqref{3.31}-\eqref{3.30} and the notation conventions~\eqref{3.26.N} and,
\begin{equation}\label{3.27.N}
\left\{\ \boldsymbol{A},\ \boldsymbol{B}\ \right\}_{\mathsmaller{\bullet}} = \boldsymbol{A}\cdot\boldsymbol{B} + \boldsymbol{B}\cdot\boldsymbol{A}\,,
\end{equation}
the orbital angular momentum~\eqref{3.26bis} is recast as,
\begin{align}\label{3.27}
&\boldsymbol{L} = \frac{1}{2}\,\Big\{\ \mathsf{I}\left(Q^{\,\boldsymbol{.}}\right),\ \boldsymbol{\Omega}\ \Big\}_{\mathsmaller{\bullet}} + \frac{1}{2}\,\sum_{\mu = 1}^{N}\,\left[\ Q^{\alpha}\,\boldsymbol{X}_{\mu\alpha},\ P_{\beta}\,\boldsymbol{X}^{\beta}_{\mu}\ \right]_{\times}\nonumber\\
&\phantom{\boldsymbol{L} =} + \frac{1}{2}\,\sum_{\nu = 1}^{n}\,\left[\ \boldsymbol{q}_{(\nu)},\ \boldsymbol{p}_{(\nu)}\ \right]_{\boldsymbol{\times}}\,,
\end{align}
which is a self-adjoint operator. Since the orbital angular momentum operator commutes with the position and momentum operators, as shown explicitly below, it is convenient to recast the angular rotation rate $\boldsymbol{\Omega}$ in terms of $\boldsymbol{L}$. In order to do so, we define the molecular orbital angular momentum $\boldsymbol{\mathcal{L}}$, the deformation orbital angular momentum $\boldsymbol{\mathscr{L}}$ and the electronic orbital angular momentum $\boldsymbol{\ell}$ respectively as,
\begin{equation}\label{3.29}
\begin{split}
&\boldsymbol{\mathcal{L}} = \frac{1}{2}\,\Big\{\ \mathsf{I}\left(Q^{\,\boldsymbol{.}}\right),\ \boldsymbol{\Omega}\ \Big\}_{\mathsmaller{\bullet}}\,,\\
&\boldsymbol{\mathscr{L}} = \frac{1}{2}\,\sum_{\mu = 1}^{N}\,\left[\ Q^{\alpha}\,\boldsymbol{X}_{\mu\alpha},\ P_{\beta}\,\boldsymbol{X}^{\beta}_{\mu}\ \right]_{\times}\,,\\
&\boldsymbol{\ell} = \frac{1}{2}\,\sum_{\nu = 1}^{n}\,\left[\ \boldsymbol{q}_{(\nu)},\ \boldsymbol{p}_{(\nu)}\ \right]_{\boldsymbol{\times}}\,.
\end{split}
\end{equation}

Using the definitions~\eqref{3.29} and the fact that the inertia tensor $\mathsf{I}\left(Q^{\,\boldsymbol{.}}\right)$ commutes with the operators $\boldsymbol{L}$, $\boldsymbol{q}_{(\nu)}$ and $\boldsymbol{p}_{(\nu)}$, the inversion of the expression~\eqref{3.27} yields the rotation rate, i.e.
\begin{equation}\label{3.29ter}
\boldsymbol{\Omega} = \frac{1}{2}\,\left\{\ \mathsf{I}\left(Q^{\,\boldsymbol{.}}\right)^{-1},\ \boldsymbol{\mathcal{L}}\ \right\}_{\mathsmaller{\bullet}}\,,
\end{equation}
where
\begin{equation}\label{3.29quad}
\boldsymbol{\mathcal{L}} = \boldsymbol{L}-\,\boldsymbol{\mathscr{L}}-\,\boldsymbol{\ell}\,.
\end{equation}

As shown in Appendix~\ref{A.8}, the commutation relations involving the orbital angular momentum operator $\boldsymbol{L}$ are obtained using the expression~\eqref{3.27} and the commutation relations~\eqref{3.30.1}.

The orbital angular momentum operator $\boldsymbol{L}$ commutes with the configuration and momentum operators $Q^{\alpha}$, $P_{\alpha}$, $\boldsymbol{q}_{(\nu)}$ and $\boldsymbol{p}_{(\nu)}$, i.e.
\begin{equation}\label{3.34.1}
\begin{split}
&\left[\ \boldsymbol{L},\ Q^{\alpha}\ \right] = \boldsymbol{0}\,,\\
&\left[\ \boldsymbol{L},\ P_{\alpha}\ \right] = \boldsymbol{0}\,,\\
&\left[\ \boldsymbol{L},\ \boldsymbol{e}^{k}\cdot\boldsymbol{q}_{(\nu)}\ \right] = \boldsymbol{0}\,,\\
&\left[\ \boldsymbol{L},\ \boldsymbol{e}_{k}\cdot\boldsymbol{p}_{(\nu)}\ \right] = \boldsymbol{0}\,.
\end{split}
\end{equation}
The orbital angular momentum operator $\boldsymbol{L}$ does not commute with the operators $\boldsymbol{\omega}$, $\boldsymbol{\Omega}$ and $\boldsymbol{L}$, i.e.
\begin{align}\label{3.35.1}
&\left[\ \boldsymbol{L},\ \boldsymbol{e}^{j}\cdot\boldsymbol{\omega}\ \right] = -\,i\hbar\ \boldsymbol{m}^{(j)}\left(\boldsymbol{\omega}\right)\,,\nonumber\\
&\left[\ \boldsymbol{L},\ \boldsymbol{e}^{j}\cdot\boldsymbol{\Omega}\ \right] = i\hbar\,\left(\boldsymbol{e}^{j}\cdot\mathsf{I}\left(Q^{\,\boldsymbol{.}}\right)^{-1}\cdot\boldsymbol{e}^{k}\right)\left(\boldsymbol{e}_{k}\times\boldsymbol{L}\right)\,,\nonumber\\
&\left[\ \boldsymbol{L},\ \boldsymbol{e}_{j}\cdot\boldsymbol{L}\ \right] = i\hbar\,\left(\boldsymbol{e}_{j}\times\boldsymbol{L}\right)\,.
\end{align}
The third commutation relation~\eqref{3.35.1} implies that,
\begin{equation}\label{3.35.4}
\begin{split}
&\left[\ \boldsymbol{L}^2,\ \boldsymbol{e}_{j}\cdot\boldsymbol{L}\ \right] = 0\,,\\
&\left[\ \boldsymbol{e}_{j}\cdot\boldsymbol{L},\ \boldsymbol{e}_{k}\cdot\boldsymbol{L}\ \right] = i\hbar\,\delta_{\mu\nu}\left(\boldsymbol{e}_{j}\times\boldsymbol{e}_{k}\right)\cdot\boldsymbol{L}\,,
\end{split}
\end{equation}
as expected. Finally, the property~\eqref{3.39} and the commutation relation~\eqref{3.35.1} imply that the canonical commutation relations for a quantum rotation are given by,
\begin{equation}\label{3.36.1}
\left[\ \boldsymbol{n}_{(j)}\left(\boldsymbol{\omega}\right)\cdot\boldsymbol{L},\ \boldsymbol{e}^{k}\cdot{\boldsymbol{\omega}}\ \right] = -\,i\hbar\,\left(\boldsymbol{e}_{j}\cdot\boldsymbol{e}^{k}\right)\,.
\end{equation}

The internal spin observables are the nuclear spin operator $\boldsymbol{S}_{(\mu)}$ and the electronic spin operator $\boldsymbol{s}_{(\nu)}$ that are respectively defined as,
\begin{equation}\label{3.36bis}
\begin{split}
&\boldsymbol{e}_{j}\cdot\boldsymbol{S}_{(\mu)} = \left(\boldsymbol{e}^{k}\cdot\mathsf{R}\left(\boldsymbol{\omega}\right)\cdot\boldsymbol{e}_{j}\right)\,\boldsymbol{e}_{k}\cdot\boldsymbol{S}_{\mu}\,,\\
&\boldsymbol{e}_{j}\cdot\boldsymbol{s}_{(\nu)} = \left(\boldsymbol{e}^{k}\cdot\mathsf{R}\left(\boldsymbol{\omega}\right)\cdot\boldsymbol{e}_{j}\right)\,\boldsymbol{e}_{k}\cdot\boldsymbol{s}_{\nu}\,.
\end{split}
\end{equation}
As shown in Appendix~\ref{A.9} the definitions~\eqref{3.36bis}, the commutation relations~\eqref{2.3 bis} and~\eqref{2.7}, and the fact that the spins commute with the molecular orientation observable $\boldsymbol{\omega}$ imply that,
\begin{equation}\label{3.36ter}
\begin{split}
&\left[\ \boldsymbol{e}_{j}\cdot\boldsymbol{S}_{(\mu)},\ \boldsymbol{e}_{k}\cdot\boldsymbol{S}_{(\nu)}\ \right] = i\hbar\,\delta_{\mu\nu}\left(\boldsymbol{e}_{j}\times\boldsymbol{e}_{k}\right)\cdot\boldsymbol{S}_{(\mu)}\,,\\
&\left[\ \boldsymbol{e}_{j}\cdot\boldsymbol{s}_{(\mu)},\ \boldsymbol{e}_{k}\cdot\boldsymbol{s}_{(\nu)}\ \right] = i\hbar\,\delta_{\mu\nu}\left(\boldsymbol{e}_{j}\times\boldsymbol{e}_{k}\right)\cdot\boldsymbol{s}_{(\mu)}\,.
\end{split}
\end{equation}
The operators $Q^{\alpha}$, $\boldsymbol{q}_{(\nu)}$ and $\boldsymbol{\omega}$ commute with the nuclear spin operators $\boldsymbol{S}_{(\mu)}$, i.e.
\begin{align}
\label{3.43}
&\left[\ \boldsymbol{S}_{(\mu)},\ \boldsymbol{e}^{j}\cdot\boldsymbol{q}_{(\nu)}\ \right] = \boldsymbol{0}\,,\\
\label{3.44}
&\left[\ \boldsymbol{S}_{(\mu)},\ Q^{\alpha}\right] = \boldsymbol{0}\,,\\
\label{3.45}
&\left[\ \boldsymbol{S}_{(\mu)},\ \boldsymbol{e}^{j}\cdot\boldsymbol{\omega}\ \right] = \boldsymbol{0}\,,
\end{align}
and also with the electronic spin operators $\boldsymbol{s}_{(\mu)}$, i.e.
\begin{align}
\label{3.46}
&\left[\ \boldsymbol{s}_{(\mu)},\ \boldsymbol{e}^{j}\cdot\boldsymbol{q}_{(\nu)}\ \right] = \boldsymbol{0}\,,\\
\label{3.47}
&\left[\ \boldsymbol{s}_{(\mu)},\ Q^{\alpha}\ \right] = \boldsymbol{0}\,,\\
\label{3.48}
&\left[\ \boldsymbol{s}_{(\mu)},\ \boldsymbol{e}^{j}\cdot\boldsymbol{\omega}\ \right] = \boldsymbol{0}\,.
\end{align}
Moreover, as demonstrated in the Appendix~\ref{A.9}, the operators $P_{\alpha}$ and $\boldsymbol{p}_{(\nu)}$ commute with the nuclear spin operators $\boldsymbol{S}_{(\mu)}$, i.e.
\begin{align}
\label{3.49}
&\left[\ \boldsymbol{S}_{(\mu)},\ \boldsymbol{e}_{j}\cdot\boldsymbol{p}_{(\nu)}\ \right] = \boldsymbol{0}\,,\\
\label{3.50}
&\left[\ \boldsymbol{S}_{(\mu)},\ P_{\alpha}\ \right] = \boldsymbol{0}\,,
\end{align}
and also with the electronic spin operators $\boldsymbol{s}_{(\mu)}$, i.e.
\begin{align}
\label{3.51}
&\left[\ \boldsymbol{s}_{(\mu)},\ \boldsymbol{e}_{j}\cdot\boldsymbol{p}_{(\nu)}\ \right] = \boldsymbol{0}\,,\\
\label{3.52}
&\left[\ \boldsymbol{s}_{(\mu)},\ P_{\alpha}\ \right] = \boldsymbol{0}\,.
\end{align}
Finally, as shown in the Appendix~\ref{A.9}, the commutation relations between the orbital angular momentum $\boldsymbol{L}$ and the spin operators $\boldsymbol{S}_{(\mu)}$ and $\boldsymbol{s}_{(\mu)}$ are respectively given by,
\begin{equation}\label{3.53}
\begin{split}
&\left[\ \boldsymbol{S}_{(\mu)},\ \boldsymbol{e}_{j}\cdot\boldsymbol{L}\ \right] = i\hbar\,\left(\boldsymbol{e}_{j}\times\boldsymbol{S}_{(\mu)}\right)\,,\\
&\left[\ \boldsymbol{s}_{(\mu)},\ \boldsymbol{e}_{j}\cdot\boldsymbol{L}\ \right] = i\hbar\,\left(\boldsymbol{e}_{j}\times\boldsymbol{s}_{(\mu)}\right)\,.
\end{split}
\end{equation}

\section{Dynamical description of a rotating and vibrating molecule}
\label{Dynamical description of a rotating and vibrating molecule}
The observable corresponding to the kinetic energy is defined as,
\begin{equation}\label{4.0.0}
T = \sum_{\mu=1}^{N}\,\frac{\boldsymbol{P}_{\mu}^{2}}{2\,M_{\mu}}\otimes\mathbb{1}_{e} + \mathbb{1}_{\mathcal{N}}\otimes\sum_{\nu=1}^{n}\,\frac{\boldsymbol{p}_{\nu}^{2}}{2\,m}\,,
\end{equation}
As shown in Appendix~\ref{B.1}, it is recast as,
\begin{equation}\label{4.0.1}
\begin{split}
&T = \frac{\boldsymbol{\mathcal{P}}^{2}}{2\,\mathcal{M}} + \sum_{\mu=1}^{N}\frac{\boldsymbol{P}^{\prime\prime\,2}_{\mu}}{2\,M_{\mu}} + \sum_{\nu=1}^{n}\frac{\boldsymbol{p}^{\prime\prime\,2}_{\nu}}{2\,m} + \frac{\hbar^2}{8}\,\Phi_{\text{res}}\left(Q^{\,\boldsymbol{.}}\right)\,,
\end{split}
\end{equation}
where $\Phi_{\text{res}}\left(Q^{\,\boldsymbol{.}}\right)$ is a residual operator resulting from the non-commutation of the rotation operator $\mathsf{R}\left(\boldsymbol{\omega}\right)$ and the momentum operator $\boldsymbol{P}_{\mu}$ that is given by,
\begin{equation}\label{4.0.2}
\begin{split}
&\Phi_{\text{res}}\left(Q^{\,\boldsymbol{.}}\right) = \text{Tr}\left(\mathsf{I}_{00}\right)\,\text{Tr}\left(\mathsf{I}\left(Q^{\,\boldsymbol{.}}\right)^{-2}\right)\\
&-\,2\,\text{Tr}\left(\mathsf{I}_{00}\cdot\mathsf{I}\left(Q^{\,\boldsymbol{.}}\right)^{-1}\right)\,\text{Tr}\left(\mathsf{I}\left(Q^{\,\boldsymbol{.}}\right)^{-1}\right)\\ 
&+ \text{Tr}\left(\mathsf{I}_{00}\cdot\mathsf{I}\left(Q^{\,\boldsymbol{.}}\right)^{-2} -\,2\,\left(\mathsf{I}_{\alpha}\cdot\mathsf{I}\left(Q^{\,\boldsymbol{.}}\right)^{-1}\right)^2\right)\\
&+2\,\text{Tr}\left(\mathsf{I}\left(Q^{\,\boldsymbol{.}}\right)^{-1}\cdot\mathsf{I}_{\alpha}\cdot \mathsf{I}\left(Q^{\,\boldsymbol{.}}\right)^{-2}\cdot\mathsf{I}^{\alpha}_{0\beta}\,Q^{\beta}\right)\,,
\end{split}
\end{equation}
and the rank-$2$ tensors $\mathsf{I}_{00}$ and $\mathsf{I}^{\alpha}_{0\beta}$ are defined respectively as,
\begin{align}
\label{4.0.3}
&\mathsf{I}_{00} = \sum_{\mu=1}^{N}\,M_{\mu}\,\boldsymbol{R}^{(0)}_{\mu}\,\boldsymbol{R}^{(0)}_{\mu}\,,\\
\label{4.0.4}
&\mathsf{I}^{\alpha}_{0\beta} = \sum_{\mu,\nu=1}^{N}\!\boldsymbol{R}^{(0)}_{\mu}\!\left(\left(\boldsymbol{R}^{(0)}_{\mu}\cdot\boldsymbol{X}^{\alpha}_{\nu}\right)\boldsymbol{X}_{\nu\beta} -\left(\boldsymbol{R}^{(0)}_{\mu}\cdot\boldsymbol{X}_{\nu\beta}\right)\boldsymbol{X}^{\alpha}_{\nu}\right)\,
\end{align}
As shown in Appendix~\ref{B.1}, using the appropriate commutation relations, the kinetic energy~\eqref{4.0.1} is recast in terms of the internal observables as,
\begin{equation}\label{4.1}
T = \frac{\boldsymbol{\mathcal{P}}^{2}}{2\,\mathcal{M}} + \frac{1}{2}\,\boldsymbol{\Omega}\cdot\mathsf{I}_{0}\cdot\boldsymbol{\Omega} +\frac{1}{2}\,P_{\alpha}\,P^{\alpha} + \sum_{\nu=1}^{n}\,\frac{\boldsymbol{p}_{(\nu)}^2}{2\,m} + \frac{\hbar^2}{8}\,\Phi_{\text{res}}\left(Q^{\,\boldsymbol{.}}\right)\,.
\end{equation}
Finally, using the definition~\eqref{3.29ter}, it is useful to recast the kinetic energy~\eqref{4.1} explicitly in terms of the orbital angular momentum, i.e.
\begin{align}\label{4.1.sec}
&T = \frac{\boldsymbol{\mathcal{P}}^{2}}{2\,\mathcal{M}} + \frac{1}{8}\,\left\{\ \boldsymbol{\mathcal{L}},\ \mathsf{I}\left(Q^{\,\boldsymbol{.}}\right)^{-1}\ \right\}_{\mathsmaller{\bullet}}\cdot\mathsf{I}_{0}\cdot\left\{\ \mathsf{I}\left(Q^{\,\boldsymbol{.}}\right)^{-1},\ \boldsymbol{\mathcal{L}}\ \right\}_{\mathsmaller{\bullet}}\nonumber\\
&\phantom{T = }+\frac{1}{2}\,P_{\alpha}\,P^{\alpha} + \sum_{\nu=1}^{n}\,\frac{\boldsymbol{p}_{(\nu)}^2}{2\,m} + \frac{\hbar^2}{8}\,\Phi_{\text{res}}\left(Q^{\,\boldsymbol{.}}\right)\,.
\end{align}
The expression~\eqref{4.1.sec} of the kinetic energy $T$ separates the rotational and vibrational degrees of freedom. It is a consequence of the Eckart conditions~\eqref{3.16},~\eqref{3.17} and~\eqref{3.17bis} and of the physical definition~\eqref{3.39 ter} of the rotation operator $\mathsf{R}\left(\boldsymbol{\omega}\right)$. It is the quantum counterpart of the expression derived by Jellinek and Li~\cite{Jellinek:1989,Jellinek:1990} and extended by Essen~\cite{Essen:1993} in a classical framework.

The Hamiltonian~\eqref{2.11} is expressed in terms of the kinetic energy as,
\begin{align}\label{4.12}
&H = T + V_{\mathcal{N}-\mathcal{N}}\left(Q^{\,\boldsymbol{.}}\right) + V_{e-e}\,(\boldsymbol{q}_{(\boldsymbol{.})}) + V_{\mathcal{N}-e}\left(Q^{\,\boldsymbol{.}},\boldsymbol{q}_{(\boldsymbol{.})}\right) \nonumber\\
&\phantom{H = } + V_{\mathcal{N}}^{SO}(Q^{\,\boldsymbol{.}},P_{\,\boldsymbol{.}},\boldsymbol{L},\boldsymbol{S}_{(\boldsymbol{.})}) + V_{e}^{SO}(\boldsymbol{q}_{(\boldsymbol{.})},\boldsymbol{p}_{(\boldsymbol{.})},\boldsymbol{s}_{(\boldsymbol{.})}) \nonumber\\
&\phantom{H = } + V_{\mathcal{N}-e}^{SO}(Q^{\,\boldsymbol{.}},P_{\,\boldsymbol{.}},\boldsymbol{L},\boldsymbol{S}_{(\boldsymbol{.})},\boldsymbol{q}_{(\boldsymbol{.})},\boldsymbol{p}_{(\boldsymbol{.})},\boldsymbol{s}_{(\boldsymbol{.})})
\end{align}
Taking into account the fact that a norm is invariant under rotation and using the definitions~\eqref{2.13},~\eqref{2.16} and \eqref{3.14}, the nuclear, electronic and interaction Coulomb potentials are recast in terms of the internal observables respectively as,
\begin{align}\label{4.13}
&V_{\mathcal{N}-\mathcal{N}}(Q^{\,\boldsymbol{.}})\nonumber\\
&= \frac{e^{2}}{8\pi\varepsilon_{0}}\!\sum_{\substack{\mu,\nu=1\\ \mu\neq\nu}}^{N}
\frac{Z_{\mu}\,Z_{\nu}}{\Vert\,\left(\boldsymbol{R}^{(0)}_{\mu}-\,\boldsymbol{R}^{(0)}_{\nu}\right)\,\mathbb{1} + Q^{\alpha}\,\left(\boldsymbol{Y}_{\mu\alpha}-\,\boldsymbol{Y}_{\nu\alpha}\right)\,\Vert}\,,\nonumber\\
&V_{e-e}(\boldsymbol{q}_{(\boldsymbol{.})}) = \frac{e^{2}}{8\pi\varepsilon_{0}}\!\sum_{\substack{\mu,\nu=1\\ \mu\neq\nu}}^{n}\frac{1}{\Vert\,\boldsymbol{\bar{q}}_{(\mu)}-\,\boldsymbol{\bar{q}}_{(\nu)}\,\Vert}\,,\\
&V_{\mathcal{N}-e}(Q^{\,\boldsymbol{.}},\boldsymbol{q}_{(\boldsymbol{.})})\nonumber\\
&= -\,\frac{e^{2}}{4\pi\varepsilon_{0}}\!\sum_{\mu=1}^{N}\sum_{\nu=1}^{n} \,\frac{Z_{\mu}}{\Vert\,\boldsymbol{R}^{(0)}_{\mu}\,\mathbb{1} + Q^{\alpha}\,\boldsymbol{Y}_{\mu\alpha}-\,\boldsymbol{\bar{q}}_{(\nu)}\,\Vert}\,,\nonumber
\end{align}
where the orthogonal basis vector $\boldsymbol{Y}_{\mu\alpha}$ is related to the orthonormal basis vector $\boldsymbol{X}_{\mu\alpha}$ by,
\begin{equation}\label{4.XX}
\boldsymbol{X}_{\mu\alpha} = \sqrt{M_{\mu}}\,\boldsymbol{Y}_{\mu\alpha}\,,
\end{equation}
and the operator $\boldsymbol{\bar{q}}_{(\nu)}\equiv\boldsymbol{r}^{\prime\prime}_{(\nu)}$ is a function of the operators $\boldsymbol{q}_{(\boldsymbol{.})}$ according to the relation~\eqref{3.15}.

The potential energy operator associated to the spin-orbit coupling between the nuclei is given by,  
\begin{equation}\label{4.20.0}
V_{\mathcal{N}}^{SO}(Q^{\,\boldsymbol{.}},P_{\,\boldsymbol{.}},\boldsymbol{L},\boldsymbol{S}_{(\boldsymbol{.})}) = -\,\sum_{\mu=1}^{N}\,\gamma_{\mu}\,\boldsymbol{S}_{(\mu)}\cdot\boldsymbol{B}^{\,\mathcal{N}}_{(\mu)}(Q^{\,\boldsymbol{.}},P_{\,\boldsymbol{.}},\boldsymbol{L})\,,
\end{equation}
where $\gamma_{\mu}>0$ is the gyromagnetic ratio of the nuclei $\mu$ and $\boldsymbol{B}^{\,\mathcal{N}}_{(\mu)}(Q^{\,\boldsymbol{.}},P_{\,\boldsymbol{.}},\boldsymbol{L})$ is a pseudo-vectorial operator corresponding to the internal magnetic field exerted on the nucleus $\mu$ and generated by the relative motion of the other nuclei. Similarly, the potential energy operator associated to the spin-orbit coupling between the electrons yields,  
\begin{equation}\label{4.20}
V_{e}^{SO}(\boldsymbol{q}_{(\boldsymbol{.})},\boldsymbol{p}_{(\boldsymbol{.})},\boldsymbol{s}_{(\boldsymbol{.})}) = -\,\sum_{\nu=1}^{n}\,\gamma_{e}\,\boldsymbol{s}_{(\nu)}\cdot\boldsymbol{B}^{\,e}_{(\nu)}(\boldsymbol{q}_{(\boldsymbol{.})},\boldsymbol{p}_{(\boldsymbol{.})})\,,
\end{equation}
where $\gamma_{e}<0$ is the gyromagnetic ratio of the electron and $\boldsymbol{B}^{\,e}_{(\nu)}(\boldsymbol{q}_{(\boldsymbol{.})},\boldsymbol{p}_{(\boldsymbol{.})})$ is a pseudo-vectorial operator corresponding to the internal magnetic field exerted on the electron $\nu$ and generated by the relative motion of the other electrons. Finally, the potential energy operator associated to the spin-orbit coupling between the nuclei and the electrons is given by,
\begin{align}\label{4.27}
&V_{\mathcal{N}-e}^{SO}(Q^{\,\boldsymbol{.}},P_{\,\boldsymbol{.}},\boldsymbol{L},\boldsymbol{S}_{(\boldsymbol{.})},\boldsymbol{q}_{(\boldsymbol{.})},\boldsymbol{p}_{(\boldsymbol{.})},\boldsymbol{s}_{(\boldsymbol{.})})\\
&=-\,\sum_{\nu=1}^{n}\,\gamma_{e}\,\boldsymbol{s}_{(\nu)}\cdot\boldsymbol{B}^{\,\mathcal{N}-e}_{(\nu)}(Q^{\,\boldsymbol{.}},P_{\,\boldsymbol{.}},\boldsymbol{L})\nonumber\\
&\phantom{=} -\,\sum_{\mu=1}^{N}\,\gamma_{\mu}\,\boldsymbol{S}_{(\mu)}\cdot\boldsymbol{B}^{\,e-\mathcal{N}}_{(\mu)}(\boldsymbol{q}_{(\boldsymbol{.})},\boldsymbol{p}_{(\boldsymbol{.})})\,,\nonumber
\end{align}
where $\boldsymbol{B}^{\,\mathcal{N}-e}_{(\nu)}(Q^{\,\boldsymbol{.}},P_{\,\boldsymbol{.}},\boldsymbol{L})$ is a pseudo-vectorial operator corresponding to the internal magnetic field exerted on the electron $\nu$ and generated by the relative motion of the nuclei and $\boldsymbol{B}^{\,e-\mathcal{N}}_{(\mu)}(\boldsymbol{q}_{(\boldsymbol{.})},\boldsymbol{p}_{(\boldsymbol{.})})$ is a pseudo-vectorial operator corresponding to the internal magnetic field exerted on the nucleus $\mu$ and generated by the relative motion of the electrons.

\section{Molecular ground state and vibrational modes}
\label{Molecular ground state and vibrational modes}

A molecular system is described by bound states defined in the neighbourhood of the ground state in the Hilbert space~\eqref{2.1}. For such states, the vibrational degrees of freedom are sufficiently small to allow a series expansion in terms of the deformation operators $Q^{\alpha}$~\cite{Wilson:1936}. To second-order, the series expansion of the Coulomb potentials read,
\begin{align}
\label{4.14}
&V_{\mathcal{N}-\mathcal{N}}\left(Q^{\,\boldsymbol{.}}\right) = V_{\mathcal{N}-\mathcal{N}\,(0)}\,\mathbb{1} + V_{\mathcal{N}-\mathcal{N}\,(\alpha)}\,Q^{\alpha}\nonumber\\
&+ \frac{1}{2}\,V_{\mathcal{N}-\mathcal{N}\,(\alpha\beta)}\,
Q^{\alpha}\,Q^{\beta} + \mathcal{O}\left({Q^{\,\boldsymbol{.}}}^{3}\right)\,,\\
\label{4.23}
&V_{\mathcal{N}-e}\left(Q^{\,\boldsymbol{.}},\boldsymbol{q}_{(\boldsymbol{.})}\right) = V_{\mathcal{N}-e\,(0)}\,(\boldsymbol{q}_{(\boldsymbol{.})})\,\mathbb{1} + V_{\mathcal{N}-e\,(\alpha)}\,(\boldsymbol{q}_{(\boldsymbol{.})})\,Q^{\alpha}\nonumber\\
&+ \frac{1}{2}\,V_{\mathcal{N}-e\,(\alpha\beta)}\,(\boldsymbol{q}_{(\boldsymbol{.})})\,Q^{\alpha}\,Q^{\beta} + \mathcal{O}\left({Q^{\,\boldsymbol{.}}}^{3}\right)\,.
\end{align}
To establish the molecular equilibrium conditions defining the ground state, we neglect the contributions due to the spin-orbit couplings. Moreover, the contribution due to the residual term $\Phi_{\text{res}}\left(Q^{\,\boldsymbol{.}}\right)$ in the kinetic energy operator~\eqref{4.1.sec} is proportional to $\hbar^2$ and can be neglected also. Furthermore, the effects of the deformation modes $Q^{\alpha}$ on the molecular inertia tensor $\mathsf{I}\left(Q^{\,\boldsymbol{.}}\right)$ are second-order contributions that we neglect as well. Thus, to zeroth-order in $Q^{\alpha}$ the inertia tensor $\mathsf{I}\left(Q^{\,\boldsymbol{.}}\right)$ reduces,
\begin{equation}\label{5.0}
\mathsf{I}\left(Q^{\,\boldsymbol{.}}\right) = \mathsf{I}_{0} + \mathcal{O}\left(Q^{\,\boldsymbol{.}}\right)\,.
\end{equation}
Using the definitions~\eqref{3.29} and~\eqref{3.29quad}, and the relation~\eqref{5.0}, the rotational part of the Hamiltonian $H$ in equation~\eqref{4.1.sec} is explicitly recast to zeroth-order in $Q^{\alpha}$ as,
\begin{align}\label{5.0.1}
&\frac{1}{8}\,\left\{\ \boldsymbol{\mathcal{L}},\ \mathsf{I}\left(Q^{\,\boldsymbol{.}}\right)^{-1}\ \right\}_{\mathsmaller{\bullet}}\cdot\mathsf{I}_{0}\cdot\left\{\ \mathsf{I}\left(Q^{\,\boldsymbol{.}}\right)^{-1},\ \boldsymbol{\mathcal{L}}\ \right\}_{\mathsmaller{\bullet}}\\
&= \frac{1}{2}\,\boldsymbol{L}\cdot\mathsf{I}_{0}^{-1}\cdot\boldsymbol{L} + \frac{1}{2}\,\boldsymbol{\ell}\cdot\mathsf{I}_{0}^{-1}\cdot\boldsymbol{\ell} -\,\boldsymbol{L}\cdot\mathsf{I}_{0}^{-1}\cdot\boldsymbol{\ell} + \mathcal{O}\left(Q^{\,\boldsymbol{.}}\right)\,.\nonumber
\end{align}
According to the relations~\eqref{4.1}, \eqref{4.12} \eqref{4.14}, \eqref{4.23} and~\eqref{5.0}, the Hamiltonian $H$ is expanded in terms of the deformation operators $Q^{\alpha}$ as,
\begin{equation}\label{5.1}
H = H_{(0)} + H_{(\alpha)}\,Q^{\alpha} + \frac{1}{2}\,H_{(\alpha\beta)}\,Q^{\alpha}Q^{\beta} + \mathcal{O}\left({Q^{\,\boldsymbol{.}}}^{3}\right)\,,
\end{equation}
where
\begin{align}\label{5.2}
&H_{(0)} = \frac{\boldsymbol{\mathcal{P}}^{2}}{2\,\mathcal{M}} + \frac{1}{2}\,P_{\alpha}\,P^{\alpha} + \frac{1}{2}\,\boldsymbol{L}\cdot\mathsf{I}_{0}^{-1}\cdot\boldsymbol{L} +  V_{\mathcal{N}-\mathcal{N}\,(0)}\,\mathbb{1}\,,\nonumber\\
&\phantom{H_{(0)} = }+ \sum_{\nu=1}^{n}\,\frac{\boldsymbol{p}_{(\nu)}^2}{2\,m} + \frac{1}{2}\,\boldsymbol{\ell}\cdot\mathsf{I}_{0}^{-1}\cdot\boldsymbol{\ell} + V_{e-e}\,(\boldsymbol{q}_{(\boldsymbol{.})})\,,\nonumber\\
&\phantom{H_{(0)} = } -\,\boldsymbol{L}\cdot\mathsf{I}_{0}^{-1}\cdot\boldsymbol{\ell} + V_{\mathcal{N}-e\,(0)}\,(\boldsymbol{q}_{(\boldsymbol{.})})\,,\\
&H_{(\alpha)} = V_{\mathcal{N}-\mathcal{N}\,(\alpha)} + V_{\mathcal{N}-e\,(\alpha)}\,(\boldsymbol{q}_{(\boldsymbol{.})})
\vphantom{\frac{1}{2}\,\boldsymbol{\ell}\cdot\mathsf{I}_{0}^{-1}\cdot\boldsymbol{\ell}}\,,\nonumber\\
&H_{(\alpha\beta)} = V_{\mathcal{N}-\mathcal{N}\,(\alpha\beta)} + V_{\mathcal{N}-e\,(\alpha\beta)}\,(\boldsymbol{q}_{(\boldsymbol{.})})
\vphantom{\frac{\boldsymbol{\mathcal{P}}^{2}}{2\,\mathcal{M}}}\,.\nonumber
\end{align}
In order to ensure that the molecular dynamics occurs in the neighbourhood the equilibrium ground state, we vary the energy $E = \langle\,H\,\rangle$ of the system with respect to the deformation modes $Q^{\alpha}$, where the brackets denote the expectation value taken on the Hilbert space~\eqref{2.1} of the molecular system. At equilibrium, the density matrix commutes with the Hamiltonian~\eqref{4.12}. We assume that this is also the case in the neighbourhood of the equilibrium state. Thus, to first-order, the variation of the energy yields the equilibrium condition, i.e.
\begin{equation}\label{5.3}
V_{\mathcal{N}-\mathcal{N}\,(\alpha)} + \Big\langle\,V_{\mathcal{N}-e\,(\alpha)}\,(\boldsymbol{q}_{(\boldsymbol{.})})\,\Big\rangle = 0\,,
\end{equation}
which defines the ground state of the molecular system. Using the definitions~\eqref{4.13}, the series expansions~\eqref{4.14} and~\eqref{4.23}, and the fact that molecular vibration modes $\boldsymbol{X}_{\mu\alpha}$ are linearly independent, the condition~\eqref{5.3} implies that, 
\begin{equation}\label{5.5}
\begin{split}
&\frac{e^{2}}{4\pi\varepsilon_{0}}\,\sum_{\substack{\nu=1\\ \mu\neq\nu}}^{N}\,Z_{\nu}\,\frac{\boldsymbol{R}^{(0)}_{\mu}-\,\boldsymbol{R}^{(0)}_{\nu}}{\Vert\boldsymbol{R}^{(0)}_{\mu}-\,\boldsymbol{R}^{(0)}_{\nu}\Vert^3}\\ &-\,\frac{e^{2}}{4\pi\varepsilon_{0}}\,\sum_{\nu=1}^{n}\,\bigg\langle\,\frac{\boldsymbol{R}^{(0)}_{\mu}\mathbb{1} -\,\boldsymbol{\bar{q}}_{(\nu)}}{\Vert\boldsymbol{R}^{(0)}_{\mu}\mathbb{1} -\,\boldsymbol{\bar{q}}_{(\nu)}\Vert^3}\,\bigg\rangle = \boldsymbol{0}\,.
\end{split}
\end{equation}
The first term in the condition~\eqref{5.5} represents the classical Coulomb force exerted by the nuclei on the nucleus $\mu$. The second term in the condition~\eqref{5.5} represents the expectation value of the quantum Coulomb force exerted by the electrons on the nucleus $\mu$. The equilibrium condition~\eqref{5.5} implies that the resulting force exerted on the nucleus $\mu$ vanishes at the equilibrium. 

The $3N-6$ equilibrium conditions~\eqref{5.5} imposed on all the vibrations modes $\boldsymbol{X}_{\mu\alpha}$, the $3$ conditions~\eqref{3.16} imposed on the origin of the coordinate system and the $3$ conditions~\eqref{3.16bis} imposed on the orientation of the coordinate system fully determine the $3N$ degrees of freedom corresponding to the positions of the nuclei.

The molecular deformation modes $\boldsymbol{X}_{\mu\alpha}$ and $\boldsymbol{X}_{\nu\beta}$ are orthogonal if $\alpha\neq\beta$ according to the condition~\eqref{3.18}. Moreover, to second-order with respect to the deformation $Q^{\alpha}$, the deformation modes are decoupled. Thus, the variation of the energy $E = \langle\,H\,\rangle$ with respect to the deformation $Q^{\alpha}$ yields,
\begin{equation}\label{5.6}
V_{\mathcal{N}-\mathcal{N}\,(\alpha\beta)} + \Big\langle\,V_{\mathcal{N}-e\,(\alpha\beta)}\,(\boldsymbol{q}_{(\boldsymbol{.})})\,\Big\rangle = 0\,,\quad\forall\ \alpha\neq\beta\,.
\end{equation}
As shown in Appendix~\ref{Vibrational modes}, the diagonal components, i.e. $\alpha=\beta$, of the Hessian matrix of the Hamiltonian~\eqref{4.12} yield the square of the angular frequency of the molecular vibration eigenmodes, i.e.
\begin{equation}\label{5.21}
\omega^2_{\alpha} = V_{\mathcal{N}-\mathcal{N}\,(\alpha\alpha)} + \Big\langle\,V_{\mathcal{N}-e\,(\alpha\alpha)}\,(\boldsymbol{q}_{(\boldsymbol{.})})\,\Big\rangle > 0\,,
\end{equation}
which ensures that $\omega_{\alpha}>0$ in the neighbourhood of the molecular ground state.

\section{Conclusion}
\label{Conclusion}

A rigorous quantum treatment of a molecular system includes kinetic couplings between nuclei and electrons and leads to quantum deviations in the commutation relations between the position and momentum operators. These deviations are proportional to the ratio of the electron mass to the molecular mass and to the ratio of a specific nuclear mass to the molecular mass. Thus, these deviations are larger for smaller molecules.

In this quantum description, the vibrational and rotational degrees of freedom are treated in a genuine quantum framework. Since quantum nonlocality forbids the existence of a center of mass frame and a molecular rest frame, we defined ``relative'' position and momentum observables, which are the quantum equivalent of the relative position and momentum variables expressed with respect to the centre of mass frame in a classical framework. In order to define ``rest'' position and momentum observables expressed with respect to a rotating molecule we defined a molecular rotation operator that is function of a molecular orientation operator. Then, we defined internal observables that account for the quantum degrees of freedom of the molecular system. These observables are the deformation and orientation operators of the molecule as well as a the position and the momentum operators of the electrons.

In a rigorous quantum description of a molecule, the molecular orientation and rotation are described by operators. As a result the ``rest'' momentum operators do not commute with the molecular orientation and rotation operators. This generates an additional contribution proportional to the residual term $\Phi_{\text{res}}\left(Q^{\,\boldsymbol{.}}\right)$ in the expression of the molecular Hamiltonian and leads to canonical rotational commutation relations~\eqref{3.36.1} between the total angular momentum operator $\boldsymbol{L}$ and the orientation operator $\boldsymbol{\omega}$. The molecular Hamiltonian satisfies a ``molecular correspondence principle'' : replacing the operators describing physical observables by variables leads to a classical molecular Hamiltonian where the residual term $\Phi_{\text{res}}\left(Q^{\,\boldsymbol{.}}\right)$ vanishes.

The rigorous quantum description of the dynamics of a rotating and vibrating molecule presented in this publication, is a prelude to the study of quantum dissipation at the molecular level~\cite{Breuer:2007}. In order to describe molecular dissipation, the quantum statistical framework provided by the quantum master equations needs to be introduced. In such a framework, where certain internal observables are treated as a statistical bath that is weakly coupled and weakly correlated to the other internal observables representing the system of interest~\cite{Reuse:2003}. The quantum master equations of the molecular system are expected to lead to dissipative couplings between the rotational, vibrational and magnetic quantum modes and to describe molecular dissipative phenomena such as molecular magnetism~\cite{Islam:2014,Qian:2014}.


\appendix

\section{Killing form of the rotation algebra}
\label{A.2}

In this appendix, we determine the explicit expression of the Killing form $\boldsymbol{n}_{(j)}\left(\boldsymbol{\omega}\right)\cdot\boldsymbol{n}_{(\ell)}\left(\boldsymbol{\omega}\right)$ of the rotation algebra. 

The rotation group action~\eqref{3.23bis} implies that,
\begin{equation}\label{A.2.3prime}
\begin{split}
&\text{Tr}\,\Big(\left(\boldsymbol{e}_j\cdot\boldsymbol{\mathsf{G}}^{T}\right)\,\left(\boldsymbol{e}_k\cdot\boldsymbol{\mathsf{G}}\right)\Big) = \boldsymbol{e}^i\cdot\Big(\left(\boldsymbol{e}_j\cdot\boldsymbol{\mathsf{G}}^{T}\right)\,\left(\boldsymbol{e}_k\cdot\boldsymbol{\mathsf{G}}\right)\Big)\,\boldsymbol{e}_i\\
&=-\,\boldsymbol{e}^i\cdot\left(\boldsymbol{e}_j\cdot\boldsymbol{\mathsf{G}}\right)\Big(\left(\boldsymbol{e}_k\cdot\boldsymbol{\mathsf{G}}\right)\,\boldsymbol{e}_i\Big) = -\,\boldsymbol{e}^i\cdot\Big(\boldsymbol{e}_j\times\left(\boldsymbol{e}_k\times\boldsymbol{e}_i\right)\Big)\\
&=\left(\boldsymbol{e}_j\cdot\boldsymbol{e}_k\right)\cdot\left(\boldsymbol{e}^i\cdot\boldsymbol{e}_i\right)-\,\left(\boldsymbol{e}_j\cdot\boldsymbol{e}_i\right)\left(\boldsymbol{e}^i\cdot\boldsymbol{e}_k\right) = 2\,\left(\boldsymbol{e}_j\cdot\boldsymbol{e}_k\right)\,.
\end{split}
\end{equation}
Thus, using the identity~\eqref{A.2.3prime} and the definitions~\eqref{A.2.1bis} and~\eqref{3.38}, the Killing form is expressed as,
\begin{equation}\label{A.2.3bis}
\begin{split}
&\boldsymbol{n}_{(j)}\left(\boldsymbol{\omega}\right)\cdot\boldsymbol{n}_{(\ell)}\left(\boldsymbol{\omega}\right)\\
&=\frac{1}{2}\ \text{Tr}\,\Big(\!\left(\boldsymbol{n}_{(j)}\left(\boldsymbol{\omega}\right)\cdot\boldsymbol{\mathsf{G}}^{\,T}\right)\cdot\left(\boldsymbol{n}_{(\ell)}\left(\boldsymbol{\omega}\right)\cdot\boldsymbol{\mathsf{G}}\right)\!\Big)\\
&=\frac{1}{2}\,\text{Tr}\,\Big(\!\left(
\boldsymbol{e}_j\cdot\partial_{\boldsymbol{\omega}}\right)\,\mathsf{R}\left(\boldsymbol{\omega}\right)^{-1}\cdot\left(\boldsymbol{e}_{\ell}\cdot\partial_{\boldsymbol{\omega}}\right)\,\mathsf{R}\left(\boldsymbol{\omega}\right)\!\Big)\,.
\end{split}
\end{equation}
Moreover, the rotation group action~\eqref{3.23bis} implies that $\forall\,\boldsymbol{x}\in\mathbb{R}^3$,
\begin{equation}\label{A.2.3.0}
\begin{split}
&\left(\boldsymbol{\omega}\cdot\boldsymbol{\mathsf{G}}\right)^2\boldsymbol{x} = \boldsymbol{\omega}\times\left(\boldsymbol{\omega}\times\boldsymbol{x}\right) = -\,\boldsymbol{\omega}^2\,\boldsymbol{x} + \boldsymbol{\omega}\left(\boldsymbol{\omega}\cdot\boldsymbol{x}\right)  \\
&\phantom{\left(\boldsymbol{\omega}\cdot\boldsymbol{\mathsf{G}}\right)^2\boldsymbol{x} }= -\,\boldsymbol{\omega}^2\left(\mathbb{1} -\,\mathsf{P}_{\boldsymbol{\omega}}\right)\,\boldsymbol{x}\,,
\end{split}
\end{equation}
where the self-adjoint projection operator $\mathsf{P}_{\boldsymbol{\omega}}$ satisfies the following identity $\forall\,n\in\mathbb{N}^{\star}$,
\begin{equation}\label{A.2.3.2}
\mathsf{P}_{\boldsymbol{\omega}}^2 = \mathsf{P}_{\boldsymbol{\omega}} \qquad\Rightarrow\qquad \left(\mathbb{1} -\,\mathsf{P}_{\boldsymbol{\omega}}\right)^n = \mathbb{1} -\,\mathsf{P}_{\boldsymbol{\omega}}\,.
\end{equation}
The projection operator~\eqref{3.39.def} is orthogonal to the rotation group action~\eqref{3.23bis}, i.e.
\begin{equation}\label{A.2.3.3}
\mathsf{P}_{\boldsymbol{\omega}}\cdot\left(\boldsymbol{\omega}\cdot\boldsymbol{\mathsf{G}}\right)\,\boldsymbol{x} = \mathsf{P}_{\boldsymbol{\omega}}\cdot\left(\boldsymbol{\omega}\times\boldsymbol{x}\right) = \frac{\boldsymbol{\omega}}{\Vert\boldsymbol{\omega}\Vert^2}\,\Big(\boldsymbol{\omega}\cdot\left(\boldsymbol{\omega}\times\boldsymbol{x}\right)\Big) = \boldsymbol{0}\,.
\end{equation}
Using the properties~\eqref{A.2.3.0},~\eqref{A.2.3.2} and~\eqref{A.2.3.3}, the rotation operator~\eqref{3.21} is recast as,
\begin{equation}\label{A.2.3.4}
\begin{split}
&\mathsf{R}\left(\boldsymbol{\omega}\right) = \mathbb{1} + \sum_{n=1}^{\infty}\frac{\left(\boldsymbol{\omega}\cdot\boldsymbol{\mathsf{G}}\right)^{2n}}{\left(2n\right)!} + \sum_{n=0}^{\infty}\frac{\left(\boldsymbol{\omega}\cdot\boldsymbol{\mathsf{G}}\right)^{2n+1}}{\left(2n+1\right)!}\\
&= \mathbb{1} + \sum_{n=1}^{\infty}\frac{\left(-1\right)^{n}\boldsymbol{\omega}^{2n}}{\left(2n\right)!}\left(\mathbb{1} -\,\mathsf{P}_{\boldsymbol{\omega}}\right)\\
&+\frac{1}{\boldsymbol{\omega}}\sum_{n=0}^{\infty}\frac{\left(-1\right)^{n}\boldsymbol{\omega}^{2n+1}}{\left(2n+1\right)!}\left(\boldsymbol{\omega}\cdot\boldsymbol{\mathsf{G}}\right)\\
&= \cos\boldsymbol{\omega} + \left(\mathbb{1} -\,\cos\boldsymbol{\omega}\right)\,\mathsf{P}_{\boldsymbol{\omega}} + \frac{\sin\boldsymbol{\omega}}{\boldsymbol{\omega}}\left(\boldsymbol{\omega}\cdot\boldsymbol{\mathsf{G}}\right)\,.
\end{split}
\end{equation}
The definition~\eqref{A.2.3.4}, the trace properties
\begin{equation}\label{A.2.3.5}
\begin{split}
&\text{Tr}\,\left(\mathsf{P}_{\boldsymbol{\omega}}\right) = 1\vphantom{\Big(\Big)}\,,\\
&\text{Tr}\,\left(\boldsymbol{\omega}\cdot\boldsymbol{\mathsf{G}}\right) = 0\,,\\
&\text{Tr}\,\left(\mathsf{P}_{\boldsymbol{\omega}^{\prime}}\cdot\mathsf{P}_{\boldsymbol{\omega}}\right) = \frac{\left(\boldsymbol{\omega^{\prime}}\cdot\boldsymbol{\omega}\right)^2}{\boldsymbol{\omega}^{\prime 2}\,\boldsymbol{\omega}^2}\,,\\
&\text{Tr}\,\Big(\!\left(\boldsymbol{\omega^{\prime}}\cdot\boldsymbol{\mathsf{G}}\right)\left(\boldsymbol{\omega}\cdot\boldsymbol{\mathsf{G}}\right)\!\Big) = -\,2\,\boldsymbol{\omega^{\prime}}\cdot\boldsymbol{\omega}\,,\\
&\text{Tr}\,\Big(\mathsf{P}_{\boldsymbol{\omega^{\prime}}}\cdot\,\left(\boldsymbol{\omega}\cdot\boldsymbol{\mathsf{G}}\right)\Big) = 0\,,
\end{split}
\end{equation}
and trigonometric identities imply that,
\begin{align}\label{A.2.3ter}
&\text{Tr}\,\Big(\mathsf{R}\left(\boldsymbol{\omega^{\prime}}\right)^{-1}\cdot\mathsf{R}\left(\boldsymbol{\omega}\right)\Big)\\
&= 4\left(\cos\frac{\boldsymbol{\omega^{\prime}}}{2}\cos\frac{\boldsymbol{\omega}}{2} + \frac{\boldsymbol{\omega^{\prime}}\,\boldsymbol{\omega}}{\boldsymbol{\omega^{\prime}}\cdot\boldsymbol{\omega}}\sin\frac{\boldsymbol{\omega^{\prime}}}{2}\sin\frac{\boldsymbol{\omega}}{2}\right)^2-\,1\,.\nonumber
\end{align}
The relations~\eqref{A.2.3bis} and~\eqref{A.2.3ter} with the help of trigonometric identities then imply in turn that,
\begin{align}\label{A.2.3quad}
&\boldsymbol{n}_{(j)}\left(\boldsymbol{\omega}\right)\cdot\boldsymbol{n}_{(\ell)}\left(\boldsymbol{\omega}\right)\nonumber\\
&=\frac{1}{2}\,\left(\boldsymbol{e}_j\cdot\partial_{\boldsymbol{\omega}}\right)\,\left(\boldsymbol{e}_{\ell}\cdot\partial_{\boldsymbol{\omega}^{\prime}}\right)\,\text{Tr}\,\left(\mathsf{R}\left(\boldsymbol{\omega^{\prime}}\right)^{-1}\cdot\mathsf{R}\left(\boldsymbol{\omega}\right)\right)\,\delta_{\boldsymbol{\omega}\boldsymbol{\omega}^{\prime}}\nonumber\\
& = 2\,\left(\boldsymbol{e}_j\cdot\partial_{\boldsymbol{\omega}}\right)\,\left(\boldsymbol{e}_{\ell}\cdot\partial_{\boldsymbol{\omega}^{\prime}}\right)\\
&\phantom{ = }\left(\cos\frac{\boldsymbol{\omega^{\prime}}}{2}\cos\frac{\boldsymbol{\omega}}{2} + \frac{\boldsymbol{\omega^{\prime}}\,\boldsymbol{\omega}}{\boldsymbol{\omega^{\prime}}\cdot\boldsymbol{\omega}}\sin\frac{\boldsymbol{\omega^{\prime}}}{2}\sin\frac{\boldsymbol{\omega}}{2}\right)^2\,\delta_{\boldsymbol{\omega}\boldsymbol{\omega}^{\prime}}\nonumber
\\
&\displaystyle{ =
\boldsymbol{e}_j\cdot\left(\frac{\boldsymbol{\omega}\,\boldsymbol{\omega}}{\Vert\boldsymbol{\omega}\Vert^2} + \left(\frac{2}{\Vert\boldsymbol{\omega}\Vert}
\,\sin\frac{\Vert\boldsymbol{\omega}\Vert}{2}\right)^{2}\left(\mathbb{1} -\,\frac{\boldsymbol{\omega}\,\boldsymbol{\omega}}{\Vert\boldsymbol{\omega}\Vert^2}\right)\right)\cdot\boldsymbol{e}_{\ell}}\,.\nonumber
\end{align}

Finally, the bilinear form~\eqref{A.2.3quad} correspond to the Killing form~\eqref{3.39.K} that is expressed in components as
\begin{equation}\label{A.2.3}
\begin{split}
&\boldsymbol{e}_j\cdot\mathsf{K}\left(\boldsymbol{\omega}\right)\cdot\boldsymbol{e}_{\ell} \equiv \boldsymbol{n}_{(j)}\left(\boldsymbol{\omega}\right)\cdot\boldsymbol{n}_{(\ell)}\left(\boldsymbol{\omega}\right)\\
&=
\boldsymbol{e}_j\cdot\Big(\mathsf{P}_{\boldsymbol{\omega}} + A\left(\mathbb{1} -\,\mathsf{P}_{\boldsymbol{\omega}}\right)\Big)\cdot\boldsymbol{e}_{\ell}\,.
\end{split}
\end{equation}
The components of the symmetric rank-$2$ tensor $\mathsf{K}\left(\boldsymbol{\omega}\right)^{-1}$ are given by
\begin{equation}\label{A.2.4}
\boldsymbol{e}^j\cdot\mathsf{K}\left(\boldsymbol{\omega}\right)^{-1}\cdot\boldsymbol{e}^{\ell} = \boldsymbol{e}^j\cdot\Big(\mathsf{P}_{\boldsymbol{\omega}} + A^{-1}\left(\mathbb{1} -\,\mathsf{P}_{\boldsymbol{\omega}}\right)\Big)\cdot\boldsymbol{e}^{\ell}\,.
\end{equation}
The expressions~\eqref{A.2.3} and~\eqref{A.2.4} imply that,
\begin{equation}\label{A.2.5}
\begin{split}
&\mathsf{K}\left(\boldsymbol{\omega}\right)\cdot\boldsymbol{\omega} = \boldsymbol{\omega}\,,\\
&\mathsf{K}\left(\boldsymbol{\omega}\right)^{-1}\cdot\boldsymbol{\omega} = \boldsymbol{\omega}\,.
\end{split}
\end{equation}

The operator $\boldsymbol{m}^{(j)}\left(\boldsymbol{\omega}\right)$ is defined as the dual of the operator $\boldsymbol{n}_{(\ell)}\left(\boldsymbol{\omega}\right)$, i.e.
\begin{equation}\label{A.2.7}
\boldsymbol{m}^{(j)}\left(\boldsymbol{\omega}\right) \equiv \left(\boldsymbol{e}^j\cdot\mathsf{K}\left(\boldsymbol{\omega}\right)^{-1}\cdot\boldsymbol{e}^{\ell}\right)\boldsymbol{n}_{(\ell)}\left(\boldsymbol{\omega}\right)\,.
\end{equation}
The expressions~\eqref{A.2.3} and~\eqref{A.2.7} yield the duality condition~\eqref{3.39}, i.e.
\begin{align}\label{A.2.8}
&\boldsymbol{m}^{(j)}\left(\boldsymbol{\omega}\right)\cdot\boldsymbol{n}_{(\ell)}\left(\boldsymbol{\omega}\right)\nonumber\\
&=\left(\boldsymbol{e}^j\cdot\mathsf{K}\left(\boldsymbol{\omega}\right)^{-1}\cdot\boldsymbol{e}^{k}\right)\,\boldsymbol{n}_{(k)}\left(\boldsymbol{\omega}\right)\cdot\boldsymbol{n}_{(\ell)}\left(\boldsymbol{\omega}\right)\\
&=\left(\boldsymbol{e}^j\cdot\mathsf{K}\left(\boldsymbol{\omega}\right)^{-1}\cdot\boldsymbol{e}^{k}\right)\,\left(\boldsymbol{e}_k\cdot\mathsf{K}\left(\boldsymbol{\omega}\right)\cdot\boldsymbol{e}_{\ell}\right) = \boldsymbol{e}^j\cdot\boldsymbol{e}_{\ell}\,.\nonumber
\end{align}
The definitions~\eqref{3.21} and~\eqref{3.38} imply that
\begin{align}\label{A.2.9.1}
&\left(\boldsymbol{e}^j\cdot\boldsymbol{\omega}\right)\,\boldsymbol{n}_{(j)}\left(\boldsymbol{\omega}\right)\cdot\boldsymbol{\mathsf{G}}\nonumber\\
&= \left(\boldsymbol{e}^j\cdot\boldsymbol{\omega}\right)\,\Big(\mathsf{R}\left(\boldsymbol{\omega}\right)^{-1}\cdot\left(\boldsymbol{e}_j\cdot\partial_{\boldsymbol{\omega}}\right)\,\mathsf{R}\left(\boldsymbol{\omega}\right)\Big)\\
&= \mathsf{R}\left(\boldsymbol{\omega}\right)^{-1}\cdot\left(\boldsymbol{\omega}\cdot\partial_{\boldsymbol{\omega}}\right)\,\mathsf{R}\left(\boldsymbol{\omega}\right) = \mathsf{R}\left(\boldsymbol{\omega}\right)^{-1}\cdot\left(\boldsymbol{\omega}\cdot\boldsymbol{\mathsf{G}}\right)\cdot\mathsf{R}\left(\boldsymbol{\omega}\right)\nonumber\\
&= \boldsymbol{\omega}\cdot\boldsymbol{\mathsf{G}}\,,\vphantom{\Big(\Big)}\nonumber
\end{align}
which yields the identity,
\begin{equation}\label{A.2.9.2}
\left(\boldsymbol{e}^j\cdot\boldsymbol{\omega}\right)\, \boldsymbol{n}_{(j)}\left(\boldsymbol{\omega}\right) = \boldsymbol{\omega}\,.
\end{equation}
Moreover, the Killing form~\eqref{A.2.3} and the identity~\eqref{A.2.9.2} then imply that,
\begin{align}\label{A.2.9.3}
&\boldsymbol{\omega}\cdot\boldsymbol{n}_{(\ell)}\left(\boldsymbol{\omega}\right) = \left(\boldsymbol{e}^j\cdot\boldsymbol{\omega}\right)\, \boldsymbol{n}_{(j)}\left(\boldsymbol{\omega}\right)\cdot\boldsymbol{n}_{(\ell)}\left(\boldsymbol{\omega}\right)\nonumber\\
&\phantom{\boldsymbol{\omega}\cdot\boldsymbol{n}_{(\ell)}\left(\boldsymbol{\omega}\right) }= \left(\boldsymbol{e}^j\cdot\boldsymbol{\omega}\right)\,\left(\boldsymbol{e}_j\cdot\mathsf{K}\left(\boldsymbol{\omega}\right)\cdot\boldsymbol{e}_{\ell}\right) =  \boldsymbol{\omega}\cdot\boldsymbol{e}_{\ell}\,.
\end{align}
Using the definition~\eqref{A.2.7}, the first property~\eqref{A.2.5} yields,
\begin{align}\label{A.2.9.4}
&\left(\boldsymbol{e}_j\cdot\boldsymbol{\omega}\right)\, \boldsymbol{m}^{(j)}\left(\boldsymbol{\omega}\right) =  \left(\boldsymbol{e}_j\cdot\boldsymbol{\omega}\right)\,\left(\boldsymbol{e}^j\cdot\mathsf{K}\left(\boldsymbol{\omega}\right)^{-1}\cdot\boldsymbol{e}^{\ell}\right)\, \boldsymbol{n}_{(\ell)}\left(\boldsymbol{\omega}\right)\nonumber\\
&\phantom{\left(\boldsymbol{e}_j\cdot\boldsymbol{\omega}\right)\, \boldsymbol{m}^{(j)}\left(\boldsymbol{\omega}\right) }= \left(\boldsymbol{e}^{\ell}\cdot\boldsymbol{\omega}\right)\,\boldsymbol{n}_{(\ell)}\left(\boldsymbol{\omega}\right) = \boldsymbol{\omega}\,.
\end{align}
Moreover, the identity~\eqref{A.2.9.4} and the Killing form~\eqref{A.2.3} then imply that,
\begin{align}\label{A.2.9.5}
&\boldsymbol{\omega}\cdot\boldsymbol{m}^{(j)}\left(\boldsymbol{\omega}\right) = 
\boldsymbol{\omega}\cdot\left(\boldsymbol{e}^j\cdot\mathsf{K}\left(\boldsymbol{\omega}\right)^{-1}\cdot\boldsymbol{e}^{\ell}\right)\, \boldsymbol{n}_{(\ell)}\left(\boldsymbol{\omega}\right)\\
&\phantom{\boldsymbol{\omega}\cdot\boldsymbol{m}^{(j)}\left(\boldsymbol{\omega}\right)} = \left(\boldsymbol{e}^j\cdot\mathsf{K}\left(\boldsymbol{\omega}\right)^{-1}\cdot\boldsymbol{e}^{\ell}\right)\,\left(\boldsymbol{\omega}\cdot\boldsymbol{e}_{\ell}\right) = \boldsymbol{e}^j\cdot\boldsymbol{\omega}\,.\nonumber
\end{align}

\section{Commutation relations of the momentum and orientation operators}
\label{A.3}
In this appendix, we determine the commutations relations of the momentum operators $\boldsymbol{P}^{\prime\prime}_{\mu}$ and $\boldsymbol{p}^{\prime\prime}_{\nu}$ with the orientation $\boldsymbol{\omega}$. 

In order to determine these relations, we use the Baker-Campbell-Hausdorff formulas, i.e.
\begin{equation}\label{A.2.12bis}
\begin{split}
&e^{\boldsymbol{X}}\,\boldsymbol{Y}\,e^{\boldsymbol{-\,X}} = \sum_{k=0}^{\infty} \frac{1}{k!}\,\left[\ \boldsymbol{X},\ \boldsymbol{Y}\ \right]_{\,k}\,,\\
&e^{\boldsymbol{-\,X}}\,\boldsymbol{Y}\,e^{\boldsymbol{X}} = \sum_{k=0}^{\infty} \frac{\left(-\,1\right)^k}{k!}\,\left[\ \boldsymbol{X},\ \boldsymbol{Y}\ \right]_{\,k}\,,
\end{split}
\end{equation}
where
\begin{equation}\label{A.2.12ter}
\begin{split}
&\left[\ \boldsymbol{X},\ \boldsymbol{Y}\ \right]_{\,k} = \left[\ \boldsymbol{X},\ \left[\ \boldsymbol{X},\ \boldsymbol{Y}\ \right]\,\right]_{\,k-1}\,,\\
&\left[\ \boldsymbol{X},\ \boldsymbol{Y}\ \right]_{\,0} = \boldsymbol{Y}\,. 
\end{split}
\end{equation}
According to the properties~\eqref{3.21},~\eqref{A.2.1bis} and the formulas~\eqref{A.2.12bis},
\begin{align}\label{A.2.12quad}
&\left[\ \boldsymbol{e}_{\ell}\cdot\boldsymbol{P}^{\prime}_{\nu},\ \mathsf{R}\left(\boldsymbol{\omega}\right)\ \right] \nonumber\vphantom{\Big(\Big)}\\
&= \mathsf{R}\left(\boldsymbol{\omega}\right)\cdot\Big(\exp{(-\,\boldsymbol{\omega}\cdot\boldsymbol{\mathsf{G}})}\left(\boldsymbol{e}_{\ell}\cdot\boldsymbol{P}^{\prime}_{\nu}\right)\exp{(\boldsymbol{\omega}\cdot\boldsymbol{\mathsf{G}})}-\,\boldsymbol{e}_{\ell}\cdot\boldsymbol{P}^{\prime}_{\nu}\Big)\nonumber
\\
&= \mathsf{R}\left(\boldsymbol{\omega}\right)\cdot\sum_{k=1}^{\infty}\frac{(-\,1)^{k}}{k!}\left[\ \boldsymbol{\omega}\cdot\boldsymbol{\mathsf{G}},\  \boldsymbol{e}_{\ell}\cdot\boldsymbol{P}^{\prime}_{\nu}\ \right]_{\,k}
\\
&= \mathsf{R}\left(\boldsymbol{\omega}\right)\cdot\!\left[\ \boldsymbol{e}^j\cdot\boldsymbol{\omega},\ \boldsymbol{e}_{\ell}\cdot\boldsymbol{P}^{\prime}_{\nu}\ \right]\!\sum_{k=1}^{\infty}\frac{(-\,1)^{k}}{k!}\left[\ \boldsymbol{\omega}\cdot\boldsymbol{\mathsf{G}},\  \boldsymbol{e}_j\cdot\boldsymbol{\mathsf{G}}\ \right]_{\,k-1}\nonumber
\end{align}
where the $k=0$ term cancels out on the third line and the orientation operator is a function of the rest positions of the nuclei $\boldsymbol{R}^{(0)}_{\mu}$ and the deformation operators $\boldsymbol{R}^{\prime}_{\mu}$. Using the relation,
\begin{equation}\label{A.2.12pet}
\begin{split}
&\left[\ \boldsymbol{e}_j\cdot\partial_{\boldsymbol{\omega}},\ \boldsymbol{\omega}\cdot\boldsymbol{\mathsf{G}}\ \right]\,\boldsymbol{x}\\
&= \left(\boldsymbol{e}_j\cdot\partial_{\boldsymbol{\omega}}\right)\,\left(\boldsymbol{\omega}\times\boldsymbol{x}\right) -\,\boldsymbol{\omega}\times\left(\boldsymbol{e}_j\cdot\partial_{\boldsymbol{\omega}}\right)\,\boldsymbol{x}\\
&= \left(\boldsymbol{e}_j\cdot\partial_{\boldsymbol{\omega}}\right)\,\boldsymbol{\omega}\times\boldsymbol{x} = \boldsymbol{e}_j\times\boldsymbol{x} = \left(\boldsymbol{e}_j\cdot\boldsymbol{\mathsf{G}}\right)\,\boldsymbol{x}\,,
\end{split}
\end{equation}
the Baker-Campbell-Hausdorff formula~\eqref{A.2.12bis} and the group identity~\eqref{3.38}, the commutator~\eqref{A.2.12quad} is recast as,
\begin{align}\label{A.3.1}
&\left[\ \boldsymbol{e}_{\ell}\cdot\boldsymbol{P}^{\prime}_{\nu},\ \mathsf{R}\left(\boldsymbol{\omega}\right)\ \right] \nonumber\\
&=-\,\mathsf{R}\left(\boldsymbol{\omega}\right)\cdot\left[\ \boldsymbol{e}^j\cdot\boldsymbol{\omega},\ \boldsymbol{e}_{\ell}\cdot\boldsymbol{P}^{\prime}_{\nu}\ \right]\!\sum_{k=0}^{\infty}\frac{(-1)^{k}}{k!}\!\left[\ \boldsymbol{\omega}\cdot\boldsymbol{\mathsf{G}},\ \boldsymbol{e}_j\cdot\partial_{\boldsymbol{\omega}}\ \right]_{\,k}\nonumber
\\
&\vphantom{\sum_{k=0}^{\infty}\frac{(-1)^{k}}{k!}} = \mathsf{R}\left(\boldsymbol{\omega}\right)\cdot\left[\ \boldsymbol{e}_{\ell}\cdot\boldsymbol{P}^{\prime}_{\nu},\ \boldsymbol{e}^j\cdot\boldsymbol{\omega}\ \right]\nonumber\\
&\vphantom{\sum_{k=0}^{\infty}\frac{(-1)^{k}}{k!}}\phantom{= } \cdot\exp\left(-\,\boldsymbol{\omega}\cdot\boldsymbol{\mathsf{G}}\right)\,\left(\boldsymbol{e}_j\cdot\partial_{\boldsymbol{\omega}}\right)
\,\exp\left(\boldsymbol{\omega}\cdot\boldsymbol{\mathsf{G}}\right)\nonumber\\
&\vphantom{\sum_{k=0}^{\infty}\frac{(-1)^{k}}{k!}} = \mathsf{R}\left(\boldsymbol{\omega}\right)\cdot\left[\ \boldsymbol{e}_{\ell}\cdot\boldsymbol{P}^{\prime}_{\nu},\ \boldsymbol{e}^j\cdot\boldsymbol{\omega}\ \right]\, \mathsf{R}\left(\boldsymbol{\omega}\right)^{-1}\cdot\left(\boldsymbol{e}_j\cdot\partial_{\boldsymbol{\omega}}\right)\,\mathsf{R}\left(\boldsymbol{\omega}\right)\nonumber\\
&\vphantom{\sum_{k=0}^{\infty}\frac{(-1)^{k}}{k!}} = \left[\ \boldsymbol{e}_{\ell}\cdot\boldsymbol{P}^{\prime}_{\nu},\ \boldsymbol{e}^j\cdot\boldsymbol{\omega}\ \right]\,\mathsf{R}\left(\boldsymbol{\omega}\right)\cdot\left(\boldsymbol{n}_{(j)}\left(\boldsymbol{\omega}\right)\cdot\boldsymbol{\mathsf{G}}\right)\,.
\end{align}
Performing a similar calculation using the properties~\eqref{A.2.1bis} and~\eqref{3.38} yields,
\begin{align}
\label{A.3.1bis}
&\left[\ \boldsymbol{e}_{\ell}\cdot\boldsymbol{P}^{\prime}_{\nu},\ \mathsf{R}\left(\boldsymbol{\omega}\right)^{-1}\ \right]\\
&= -\,\left[\ \boldsymbol{e}_{\ell}\cdot\boldsymbol{P}^{\prime}_{\nu},\ \boldsymbol{e}^j\cdot\boldsymbol{\omega}\ \right]\,\left(\boldsymbol{n}_{(j)}\left(\boldsymbol{\omega}\right)\cdot\boldsymbol{\mathsf{G}}\right)\cdot\mathsf{R}\left(\boldsymbol{\omega}\right)^{-1}\,,\nonumber\\
&\nonumber\\
\label{A.3.2quart}
&\left[\ \boldsymbol{e}_{\ell}\cdot\boldsymbol{P}^{\prime\prime}_{\nu},\ \mathsf{R}\left(\boldsymbol{\omega}\right)\ \right]\\
&= \left[\ \boldsymbol{e}_{\ell}\cdot\boldsymbol{P}^{\prime\prime}_{\nu},\ \boldsymbol{e}^j\cdot\boldsymbol{\omega}\ \right]\,\mathsf{R}\left(\boldsymbol{\omega}\right)\cdot\left(\boldsymbol{n}_{(j)}\left(\boldsymbol{\omega}\right)\cdot\boldsymbol{\mathsf{G}}\right)\,,\nonumber\\
&\nonumber\\
\label{A.3.2ter}
&\left[\ \boldsymbol{e}_{\ell}\cdot\boldsymbol{P}^{\prime\prime}_{\nu},\ \mathsf{R}\left(\boldsymbol{\omega}\right)^{-1}\ \right]\\
&= -\,\left[\ \boldsymbol{e}_{\ell}\cdot\boldsymbol{P}^{\prime\prime}_{\nu},\ \boldsymbol{e}^j\cdot\boldsymbol{\omega}\ \right]\,\left(\boldsymbol{n}_{(j)}\left(\boldsymbol{\omega}\right)\cdot\boldsymbol{\mathsf{G}}\right)\cdot\mathsf{R}\left(\boldsymbol{\omega}\right)^{-1}\,.\nonumber
\end{align}
The condition~\eqref{3.39 ter} implies that,
\begin{equation}\label{A.3.2pent}
\left[\ \boldsymbol{e}_{\ell}\cdot\boldsymbol{P}^{\prime}_{\nu},\ \sum_{\mu=1}^{N} M_{\mu}\left(\mathsf{R}\left(\boldsymbol{\omega}\right)\cdot\boldsymbol{R}^{(0)}_{\mu}\right)\times\boldsymbol{R}^{\prime}_{\mu}\ \right] = \mathsf{0}\,,
\end{equation}
which is expanded as,
\begin{equation}\label{A.3.2hex}
\begin{split}
&\sum_{\mu=1}^{N} M_{\mu}\left(\left[\ \boldsymbol{e}_{\ell}\cdot\boldsymbol{P}^{\prime}_{\nu},\ \mathsf{R}\left(\boldsymbol{\omega}\right)\ \right]\cdot\boldsymbol{R}^{(0)}_{\mu}\right)\times\boldsymbol{R}^{\prime}_{\mu}\\
&+ \sum_{\mu=1}^{N} M_{\mu}\left(\mathsf{R}\left(\boldsymbol{\omega}\right)\cdot\boldsymbol{R}^{(0)}_{\mu}\right)\times\left[\ \boldsymbol{e}_{\ell}\cdot\boldsymbol{P}^{\prime}_{\nu},\ \boldsymbol{R}^{\prime}_{\mu}\ \right] = \mathsf{0}\,.
\end{split}
\end{equation}
The commutator~\eqref{A.3.1} and the rotation group action~\eqref{3.23bis} imply that,
\begin{equation}\label{A.3.2hep}
\begin{split}
&\sum_{\mu=1}^{N} M_{\mu}\left(\left[\ \boldsymbol{e}_{\ell}\cdot\boldsymbol{P}^{\prime}_{\nu},\ \mathsf{R}\left(\boldsymbol{\omega}\right)\ \right]\cdot\boldsymbol{R}^{(0)}_{\mu}\right)\times\boldsymbol{R}^{\prime}_{\mu}\\
&= \left[\ \boldsymbol{e}_{\ell}\cdot\boldsymbol{P}^{\prime}_{\nu},\ \boldsymbol{e}^j\cdot\boldsymbol{\omega}\ \right]\\
&\phantom{=} \cdot\sum_{\mu=1}^{N} M_{\mu}\left( \mathsf{R}\left(\boldsymbol{\omega}\right)\cdot\left(\boldsymbol{n}_{(j)}\left(\boldsymbol{\omega}\right)\cdot\boldsymbol{\mathsf{G}}\right) \boldsymbol{R}^{(0)}_{\mu}\right)\times\boldsymbol{R}^{\prime}_{\mu}\\
&= \left[\ \boldsymbol{e}_{\ell}\cdot\boldsymbol{P}^{\prime}_{\nu},\ \boldsymbol{e}^j\cdot\boldsymbol{\omega}\ \right]\\
&\phantom{=} \cdot\sum_{\mu=1}^{N} M_{\mu}\left( \mathsf{R}\left(\boldsymbol{\omega}\right)\cdot\left(\boldsymbol{n}_{(j)}\left(\boldsymbol{\omega}\right)\times\boldsymbol{R}^{(0)}_{\mu}\right)\right)\times\boldsymbol{R}^{\prime}_{\mu}\,,
\end{split}
\end{equation}
the commutation relations~\eqref{3.13} and the constraint~\eqref{3.16} imply that,
\begin{align}\label{A.3.2oct}
&\sum_{\mu=1}^{N} M_{\mu}\left(\mathsf{R}\left(\boldsymbol{\omega}\right)\cdot\boldsymbol{R}^{(0)}_{\mu}\right)\times\left[\ \boldsymbol{e}_{\ell}\cdot\boldsymbol{P}^{\prime}_{\nu},\ \boldsymbol{R}^{\prime}_{\mu}\ \right]\nonumber
\\
&= -\,i\hbar\ \sum_{\mu=1}^{N} M_{\mu}\left(\mathsf{R}\left(\boldsymbol{\omega}\right)\cdot\boldsymbol{R}^{(0)}_{\mu}\right)\times\left(\delta_{\nu\mu}-\frac{M_{\nu}}{\mathcal{M}}\right)\,\boldsymbol{e}_{\ell}\nonumber\\
& = -\,i\hbar\, M_{\nu}\,\left(\mathsf{R}\left(\boldsymbol{\omega}\right)\cdot\boldsymbol{R}^{(0)}_{\nu}\right)\times\boldsymbol{e}_{\ell}\,.
\end{align}
The identities~\eqref{A.3.2hex},~\eqref{A.3.2hep} and~\eqref{A.3.2oct} yield,
\begin{align}\label{A.3.4}
&\left[\ \boldsymbol{e}_{\ell}\cdot\boldsymbol{P}^{\prime}_{\nu},\ \boldsymbol{e}^j\cdot\boldsymbol{\omega}\ \right]\!\sum_{\mu=1}^{N} M_{\mu}\!\left( \mathsf{R}\left(\boldsymbol{\omega}\right)\!\cdot\!\left(\boldsymbol{n}_{(j)}\left(\boldsymbol{\omega}\right)\!\times\!\boldsymbol{R}^{(0)}_{\mu}\right)\right)\!\times\!\boldsymbol{R}^{\prime}_{\mu}\nonumber\\
&= i\hbar\, M_{\nu}\,\left(\mathsf{R}\left(\boldsymbol{\omega}\right)\cdot\boldsymbol{R}^{(0)}_{\nu}\right)\times\boldsymbol{e}_{\ell}\,.
\end{align}
Multiplying the commutation relation~\eqref{A.3.4} by $\boldsymbol{e}^{\ell}\cdot\mathsf{R}\left(\boldsymbol{\omega}\right)\cdot\boldsymbol{e}_{k}$ and using the relation~\eqref{3.13ter} between $\boldsymbol{P}^{\prime}_{\mu}$ and $\boldsymbol{P}^{\prime\prime}_{\mu}$ yields,
\begin{align}\label{A.3.4 prime}
&\left[\ \boldsymbol{e}_{\ell}\cdot\boldsymbol{P}^{\prime\prime}_{\nu},\ \boldsymbol{e}^j\cdot\boldsymbol{\omega}\ \right]\!\sum_{\mu=1}^{N} M_{\mu}\!\left( \mathsf{R}\left(\boldsymbol{\omega}\right)\!\cdot\!\left(\boldsymbol{n}_{(j)}\left(\boldsymbol{\omega}\right)\!\times\!\boldsymbol{R}^{(0)}_{\mu}\right)\right)\!\times\!\boldsymbol{R}^{\prime}_{\mu}\nonumber\\
&= i\hbar\, M_{\nu}\,\left(\mathsf{R}\left(\boldsymbol{\omega}\right)\cdot\boldsymbol{R}^{(0)}_{\nu}\right)\times\left(\mathsf{R}\left(\boldsymbol{\omega}\right)\cdot\boldsymbol{e}_{\ell}\right)\\
&= i\hbar\, M_{\nu}\,\mathsf{R}\left(\boldsymbol{\omega}\right)\cdot\left(\boldsymbol{R}^{(0)}_{\nu}\times\boldsymbol{e}_{\ell}\right)\,.\nonumber
\end{align}
The relation~\eqref{3.13bis} between $\boldsymbol{R}^{\prime}_{\mu}$ and $\boldsymbol{R}^{\prime\prime}_{\mu}$ implies that,
\begin{equation}\label{A.3.4 second}
\begin{split}
&\left(\mathsf{R}\left(\boldsymbol{\omega}\right)\cdot\left(\boldsymbol{n}_{(j)}\left(\boldsymbol{\omega}\right)\times\boldsymbol{R}^{(0)}_{\mu}\right)\right)\times\boldsymbol{R}^{\prime}_{\mu}\\
&= \left(\mathsf{R}\left(\boldsymbol{\omega}\right)\cdot\left(\boldsymbol{n}_{(j)}\left(\boldsymbol{\omega}\right)\times\boldsymbol{R}^{(0)}_{\mu}\right)\right)\times\left(\mathsf{R}\left(\boldsymbol{\omega}\right)\cdot\boldsymbol{R}^{\prime\prime}_{\mu}\right)\\
&= \mathsf{R}\left(\boldsymbol{\omega}\right)\cdot\left(\left(\boldsymbol{n}_{(j)}\left(\boldsymbol{\omega}\right)\times\boldsymbol{R}^{(0)}_{\mu}\right)\times\boldsymbol{R}^{\prime\prime}_{\mu}\right)\,.
\end{split}
\end{equation}
The identities~\eqref{A.3.4 prime} and~\eqref{A.3.4 second} imply in turn that,
\begin{align}\label{A.3.4bis}
&\left[\ \boldsymbol{e}_{\ell}\cdot\boldsymbol{P}^{\prime\prime}_{\nu},\ \boldsymbol{e}^j\cdot\boldsymbol{\omega}\ \right]\,\sum_{\mu=1}^{N} M_{\mu}\left(\boldsymbol{n}_{(j)}\left(\boldsymbol{\omega}\right)\times\boldsymbol{R}^{(0)}_{\mu}\right)\times\boldsymbol{R}^{\prime\prime}_{\mu}\nonumber\\
&= i\hbar\, M_{\nu}\,\boldsymbol{R}^{(0)}_{\nu}\times\boldsymbol{e}_{\ell}\,.
\end{align}

Now, by analogy with the commutation relation~\eqref{A.3.1},
\begin{equation}\label{A.3.4ter}
\left[\ \boldsymbol{e}_{\ell}\cdot\boldsymbol{p}^{\prime}_{\nu},\ \mathsf{R}\left(\boldsymbol{\omega}\right)\ \right] = \left[\ \boldsymbol{e}_{\ell}\cdot\boldsymbol{p}^{\prime}_{\nu},\ \boldsymbol{e}^j\cdot\boldsymbol{\omega}\ \right]\,\mathsf{R}\left(\boldsymbol{\omega}\right)\cdot\left(\boldsymbol{n}_{(j)}\left(\boldsymbol{\omega}\right)\!\cdot\!\boldsymbol{\mathsf{G}}\right)
\end{equation}
The condition~\eqref{3.39 ter} implies that,
\begin{equation}\label{A.3.4 quattro}
\left[\ \boldsymbol{e}_{\ell}\cdot\boldsymbol{p}^{\prime}_{\nu},\ \sum_{\mu=1}^{N} M_{\mu}\left(\mathsf{R}\left(\boldsymbol{\omega}\right)\cdot\boldsymbol{R}^{(0)}_{\mu}\right)\times\boldsymbol{R}^{\prime}_{\mu}\ \right] = \mathsf{0}\,,
\end{equation}
which is expanded as,
\begin{equation}\label{A.3.4pent}
\begin{split}
&\sum_{\mu=1}^{N} M_{\mu}\left(\left[\ \boldsymbol{e}_{\ell}\cdot\boldsymbol{p}^{\prime}_{\nu},\ \mathsf{R}\left(\boldsymbol{\omega}\right)\ \right]\cdot\boldsymbol{R}^{(0)}_{\mu}\right)\times\boldsymbol{R}^{\prime}_{\mu}\\
&+ \sum_{\mu=1}^{N} M_{\mu}\left(\mathsf{R}\left(\boldsymbol{\omega}\right)\cdot\boldsymbol{R}^{(0)}_{\mu}\right)\times\left[\ \boldsymbol{e}_{\ell}\cdot\boldsymbol{p}^{\prime}_{\nu},\ \boldsymbol{R}^{\prime}_{\mu}\ \right] = \mathsf{0}\,.
\end{split}
\end{equation}
The commutator~\eqref{A.3.4ter} and the rotation group action~\eqref{3.23bis} imply that,
\begin{equation}\label{A.3.4hex}
\begin{split}
&\sum_{\mu=1}^{N} M_{\mu}\left(\left[\ \boldsymbol{e}_{\ell}\cdot\boldsymbol{p}^{\prime}_{\nu},\ \mathsf{R}\left(\boldsymbol{\omega}\right)\ \right]\cdot\boldsymbol{R}^{(0)}_{\mu}\right)\times\boldsymbol{R}^{\prime}_{\mu}\\
&= \left[\ \boldsymbol{e}_{\ell}\cdot\boldsymbol{p}^{\prime}_{\nu},\ \boldsymbol{e}^j\cdot\boldsymbol{\omega}\ \right]\\
&\phantom{= }\cdot\sum_{\mu=1}^{N} M_{\mu}\left( \mathsf{R}\left(\boldsymbol{\omega}\right)\cdot\left(\boldsymbol{n}_{(j)}\left(\boldsymbol{\omega}\right)\cdot\boldsymbol{\mathsf{G}}\right) \boldsymbol{R}^{(0)}_{\mu}\right)\times\boldsymbol{R}^{\prime}_{\mu}\\
&= \left[\ \boldsymbol{e}_{\ell}\cdot\boldsymbol{p}^{\prime}_{\nu},\ \boldsymbol{e}^j\cdot\boldsymbol{\omega}\ \right]\\
&\phantom{= }\cdot\sum_{\mu=1}^{N} M_{\mu}\left( \mathsf{R}\left(\boldsymbol{\omega}\right)\cdot\left(\boldsymbol{n}_{(j)}\left(\boldsymbol{\omega}\right)\times\boldsymbol{R}^{(0)}_{\mu}\right)\right)\times\boldsymbol{R}^{\prime}_{\mu}\,,
\end{split}
\end{equation}
the commutations relations~\eqref{3.13} and the constraint~\eqref{3.16} imply that,
\begin{align}\label{A.3.4hep}
&\sum_{\mu=1}^{N} M_{\mu}\left(\mathsf{R}\left(\boldsymbol{\omega}\right)\cdot\boldsymbol{R}^{(0)}_{\mu}\right)\times\left[\ \boldsymbol{e}_{\ell}\cdot\boldsymbol{p}^{\prime}_{\nu},\ \boldsymbol{R}^{\prime}_{\mu}\ \right]
\nonumber\\
&= -\,i\hbar\ \sum_{\mu=1}^{N} M_{\mu}\left(\mathsf{R}\left(\boldsymbol{\omega}\right)\cdot\boldsymbol{R}^{(0)}_{\mu}\right)\times\left(\frac{m}{\mathcal{M}}\,\boldsymbol{e}_{\ell}\right) = 0\,.
\end{align}
The identities~\eqref{A.3.4pent},~\eqref{A.3.4hex} and~\eqref{A.3.4hep} yield,
\begin{equation}\label{A.3.4quad}
\begin{split}
&\left[\ \boldsymbol{e}_{\ell}\cdot\boldsymbol{p}^{\prime}_{\nu},\ \boldsymbol{e}^j\cdot\boldsymbol{\omega}\ \right]\\
&\cdot\sum_{\mu=1}^{N} M_{\mu}\left( \mathsf{R}\left(\boldsymbol{\omega}\right)\cdot\left(\boldsymbol{n}_{(j)}\left(\boldsymbol{\omega}\right)\times\boldsymbol{R}^{(0)}_{\mu}\right)\right)\times\boldsymbol{R}^{\prime}_{\mu}
= 0\,.
\end{split}
\end{equation}
Finally, by analogy with the identities~\eqref{A.3.4}-\eqref{A.3.4bis}, we obtain the last commutation relation~\eqref{3.31.1}, i.e.
\begin{equation}\label{A.3.4.0}
\left[\ \boldsymbol{e}_{\ell}\cdot\boldsymbol{p}^{\prime\prime}_{\nu},\ \boldsymbol{e}^j\cdot\boldsymbol{\omega}\ \right]= 0\,.
\end{equation}

\section{Commutation relations of the position and momentum operators}
\label{A.4}
In this appendix, we determine the commutations relations of the momentum operator $\boldsymbol{P}^{\prime\prime}_{\mu}$ with the position operators $\boldsymbol{R}^{\prime\prime}_{\nu}$ and $\boldsymbol{r}^{\prime\prime}_{\nu}$, and the momentum operators $\boldsymbol{P}^{\prime\prime}_{\nu}$ and $\boldsymbol{p}^{\prime\prime}_{\nu}$ respectively. 

Using the relations~\eqref{3.13bis} and~\eqref{3.13ter}, the commutation relation yields,
\begin{align}\label{A.4.10}
&\left[\ \boldsymbol{e}_j\cdot\boldsymbol{P}^{\prime\prime}_{\mu},\  \boldsymbol{e}^k\cdot\boldsymbol{r}^{\prime\prime}_{\nu}\ \right]\nonumber\\
&= \left[\ \left(\mathsf{R}\left(\boldsymbol{\omega}\right)\cdot\boldsymbol{e}_j\right)\cdot\boldsymbol{P}^{\prime}_{\mu},\  \left(\boldsymbol{e}^k\cdot\mathsf{R}\left(\boldsymbol{\omega}\right)^{-1}\right)\cdot\boldsymbol{r}^{\prime}_{\nu}\ \right]\\
&= \left(\boldsymbol{e}^{\ell}\cdot\mathsf{R}\left(\boldsymbol{\omega}\right)\cdot\boldsymbol{e}_j\right)\left[\ \boldsymbol{e}_{\ell}\cdot\boldsymbol{P}^{\prime}_{\mu},\ \boldsymbol{e}^{k}\cdot\mathsf{R}\left(\boldsymbol{\omega}\right)^{-1}\cdot\boldsymbol{e}_{m}\ \right]\,\boldsymbol{e}^{m}\cdot\boldsymbol{r}^{\prime}_{\nu}\nonumber\\ 
&+\left(\boldsymbol{e}^{\ell}\cdot\mathsf{R}\left(\boldsymbol{\omega}\right)\cdot\boldsymbol{e}_j\right)\left(\boldsymbol{e}^{k}\cdot\mathsf{R}\left(\boldsymbol{\omega}\right)^{-1}\!\cdot\boldsymbol{e}_m\right)\left[\ \boldsymbol{e}_{\ell}\cdot\boldsymbol{P}^{\prime}_{\mu},\ \boldsymbol{e}^{m}\cdot\boldsymbol{r}^{\prime}_{\nu}\ \right]\nonumber
\end{align}
Using the commutation relations~\eqref{3.13} and~\eqref{A.3.1bis}, the definition of the group action~\eqref{3.23bis} and the relations~\eqref{3.13bis} and~\eqref{3.13ter}, the commutation relation~\eqref{A.4.10} yields the first commutation relation~\eqref{3.17quad}, i.e.
\begin{align}\label{A.4.10bis}
&\left[\ \boldsymbol{e}_j\cdot\boldsymbol{P}^{\prime\prime}_{\mu},\  \boldsymbol{e}^k\cdot\boldsymbol{r}^{\prime\prime}_{\nu}\ \right]\nonumber\\
&= -\,\left(\boldsymbol{e}^{\ell}\cdot\mathsf{R}\left(\boldsymbol{\omega}\right)\cdot\boldsymbol{e}_j\right)\,
\left[\ \boldsymbol{e}_{\ell}\cdot\boldsymbol{P}^{\prime}_{\mu},\ \boldsymbol{e}^{s}\cdot\boldsymbol{\omega}\ \right]\nonumber\\
&\cdot\boldsymbol{e}^{k}\cdot\left(\boldsymbol{n}_{(s)}\left(\boldsymbol{\omega}\right)\cdot\boldsymbol{\mathsf{G}}\right)\cdot\left(\mathsf{R}\left(\boldsymbol{\omega}\right)^{-1}\cdot\boldsymbol{e}_m\right)\left(\boldsymbol{e}^{m}\cdot\mathsf{R}\left(\boldsymbol{\omega}\right)\cdot\boldsymbol{r}^{\prime\prime}_{\nu}\right)\nonumber\\
&+ \left(\boldsymbol{e}^{\ell}\cdot\mathsf{R}\left(\boldsymbol{\omega}\right)\cdot\boldsymbol{e}_j\right)\left(\boldsymbol{e}^{k}\cdot\mathsf{R}\left(\boldsymbol{\omega}\right)^{-1}\!\cdot\boldsymbol{e}_m\right)\,i\hbar\,\left(\boldsymbol{e}_{\ell}\cdot\boldsymbol{e}^{m}\right)\,\frac{M_{\mu}}{\mathcal{M}}\,\mathbb{1}\nonumber\\
&= -\,\left[\ \boldsymbol{e}_{j}\cdot\boldsymbol{P}^{\prime\prime}_{\mu},\ \boldsymbol{e}^{s}\cdot\boldsymbol{\omega}\ \right]\,\boldsymbol{e}^{k}\cdot\left(\boldsymbol{n}_{(s)}\left(\boldsymbol{\omega}\right)\times\boldsymbol{r}^{\prime\prime}_{\nu}\right)\\
&+ i\hbar\,\left(\boldsymbol{e}_{j}\cdot\boldsymbol{e}^{k}\right)\,\frac{M_{\mu}}{\mathcal{M}}\,\mathbb{1}\,.\nonumber
\end{align}
Similarly, using the relations~\eqref{3.13bis} and~\eqref{3.13ter}, the commutation relation yields,
\begin{align}\label{A.4.11}
&\left[\ \boldsymbol{e}_j\cdot\boldsymbol{P}^{\prime\prime}_{\mu},\  \boldsymbol{e}_k\cdot\boldsymbol{p}^{\prime\prime}_{\nu}\ \right]\nonumber\\
&= \left[\ \left(\mathsf{R}\left(\boldsymbol{\omega}\right)\cdot\boldsymbol{e}_j\right)\cdot\boldsymbol{P}^{\prime}_{\mu},\  \left(\mathsf{R}\left(\boldsymbol{\omega}\right)\cdot\boldsymbol{e}_k\right)\cdot\boldsymbol{p}^{\prime}_{\nu}\ \right]\\
&= \left(\boldsymbol{e}^{\ell}\cdot\mathsf{R}\left(\boldsymbol{\omega}\right)\cdot\boldsymbol{e}_j\right)\,
\left[\ \boldsymbol{e}_{\ell}\cdot\boldsymbol{P}^{\prime}_{\mu},\ \boldsymbol{e}^{m}\cdot\mathsf{R}\left(\boldsymbol{\omega}\right)\cdot\boldsymbol{e}_{k}\ \right]\,\boldsymbol{e}_{m}\cdot\boldsymbol{p}^{\prime}_{\nu}\nonumber\\ 
&+ \left(\boldsymbol{e}^{\ell}\cdot\mathsf{R}\left(\boldsymbol{\omega}\right)\cdot\boldsymbol{e}_j\right)\,
\left(\boldsymbol{e}^{m}\cdot\mathsf{R}\left(\boldsymbol{\omega}\right)\cdot\boldsymbol{e}_{k}\right)\,
\left[\ \boldsymbol{e}_{\ell}\cdot\boldsymbol{P}^{\prime}_{\mu},\ \boldsymbol{e}_{m}\cdot\boldsymbol{p}^{\prime}_{\nu}\ \right]\,.\nonumber
\end{align}
Using the commutation relations~\eqref{3.13} and~\eqref{A.3.1}, the definition of the group action~\eqref{3.23bis} and the relation~\eqref{3.13ter}, the commutation relation~\eqref{A.4.11} yields the second commutation relation~\eqref{3.17quad}, i.e.
\begin{align}\label{A.4.11bis}
&\left[\ \boldsymbol{e}_j\cdot\boldsymbol{P}^{\prime\prime}_{\mu},\  \boldsymbol{e}_k\cdot\boldsymbol{p}^{\prime\prime}_{\nu}\ \right]\nonumber\\
&= \left(\boldsymbol{e}^{\ell}\cdot\mathsf{R}\left(\boldsymbol{\omega}\right)\cdot\boldsymbol{e}_j\right)\,
\left[\ \boldsymbol{e}_{\ell}\cdot\boldsymbol{P}^{\prime}_{\mu},\ \boldsymbol{e}^{s}\cdot\boldsymbol{\omega}\ \right]\\
&\cdot\boldsymbol{e}_{k}\cdot\left(\boldsymbol{n}_{(s)}\left(\boldsymbol{\omega}\right)\cdot\boldsymbol{\mathsf{G}}\right)\cdot\left(\boldsymbol{e}^{m}\cdot\mathsf{R}\left(\boldsymbol{\omega}\right)\right)\,\left(\mathsf{R}\left(\boldsymbol{\omega}\right)^{-1}\cdot\boldsymbol{e}_{m}\right)\cdot\boldsymbol{p}^{\prime\prime}_{\nu}\nonumber\\
&= \left[\ \boldsymbol{e}_{j}\cdot\boldsymbol{P}^{\prime\prime}_{\mu},\ \boldsymbol{e}^{s}\cdot\boldsymbol{\omega}\ \right]\,\boldsymbol{e}_{k}\cdot\left(\boldsymbol{n}_{(s)}\left(\boldsymbol{\omega}\right)\times\boldsymbol{p}^{\prime\prime}_{\nu}\right)\,.\nonumber
\end{align}
By analogy with the commutation relation~\eqref{A.4.10}, using the relations~\eqref{3.13bis} and~\eqref{3.13ter}, the commutation relation yields,
\begin{align}\label{A.4.11ter}
&\left[\ \boldsymbol{e}_j\cdot\boldsymbol{P}^{\prime\prime}_{\mu},\  \boldsymbol{e}^k\cdot\boldsymbol{R}^{\prime\prime}_{\nu}\ \right]\\
&= \left(\boldsymbol{e}^{\ell}\cdot\mathsf{R}\left(\boldsymbol{\omega}\right)\cdot\boldsymbol{e}_j\right)\,
\left[\ \boldsymbol{e}_{\ell}\cdot\boldsymbol{P}^{\prime}_{\mu},\ \boldsymbol{e}^{k}\cdot\mathsf{R}\left(\boldsymbol{\omega}\right)^{-1}\!\cdot\boldsymbol{e}_{m}\ \right]\,\boldsymbol{e}^{m}\cdot\boldsymbol{R}^{\prime}_{\nu}\nonumber\\ 
&+ \left(\boldsymbol{e}^{\ell}\cdot\mathsf{R}\left(\boldsymbol{\omega}\right)\cdot\boldsymbol{e}_j\right)\left(\boldsymbol{e}^{k}\cdot\mathsf{R}\left(\boldsymbol{\omega}\right)^{-1}\!\cdot\boldsymbol{e}_m\right)\left[\ \boldsymbol{e}_{\ell}\cdot\boldsymbol{P}^{\prime}_{\mu},\ \boldsymbol{e}^{m}\cdot\boldsymbol{R}^{\prime}_{\nu}\ \right]\nonumber
\end{align}
By analogy with the commutation relation~\eqref{A.4.10bis}, using the commutation relations~\eqref{3.13} and~\eqref{A.3.1bis}, the definition of the group action~\eqref{3.23bis} and the relations~\eqref{3.13bis} and~\eqref{3.13ter} yield the third commutation relation~\eqref{3.17quad}, i.e.
\begin{align}\label{A.4.12}
&\left[\ \boldsymbol{e}_j\cdot\boldsymbol{P}^{\prime\prime}_{\mu},\  \boldsymbol{e}^k\cdot\boldsymbol{R}^{\prime\prime}_{\nu}\ \right] = -\,i\hbar\left(\boldsymbol{e}_j\cdot\boldsymbol{e}^k\right)\,\left(\delta_{\mu\nu}-\,\frac{M_{\mu}}{\mathcal{M}}\right)\,\mathbb{1}\nonumber\\
&-\,\left[\ \boldsymbol{e}_{j}\cdot\boldsymbol{P}^{\prime\prime}_{\mu},\ \boldsymbol{e}^{s}\cdot\boldsymbol{\omega}\ \right]\,\boldsymbol{e}^{k}\cdot\left(\boldsymbol{n}_{(s)}\left(\boldsymbol{\omega}\right)\times\boldsymbol{R}^{\prime\prime}_{\nu}\right)\,. 
\end{align}
Using the relation~\eqref{3.13ter}, the commutation relation of the momentum operators is expressed as, 
\begin{align}\label{A.4.13bis}
\begin{split}
&\left[\ \boldsymbol{e}_j\cdot\boldsymbol{P}^{\prime\prime}_{\mu},\  \boldsymbol{e}_k\cdot\boldsymbol{P}^{\prime\prime}_{\nu}\ \right] =\\
&\frac{1}{2}\,\Big\{\ \boldsymbol{e}_{\ell}\cdot\boldsymbol{P}^{\prime}_{\nu},\ \left[\ \boldsymbol{e}_j\cdot\boldsymbol{P}^{\prime\prime}_{\mu},\ \boldsymbol{e}^{\ell}\cdot\mathsf{R}\left(\boldsymbol{\omega}\right)\cdot\boldsymbol{e}_{k}\ \right]\ \Big\}\\
&-\,\frac{1}{2}\,\Big\{\ \boldsymbol{e}_{\ell}\cdot\boldsymbol{P}^{\prime}_{\mu},\ \left[\ \boldsymbol{e}_{k}\cdot\boldsymbol{P}^{\prime\prime}_{\nu},\ \boldsymbol{e}^{\ell}\cdot\mathsf{R}\left(\boldsymbol{\omega}\right)\cdot\boldsymbol{e}_{j}\ \right]\ \Big\}\,.
\end{split}	
\end{align}
According to equation~\eqref{A.3.2quart}, 
\begin{align}\label{A.4.13bisbis}
&\left[\ \boldsymbol{e}_j\cdot\boldsymbol{P}^{\prime\prime}_{\mu},\ \boldsymbol{e}^{\ell}\cdot\mathsf{R}\left(\boldsymbol{\omega}\right)\cdot\boldsymbol{e}_{k}\ \right]\\
&= \left[\ \boldsymbol{e}_j\cdot\boldsymbol{P}^{\prime\prime}_{\mu},\ \boldsymbol{e}^{m}\cdot\boldsymbol{\omega}\ \right]\, \boldsymbol{e}^{\ell}\cdot\left(\mathsf{R}\left(\boldsymbol{\omega}\right)\cdot\left(\boldsymbol{n}_{(m)}\left(\boldsymbol{\omega}\right)\cdot\boldsymbol{\mathsf{G}}\right)\right)\cdot\boldsymbol{e}_{k}\,.\nonumber
\end{align}
Introducing the rank-$2$ tensorial operator,
\begin{equation}\label{A.4.13quart}
\mathsf{A}_{\mu\left(j\right)} = \left[\ \boldsymbol{e}_j\cdot\boldsymbol{P}^{\prime\prime}_{\mu},\ \boldsymbol{e}^{m}\cdot\boldsymbol{\omega}\ \right]\left(\boldsymbol{n}_{(m)}\left(\boldsymbol{\omega}\right)\cdot\boldsymbol{\mathsf{G}}\right)\,,
\end{equation}
and using the relation~\eqref{A.4.13bisbis}, the commutator~\eqref{A.4.13bis} is recast as,
\begin{align}\label{A.4.13ter}
&\left[\ \boldsymbol{e}_j\cdot\boldsymbol{P}^{\prime\prime}_{\mu},\  \boldsymbol{e}_k\cdot\boldsymbol{P}^{\prime\prime}_{\nu}\ \right] =\\
&\frac{1}{2}\,\Big\{\ \boldsymbol{e}_{\ell}\cdot\boldsymbol{P}^{\prime}_{\nu},\ \left(\boldsymbol{e}^{\ell}\cdot\mathsf{R}\left(\boldsymbol{\omega}\right)\cdot\boldsymbol{e}_{n}\right)\,\left(\boldsymbol{e}^{n}\cdot\mathsf{A}_{\mu\left(j\right)}\cdot\boldsymbol{e}_{k}\right)\ \Big\}\nonumber\\
&-\,\frac{1}{2}\,\Big\{\ \boldsymbol{e}_{\ell}\cdot\boldsymbol{P}^{\prime}_{\mu},\  \left(\boldsymbol{e}^{\ell}\cdot\mathsf{R}\left(\boldsymbol{\omega}\right)\cdot\boldsymbol{e}_{n}\right)\,\left(\boldsymbol{e}^{n}\cdot\mathsf{A}_{\nu\left(k\right)}\cdot\boldsymbol{e}_{j}\right)\ \Big\}\,.\nonumber	
\end{align}
The relation~\eqref{3.13ter} implies that the commutation relation~\eqref{A.4.13ter} reduces to,
\begin{align}\label{A.4.13quint}
&\left[\ \boldsymbol{e}_j\cdot\boldsymbol{P}^{\prime\prime}_{\mu},\  \boldsymbol{e}_k\cdot\boldsymbol{P}^{\prime\prime}_{\nu}\ \right] = \frac{1}{2}\,\Big\{\ \boldsymbol{e}_{\ell}\cdot\boldsymbol{P}^{\prime\prime}_{\nu},\ \boldsymbol{e}^{\ell}\cdot\mathsf{A}_{\mu\left(j\right)}\cdot\boldsymbol{e}_{k}\ \Big\}\nonumber\\
& -\,\frac{1}{2}\,\Big\{\ \boldsymbol{e}_{\ell}\cdot\boldsymbol{P}^{\prime\prime}_{\mu},\  \boldsymbol{e}^{\ell}\cdot\mathsf{A}_{\nu\left(k\right)}\cdot\boldsymbol{e}_{j}\ \Big\}\,.
\end{align}
Finally, using the expression~\eqref{A.4.13quart} of the rank-2 tensor $\mathsf{A}_{\mu\left(j\right)}$ and the definition of the rotation group action~\eqref{3.23bis} the commutation relation~\eqref{A.4.13quint} yields the last commutation relation~\eqref{3.17quad}, i.e.
\begin{align}\label{A.4.14}
&\left[\ \boldsymbol{e}_j\cdot\boldsymbol{P}^{\prime\prime}_{\mu},\  \boldsymbol{e}_k\cdot\boldsymbol{P}^{\prime\prime}_{\nu}\ \right] =\\
&\frac{1}{2}\,\Big\{\ \boldsymbol{e}_{\ell}\cdot\boldsymbol{P}^{\prime\prime}_{\nu},\ 
\left[\ \boldsymbol{e}_j\cdot\boldsymbol{P}^{\prime\prime}_{\mu},\ \boldsymbol{e}^{m}\cdot\boldsymbol{\omega}\ \right]\left(\boldsymbol{n}_{(m)}\left(\boldsymbol{\omega}\right)\times\boldsymbol{e}_{k}\right)\cdot\boldsymbol{e}^{\ell}\ \Big\}\nonumber\\
&-\frac{1}{2}\,\Big\{\ \boldsymbol{e}_{\ell}\cdot\boldsymbol{P}^{\prime\prime}_{\mu},\  \left[\ \boldsymbol{e}_k\cdot\boldsymbol{P}^{\prime\prime}_{\nu},\ \boldsymbol{e}^{m}\cdot\boldsymbol{\omega}\ \right]\left(\boldsymbol{n}_{(m)}\left(\boldsymbol{\omega}\right)\times\boldsymbol{e}_{j}\right)\cdot\boldsymbol{e}^{\ell}\ \Big\}\nonumber
\end{align}

\section{Internal observables}
\label{A.5}
In this appendix, we determine the expressions for the internal observables $Q^{\alpha}$, $P_{\alpha}$ and $\boldsymbol{\Omega}$. The definition~\eqref{3.14} and the constraint~\eqref{3.18} imply that
\begin{align}\label{A.5.3 bis}
&\sum_{\mu=1}^{N}\,\sqrt{M_{\mu}}\,\left(\boldsymbol{X}^{\alpha}_{\mu}\cdot\boldsymbol{R}^{\prime\prime}_{\mu}\right)
 = \sum_{\mu=1}^{N}\,\sqrt{M_{\mu}}\, \left(\boldsymbol{X}^{\alpha}_{\mu}\cdot\boldsymbol{R}^{(0)}_{\mu}\right)\,\mathbb{1}
+ Q^{\alpha}\nonumber\\
&-\,\frac{m}{M}\,\sum_{\mu=1}^{N}\sum_{\nu,\nu^{\prime}=1}^{n}\,\sqrt{M_{\mu}}\,A_{\nu\nu^{\prime}}\, \left(\boldsymbol{X}^{\alpha}_{\mu}\cdot\boldsymbol{q}_{(\nu^{\prime})}\right)\,.
\end{align}
Using the constraint~\eqref{3.17}, the identity~\eqref{A.5.3 bis} yields the expression~\eqref{3.27 prime} for the internal observable $Q^{\alpha}$, i.e.
\begin{equation}\label{A.5.4}
Q^{\alpha} = \sum_{\mu=1}^{N}\,\sqrt{M_{\mu}}\, \boldsymbol{X}^{\alpha}_{\mu}\cdot\Big(\boldsymbol{R}^{\prime\prime}_{\mu}-\,\boldsymbol{R}^{(0)}_{\mu}\,\mathbb{1}\Big)\,.
\end{equation}
Similarly, the definition~\eqref{3.25} and the constraint~\eqref{3.18} imply that, 
\begin{align}\label{A.5.1}
&\sum_{\mu=1}^{N}\,\frac{1}{\sqrt{M_{\mu}}}\,\left(\boldsymbol{X}_{\mu\alpha}\cdot\boldsymbol{P}^{\prime\prime}_{\mu}\right)
 =\nonumber\\
&\sum_{\mu=1}^{N}\,\frac{1}{\sqrt{M_{\mu}}}\, \left(\boldsymbol{X}_{\mu\alpha}\cdot\left(\boldsymbol{\Omega}\times\left(M_{\mu}\,\boldsymbol{R}^{(0)}_{\mu}\right)\right)\right)\,\mathbb{1}
+P_{\alpha}\\
&-\,\frac{m}{M}\,\sum_{\mu=1}^{N}\sum_{\nu,\nu^{\prime}=1}^{n}\,\sqrt{M_{\mu}}\,A_{\nu\nu^{\prime}}\, \left(\boldsymbol{X}_{\mu\alpha}\cdot\boldsymbol{p}_{(\nu^{\prime})}\right)\,.\nonumber
\end{align}
Using the constraints~\eqref{3.17} and~\eqref{3.17bis} the identity~\eqref{A.5.1} yields the expression~\eqref{3.27 prime} for the internal observable $P_{\alpha}$, i.e.
\begin{equation}\label{A.5.9}
P_{\alpha}=\sum_{\mu=1}^{N}\frac{1}{\sqrt{M_{\mu}}}\,\left(\boldsymbol{X}_{\mu\alpha}\cdot\boldsymbol{P}^{\prime\prime}_{\mu}\right)\,.
\end{equation}

The relation~\eqref{3.13ter}, the constraints~\eqref{3.16}-\eqref{3.17bis} and the definition~\eqref{3.32} imply that,
\begin{align}\label{A.5.9bis}
&\sum_{\mu=1}^{N}\,\boldsymbol{R}^{(0)}_{\mu}\times\boldsymbol{P}^{\prime\prime}_{\mu} = \sum_{\mu=1}^{N}\boldsymbol{R}^{(0)}_{\mu}\times\Big(\boldsymbol{\Omega}\times\left(M_{\mu}\,\boldsymbol{R}^{(0)}_{\mu}\right)\Big)\nonumber\\
&= \left(\boldsymbol{e}^{k}\cdot\boldsymbol{\Omega}\right)\,\sum_{\mu=1}^{N}\,M_{\mu}\,\boldsymbol{e}_{\ell}\cdot\left(\boldsymbol{R}^{(0)}_{\mu}\times\left(\boldsymbol{e}_{k}\times\boldsymbol{R}^{(0)}_{\mu}\right)\right)\,\boldsymbol{e}^{\ell}\nonumber\\
&= \left(\boldsymbol{e}^{k}\cdot\boldsymbol{\Omega}\right)\,\sum_{\mu=1}^{N}\,M_{\mu}\,\left(\boldsymbol{e}_{\ell}\times\boldsymbol{R}^{(0)}_{\mu}\right)\cdot\left(\boldsymbol{e}_{k}\times\boldsymbol{R}^{(0)}_{\mu}\right)\,\boldsymbol{e}^{\ell}\nonumber\\
&= \left(\boldsymbol{e}^{k}\cdot\boldsymbol{\Omega}\right)\,\left(\boldsymbol{e}_{\ell}\cdot\mathsf{I}_0\cdot\boldsymbol{e}_{k}\right)\,\boldsymbol{e}^{\ell}\,,\vphantom{\sum_{\mu=1}^{N}}
\end{align}
which implies in turn that,
\begin{equation}\label{A.5.10ter}
\sum_{\mu=1}^{N}\,\left(\boldsymbol{e}_{k}\times\boldsymbol{R}^{(0)}_{\mu}\right)\cdot\boldsymbol{P}^{\prime\prime}_{\mu} = \left(\boldsymbol{e}_{k}\cdot\mathsf{I}_0\cdot\boldsymbol{e}_{j}\right)\,\left(\boldsymbol{e}^{j}\cdot\boldsymbol{\Omega}\right)\,.
\end{equation}
Using the property~\eqref{3.32bis} of the inertia tensor $\mathsf{I}_0$, the identity~\eqref{A.5.10ter} yields the internal observable $\boldsymbol{e}^{k}\cdot\boldsymbol{\Omega}$, i.e.
\begin{equation}\label{A.5.12}
\boldsymbol{e}^{k}\cdot\boldsymbol{\Omega} = \sum_{\mu=1}^{N} \left(\frac{\boldsymbol{e}_{k}\times\boldsymbol{R}^{(0)}_{\mu}}{\boldsymbol{e}_{k}\cdot\mathsf{I}_0\cdot\boldsymbol{e}_{k}}\right)\cdot\boldsymbol{P}^{\prime\prime}_{\mu}\,.
\end{equation}

\section{Rotation operator}
\label{A.1}

In this appendix, we show that the Eckart conditions~\eqref{3.16} and~\eqref{3.17bis}, and the relations~\eqref{3.13bis} and~\eqref{3.14} imply the physical definition~\eqref{3.39 ter} of the rotation operator $\mathsf{R}\left(\boldsymbol{\omega}\right)$. 

The expression~\eqref{3.14} of the rest momentum $\boldsymbol{R}^{\prime\prime}_{\mu}$ in terms of the equilibrium position $\boldsymbol{R}^{(0)}_{\mu}$ and the rest observables $Q^{\alpha}$ and $\boldsymbol{q}_{(\nu^{\prime})}$ implies that,
\begin{equation}\label{A.1.1}
\begin{split}
&\sum_{\mu=1}^{N}\,M_{\mu}\,\boldsymbol{R}^{(0)}_{\mu}\times\boldsymbol{R}^{\prime\prime}_{\mu}\\
&= \sum_{\mu=1}^{N}\,M_{\mu}\,\left(\boldsymbol{R}^{(0)}_{\mu}\times\boldsymbol{R}^{(0)}_{\mu}\right)\,\mathbb{1}\\
&\phantom{=} + Q^{\alpha}\left(\,\sum_{\mu=1}^{N}\,\sqrt{M_{\mu}}\,\left(\boldsymbol{R}^{(0)}_{\mu}\times\boldsymbol{X}_{\mu\alpha}\right)\right)\\
&\phantom{=} -\,\left(\,\sum_{\mu=1}^{N}\,M_{\mu}\,\boldsymbol{R}^{(0)}_{\mu}\right)\times\left(\frac{m}{M}\sum_{\nu,\nu^{\prime} = 1}^{n}\!A_{\nu\nu^{\prime}}\,\boldsymbol{q}_{(\nu^{\prime})}\right)\,.
\end{split}
\end{equation}
Using the Eckart conditions~\eqref{3.16} and~\eqref{3.17bis}, the RHS of the relation~\eqref{A.1.1} vanishes, i.e.
\begin{equation}\label{A.1.2}
\sum_{\mu=1}^{N}\,M_{\mu}\,\boldsymbol{R}^{(0)}_{\mu}\times\boldsymbol{R}^{\prime\prime}_{\mu} = \boldsymbol{0}\,.
\end{equation}
Finally, using the relation~\eqref{3.13bis} the condition~\eqref{A.1.2} is recast as,
\begin{equation}\label{A.1.3}
\sum_{\mu=1}^{N}\,M_{\mu}\,\boldsymbol{R}^{(0)}_{\mu}\times\left(\mathsf{R}\left(\boldsymbol{\omega}\right)^{-1}\cdot\boldsymbol{R}^{\prime}_{\mu}\right)=\boldsymbol{0}\,,
\end{equation}
which is equivalent to the condition 
\begin{equation}\label{A.1.4}
\sum_{\mu=1}^{N}\,M_{\mu}\,\left(\mathsf{R}\left(\boldsymbol{\omega}\right)\cdot\boldsymbol{R}^{(0)}_{\mu}\right)\times\boldsymbol{R}^{\prime}_{\mu}=\boldsymbol{0}\,.
\end{equation}
The condition~\eqref{A.1.3} is the physical definition~\eqref{3.39 ter} of the rotation operator $\mathsf{R}\left(\boldsymbol{\omega}\right)$.

\section{Commutation relations of the internal observables}
\label{A.6}

In this appendix, we determine the commutation relations between the internal observables $Q^{\alpha}$, $P_{\alpha}$, $\boldsymbol{q}_{(\nu)}$, $\boldsymbol{p}_{(\nu)}$, $\boldsymbol{\omega}$ and $\boldsymbol{\Omega}$. 

Using the definition~\eqref{A.5.9} of the operator $P_{\alpha}$, the constraint~\eqref{3.17bis} and the identity~\eqref{A.3.4bis} imply that,
\begin{equation}\label{A.6.0}
\begin{split}
&\left[\ P_{\alpha},\ \boldsymbol{e}^j\cdot\boldsymbol{\omega}\ \right]\,\sum_{\mu=1}^{N} M_{\mu}\left(\boldsymbol{n}_{(j)}\left(\boldsymbol{\omega}\right)\times\boldsymbol{R}^{(0)}_{\mu}\right)\times\boldsymbol{R}^{\prime\prime}_{\mu}\\
&= i\hbar\,\sum_{\nu=1}^{N}\,\sqrt{M_{\nu}}\,\left(\boldsymbol{R}^{(0)}_{\nu}\times\boldsymbol{X}_{\nu\alpha}\right) = \boldsymbol{0}\,,
\end{split}
\end{equation}
which yields the fifth commutation relation~\eqref{3.31.1}, i.e.
\begin{equation}\label{A.6.1}
\left[\ P_{\alpha},\ \boldsymbol{e}^{j}\cdot\boldsymbol{\omega}\ \right] = 0\,.
\end{equation}
Moreover, the definition~\eqref{3.26} and the commutation relation~\eqref{A.3.4.0} imply that,
\begin{equation}\label{A.6.0before}
\left[\ \boldsymbol{e}^{k}\cdot\boldsymbol{p}_{(\nu)},\ \boldsymbol{e}^{j}\cdot\boldsymbol{\omega}\ \right] = 0\,.
\end{equation}
Using the expression~\eqref{A.5.12} for the operator $\boldsymbol{e}^{k}\cdot\boldsymbol{\Omega}$, the constraint~\eqref{3.16} and the diagonality condition~\eqref{3.32bis}, the identity~\eqref{A.3.4bis} is recast as,
\begin{equation}\label{A.6.0ter}
\begin{split}
&\left[\ \boldsymbol{e}^{k}\cdot\boldsymbol{\Omega},\ \boldsymbol{e}^j\cdot\boldsymbol{\omega}\ \right]\,\sum_{\mu=1}^{N} M_{\mu}\left(\boldsymbol{n}_{(j)}\left(\boldsymbol{\omega}\right)\times\boldsymbol{R}^{(0)}_{\mu}\right)\times\boldsymbol{R}^{\prime\prime}_{\mu}\\
&= i\hbar\,\sum_{\nu=1}^{N}\,M_{\nu}\,\left(\boldsymbol{R}^{(0)}_{\nu}\times\left(\frac{\boldsymbol{e}_{k}\times\boldsymbol{R}^{(0)}_{\mu}}{\boldsymbol{e}_{k}\cdot\mathsf{I}_0\cdot\boldsymbol{e}_{k}}\right)\right)\\
&= i\hbar\,\left(\frac{\boldsymbol{e}_{k}\cdot\mathsf{I}_0\cdot\boldsymbol{e}_{\ell}}{\boldsymbol{e}_{k}\cdot\mathsf{I}_0\cdot\boldsymbol{e}_{k}}\right)\,\boldsymbol{e}^{\ell} = i\hbar\,\boldsymbol{e}^{k}\,.
\end{split}
\end{equation}

The condition~\eqref{3.17bis} implies that,
\begin{equation}\label{A.6.0third}
\sum_{\mu=1}^{N}\, \sqrt{M_{\mu}}\,\boldsymbol{e}_{k}\cdot\left(\boldsymbol{e}_{\ell}\times\left(\boldsymbol{R}^{(0)}_{\mu}\times\boldsymbol{X}_{\mu\alpha}\right)\right) = \boldsymbol{0}\,,
\end{equation}
which is recast as, 
\begin{equation}\label{A.6.0four}
\begin{split}
&\sum_{\mu=1}^{N}\, \sqrt{M_{\mu}}\,\left(\boldsymbol{e}_{k}\cdot\boldsymbol{R}^{(0)}_{\mu}\right)\,\left(\boldsymbol{e}_{\ell}\cdot\boldsymbol{X}_{\mu\alpha}\right)\\
&= \sum_{\mu=1}^{N}\, \sqrt{M_{\mu}}\,\left(\boldsymbol{e}_{\ell}\cdot\boldsymbol{R}^{(0)}_{\mu}\right)\,\left(\boldsymbol{e}_{k}\cdot\boldsymbol{X}_{\mu\alpha}\right)\,.
\end{split}
\end{equation}
Using the identity~\eqref{A.6.0four}, the expression~\eqref{3.33} yields the property~\eqref{3.30}, i.e.
\begin{align}\label{A.6.0five}
&\boldsymbol{e}_{k}\cdot\mathsf{I}_{\alpha}\cdot\boldsymbol{e}_{\ell} = \sum_{\mu=1}^{N}\,\sqrt{M_{\mu}}\,\left(\boldsymbol{e}_{k}\times\boldsymbol{R}^{(0)}_{\mu}\right)\cdot\left(\boldsymbol{e}_{\ell}\times\boldsymbol{X}_{\mu\alpha}\right)\nonumber\\
&= \sum_{\mu=1}^{N}\,\sqrt{M_{\mu}}\,\Big(\left(\boldsymbol{R}^{(0)}_{\mu}\cdot\boldsymbol{X}_{\mu\alpha}\right)\left(\boldsymbol{e}_{k}\cdot\boldsymbol{e}_{\ell}\right)\nonumber\\
&\phantom{=\sum_{\mu=1}^{N}\,\sqrt{M_{\mu}}\,\Big(}-\,\left(\boldsymbol{e}_{k}\cdot\boldsymbol{X}_{\mu\alpha}\right)\left(\boldsymbol{e}_{\ell}\cdot\boldsymbol{R}^{(0)}_{\mu}\right)\Big)\nonumber\\
&= \sum_{\mu=1}^{N}\,\sqrt{M_{\mu}}\,\Big(\left(\boldsymbol{R}^{(0)}_{\mu}\cdot\boldsymbol{X}_{\mu\alpha}\right)\left(\boldsymbol{e}_{k}\cdot\boldsymbol{e}_{\ell}\right)\\
&\phantom{=\sum_{\mu=1}^{N}\,\sqrt{M_{\mu}}\,\Big(}-\,\left(\boldsymbol{e}_{\ell}\cdot\boldsymbol{X}_{\mu\alpha}\right)\left(\boldsymbol{e}_{k}\cdot\boldsymbol{R}^{(0)}_{\mu}\right)\Big)\nonumber\\
&= \sum_{\mu=1}^{N}\,\sqrt{M_{\mu}}\,\left(\boldsymbol{e}_{\ell}\times\boldsymbol{R}^{(0)}_{\mu}\right)\cdot\left(\boldsymbol{e}_{k}\times\boldsymbol{X}_{\mu\alpha}\right) = \boldsymbol{e}_{\ell}\cdot\mathsf{I}_{\alpha}\cdot\boldsymbol{e}_{k}\,,\nonumber
\end{align}
which shows that the tensors $\mathsf{I}_{\alpha}$ and $\mathsf{I}\left(Q^{\,\boldsymbol{.}}\right)$ are symmetric. Using the relation~\eqref{3.14}, the constraint~\eqref{3.16} and the properties~\eqref{3.17bis}, \eqref{3.31}-\eqref{3.32bis}, we obtain,
\begin{align}\label{A.6.1ter}
&\sum_{\mu=1}^{N}\,M_{\mu}\,\left(\left(\boldsymbol{n}_{(j)}\left(\boldsymbol{\omega}\right)\times\boldsymbol{R}^{(0)}_{\mu}\right)\times\boldsymbol{R}^{\prime\prime}_{\mu}\right)\cdot\boldsymbol{e}_{\ell}
\nonumber\\
&= -\,\sum_{\mu=1}^{N}\,M_{\mu}\,\left(\boldsymbol{n}_{(j)}\left(\boldsymbol{\omega}\right)\times\boldsymbol{R}^{(0)}_{\mu}\right)\cdot\left(\boldsymbol{e}_{\ell}\times\boldsymbol{R}^{\prime\prime}_{\mu}\right)
\\
&= -\,\boldsymbol{e}^{k}\cdot\boldsymbol{n}_{(j)}\left(\boldsymbol{\omega}\right)\sum_{\mu=1}^{N}\,M_{\mu}\, \left(\boldsymbol{e}_{k}\times\boldsymbol{R}^{(0)}_{\mu}\right)\left(\boldsymbol{e}_{k}\times\boldsymbol{R}^{(0)}_{\mu}\right)
\nonumber\\
&\phantom{=} -\,\boldsymbol{e}^{k}\cdot\boldsymbol{n}_{(j)}\left(\boldsymbol{\omega}\right)\sum_{\mu=1}^{N}\,\sqrt{M_{\mu}}\, \left(\boldsymbol{e}_{k}\times\boldsymbol{R}^{(0)}_{\mu}\right)\left(\boldsymbol{e}_{k}\times Q^{\alpha}\boldsymbol{X}_{\mu\alpha}\right)
\nonumber\\
&= -\,\left(\boldsymbol{e}^{k}\cdot\boldsymbol{n}_{(j)}\left(\boldsymbol{\omega}\right)\right)\nonumber\\
&\phantom{=}\cdot\Big(\left(\boldsymbol{e}_{k}\cdot\mathsf{I}_0\cdot\boldsymbol{e}_{k}\right)\left(\boldsymbol{e}^{k}\cdot\boldsymbol{e}_{\ell}\right) + Q^{\alpha}\,\left(\boldsymbol{e}_{k}\cdot\mathsf{I}_{\alpha}\cdot\boldsymbol{e}_{\ell}\right)\Big) 
\nonumber\vphantom{\sum_{\mu=1}^{N}}\\
&= -\,\boldsymbol{n}_{(j)}\left(\boldsymbol{\omega}\right)\cdot\mathsf{I}\left(Q^{\,\boldsymbol{.}}\right)\cdot\boldsymbol{e}_{\ell}\,.\nonumber
\end{align}
Using the identity~\eqref{A.6.1ter}, the identity~\eqref{A.6.0ter} is recast as, 
\begin{equation}\label{A.6.2}
\left[\ \boldsymbol{e}^{k}\cdot\boldsymbol{\Omega},\ \boldsymbol{e}^j\cdot\boldsymbol{\omega}\ \right]\,\boldsymbol{n}_{(j)}\left(\boldsymbol{\omega}\right)\cdot\mathsf{I}\left(Q^{\,\boldsymbol{.}}\right)\cdot\boldsymbol{e}_{\ell} = -\,i\hbar\,\left(\boldsymbol{e}^{k}\cdot\boldsymbol{e}_{\ell}\right)\,.
\end{equation}
Using the property~\eqref{A.2.8}, which implies that,
\begin{equation}\label{A.6.2.0}
\left(\boldsymbol{n}_{(j)}\left(\boldsymbol{\omega}\right)\cdot\mathsf{I}\left(Q^{\,\boldsymbol{.}}\right)\right)^{-1} = \mathsf{I}\left(Q^{\,\boldsymbol{.}}\right)^{-1}\cdot\boldsymbol{m}^{(j)}\left(\boldsymbol{\omega}\right)\,,
\end{equation}
the identity~\eqref{A.6.2} yields the fifth commutation relation~\eqref{3.33.1}, i.e.
\begin{equation}\label{A.6.2bis}
\left[\ \boldsymbol{e}^{k}\cdot\boldsymbol{\Omega},\ \boldsymbol{e}^j\cdot\boldsymbol{\omega}\ \right] = -\,i\hbar\,\left(\boldsymbol{e}^{k}\cdot\mathsf{I}\left(Q^{\,\boldsymbol{.}}\right)^{-1}\cdot\boldsymbol{m}^{(j)}\left(\boldsymbol{\omega}\right)\right)\,.
\end{equation}
The relation~\eqref{3.25}, the commutation relations~\eqref{A.3.4.0},~\eqref{A.6.1} and~\eqref{A.6.2bis} imply that,
\begin{align}\label{A.6.2bisbis}
&\left[\ \boldsymbol{e}_{\ell}\cdot\boldsymbol{P}^{\prime\prime}_{\mu},\ \boldsymbol{e}^j\cdot\boldsymbol{\omega}\ \right]\nonumber\\
&= \left[\ \boldsymbol{e}^{k}\cdot\boldsymbol{\Omega},\ \boldsymbol{e}^j\cdot\boldsymbol{\omega}\ \right]\left(\boldsymbol{e}_{k}\times\left(M_{\mu}\,\boldsymbol{R}^{(0)}_{\mu}\right)\right)\cdot\boldsymbol{e}_{\ell}\nonumber\\
&\phantom{= }+ \left[\ P_{\alpha},\ \boldsymbol{e}^j\cdot\boldsymbol{\omega}\ \right]\,\sqrt{M_{\mu}}\,\left(\boldsymbol{e}_{\ell}\cdot\boldsymbol{X}^{\alpha}_{\mu}\right)\\
&\phantom{= }-\sum_{\nu,\nu^{\prime}=1}^{n}A_{\nu\nu^{\prime}}\left[\ \boldsymbol{e}_{\ell}\cdot\boldsymbol{p}_{\nu^{\prime}},\ \boldsymbol{e}^j\cdot\boldsymbol{\omega}\ \right]\,\frac{M_{\mu}}{M}\nonumber\\
& = -\,i\hbar\,\left(\boldsymbol{e}^{k}\cdot\mathsf{I}\left(Q^{\,\boldsymbol{.}}\right)^{-1}\cdot\boldsymbol{m}^{(j)}\left(\boldsymbol{\omega}\right)\right)\left(\boldsymbol{e}_{k}\times\left(M_{\mu}\,\boldsymbol{R}^{(0)}_{\mu}\right)\right)\cdot\boldsymbol{e}_{\ell}\nonumber
\end{align}
Now, we establish a useful commutation relation for the dynamics. The identities~\eqref{A.2.8},~\eqref{A.3.2ter} and~\eqref{A.6.2bisbis} imply that,
\begin{align}\label{A.6.2 extra 1}
&\left[\ \boldsymbol{e}_{\ell}\cdot\boldsymbol{P}^{\prime\prime}_{\mu},\ \mathsf{R}\left(\boldsymbol{\omega}\right)^{-1}\ \right]\,\mathsf{R}\left(\boldsymbol{\omega}\right)\nonumber\\
&= -\,\left[\ \boldsymbol{e}_{\ell}\cdot\boldsymbol{P}^{\prime\prime}_{\mu},\ \boldsymbol{e}^j\cdot\boldsymbol{\omega}\ \right]\,\left(\boldsymbol{n}_{(j)}\left(\boldsymbol{\omega}\right)\cdot\boldsymbol{\mathsf{G}}\right)\\
&=i\hbar\,\left(\boldsymbol{e}^{k}\cdot\mathsf{I}\left(Q^{\,\boldsymbol{.}}\right)^{-1}\cdot\boldsymbol{m}^{(j)}\left(\boldsymbol{\omega}\right)\right)\nonumber\\
&\phantom{=i\hbar}\cdot\left(\boldsymbol{e}_{k}\times\left(M_{\mu}\,\boldsymbol{R}^{(0)}_{\mu}\right)\right)\cdot\boldsymbol{e}_{\ell}\,\left(\boldsymbol{n}_{(j)}\left(\boldsymbol{\omega}\right)\cdot\boldsymbol{\mathsf{G}}\right)\nonumber\\
&=i\hbar\,\left(\boldsymbol{e}^{k}\cdot\mathsf{I}\left(Q^{\,\boldsymbol{.}}\right)^{-1}\cdot\boldsymbol{e}^{j}\right)\left(\boldsymbol{e}_{k}\times\left(M_{\mu}\,\boldsymbol{R}^{(0)}_{\mu}\right)\right)\cdot\boldsymbol{e}_{\ell}\,\left(\boldsymbol{e}_{j}\cdot\boldsymbol{\mathsf{G}}\right)\nonumber
\end{align}
The orthogonality condition~\eqref{A.2.1bis} and the commutation relation~\eqref{A.6.2 extra 1} imply that,
\begin{align}\label{A.6.2 extra 2}
&\left[\ \boldsymbol{e}^{\ell}\cdot\boldsymbol{P}^{\prime\prime}_{\mu},\ \boldsymbol{e}^{k}\cdot\mathsf{R}\left(\boldsymbol{\omega}\right)\cdot\boldsymbol{e}_{\ell}\ \right]\,\left(\boldsymbol{e}^{m}\cdot\mathsf{R}\left(\boldsymbol{\omega}\right)^{-1}\cdot\boldsymbol{e}_{k}\right)\\
&= \left[\ \boldsymbol{e}_{\ell}\cdot\boldsymbol{P}^{\prime\prime}_{\mu},\ \boldsymbol{e}^{\ell}\cdot\mathsf{R}\left(\boldsymbol{\omega}\right)^{-1}\cdot\boldsymbol{e}_{k}\ \right]\left(\boldsymbol{e}^{k}\cdot\mathsf{R}\left(\boldsymbol{\omega}\right)\cdot\boldsymbol{e}_{n}\right)\left(\boldsymbol{e}^{n}\cdot\boldsymbol{e}^{m}\right)\nonumber\\
&=i\hbar\,\left(\boldsymbol{e}^{k}\cdot\mathsf{I}\left(Q^{\,\boldsymbol{.}}\right)^{-1}\cdot\boldsymbol{e}^{j}\right)\left(\boldsymbol{e}_{k}\times\left(M_{\mu}\,\boldsymbol{R}^{(0)}_{\mu}\right)\right)\cdot\left(\boldsymbol{e}_{j}\times\boldsymbol{e}^{m}\right)\nonumber\\
&=-\,i\hbar\,\left(\boldsymbol{e}^{k}\cdot\mathsf{I}\left(Q^{\,\boldsymbol{.}}\right)^{-1}\cdot\boldsymbol{e}^{j}\right)\left(\boldsymbol{e}_{j}\!\times\!\left(\boldsymbol{e}_{k}\!\times\!\left(M_{\mu}\,\boldsymbol{R}^{(0)}_{\mu}\right)\right)\right)\!\cdot\boldsymbol{e}^{m}\nonumber
\end{align}
The definitions~\eqref{3.15.A} and~\eqref{3.27 prime}, the commutation relation~\eqref{3.17ter} and the condition~\eqref{3.17} yield the commutation relations~\eqref{3.30.1}, i.e.
\begin{align}\label{A.6.3}
&\left[\ Q^{\alpha},\  Q^{\beta}\ \right] = 0\,,\nonumber\\	
&\left[\ \boldsymbol{e}^{j}\cdot\boldsymbol{q}_{(\nu)},\  \boldsymbol{e}^{k}\cdot\boldsymbol{q}_{(\nu^{\prime})}\ \right] = 0\,,\nonumber\\
&\left[\ \boldsymbol{e}_{j}\cdot\boldsymbol{p}_{(\nu)},\  \boldsymbol{e}_{k}\cdot\boldsymbol{p}_{(\nu^{\prime})}\ \right] = 0\,,\nonumber\\
&\left[\ \boldsymbol{e}^{j}\cdot\boldsymbol{q}_{(\nu)},\  Q^{\beta}\ \right] = 0\,,\\
&\left[\ \boldsymbol{e}_{j}\cdot\boldsymbol{p}_{(\nu)},\  Q^{\beta}\ \right]\nonumber\\
&= i\hbar\,\sum_{\nu^{\prime}=1}^{n}\sum_{\mu=1}^{N}\,\frac{m\sqrt{M_{\mu}}}{\mathcal{M}}\,A^{-1}_{\nu\nu^{\prime}}\,\left(\boldsymbol{e}_{j}\cdot\boldsymbol{X}^{\beta}_{\mu}\right)\,\mathbb{1} = 0\,,\nonumber\\
&\left[\ \boldsymbol{e}_{j}\cdot\boldsymbol{p}_{(\nu)},\  \boldsymbol{e}^{k}\cdot\boldsymbol{q}_{(\nu^{\prime})}\ \right]\nonumber\\
&= -\,i\hbar\,\sum_{\overline{\nu},\overline{\nu}^{\prime}=1}^{n}A_{\nu\overline{\nu}}^{-1}\left(\delta_{\overline{\nu}\overline{\nu}^{\prime}}-\,\frac{m}{\mathcal{M}}\right)A_{\nu^{\prime}\overline{\nu}^{\prime}}^{-1}\left(\boldsymbol{e}_{j}\cdot\boldsymbol{e}^{k}\right)\,\mathbb{1}\,.\nonumber
\end{align}
The definitions~\eqref{3.4} and~\eqref{3.15.A} yield the relation,
\begin{align}\label{A.6.3bis}
&\sum_{\overline{\nu},\overline{\nu}^{\prime}=1}^{n}\,A^{-1}_{\nu\overline{\nu}}\,\left(\delta_{\overline{\nu}\overline{\nu}^{\prime}} -\,\frac{m}{\mathcal{M}}\right)\,A^{-1}_{\nu^{\prime}\overline{\nu}^{\prime}}\nonumber\\
&= \sum_{\overline{\nu},\overline{\nu}^{\prime}=1}^{n}\,\left(\delta_{\nu\overline{\nu}} + \frac{1}{n}\,\left(\sqrt{\frac{\mathcal{M}}{M}}-\,1\right)\right)\,\left(\delta_{\overline{\nu}\overline{\nu}^{\prime}} -\,\frac{m}{\mathcal{M}}\right)\nonumber\\
&\phantom{=\sum_{\overline{\nu},\overline{\nu}^{\prime}=1}^{n}}\cdot\left(\delta_{\nu^{\prime}\overline{\nu}^{\prime}} + \frac{1}{n}\,\left(\sqrt{\frac{\mathcal{M}}{M}}-\,1\right)\right)\\
&=\left(\frac{1}{n}-\,\frac{m}{\mathcal{M}}\right)\,\left(2\,\left(\sqrt{\frac{\mathcal{M}}{M}}-\,1\right) + \left(\sqrt{\frac{\mathcal{M}}{M}}-\,1\right)^2\right)\nonumber\\
&\phantom{=}-\,\frac{m}{\mathcal{M}} + \delta_{\nu\nu^{\prime}} = \delta_{\nu\nu^{\prime}}\,,\nonumber
\end{align}
which implies that the last commutation relation~\eqref{A.6.3} yields the second canonical commutation relation~\eqref{3.32.1}, i.e.
\begin{equation}\label{A.6.4}
\left[\ \boldsymbol{e}_{j}\cdot\boldsymbol{p}_{(\nu)},\  \boldsymbol{e}^{k}\cdot\boldsymbol{q}_{(\nu^{\prime})}\ \right] = -\,i\hbar\,\left(\boldsymbol{e}_{j}\cdot\boldsymbol{e}^{k}\right)\,\delta_{\nu\nu^{\prime}}\,.
\end{equation}
The definitions~\eqref{3.15} and~\eqref{A.5.9}, the constraint~\eqref{3.17}, and the commutation relations~\eqref{A.4.10bis} and~\eqref{A.6.1} yield the fourth commutation relation~\eqref{3.31.1}, i.e.
\begin{align}\label{A.6.6}
&\left[\ P_{\alpha},\ \boldsymbol{e}^{k}\cdot\boldsymbol{q}_{(\nu^{\prime})}\ \right]\nonumber\\
&= -\,\sum_{\mu=1}^{N}\,\frac{1}{\sqrt{M_{\mu}}}\,\left[\ \boldsymbol{X}_{\mu\alpha}\cdot\boldsymbol{P}^{\prime\prime}_{\mu},\ \boldsymbol{e}^{s}\cdot\boldsymbol{\omega}\ \right]\nonumber\\
&\phantom{= -}\cdot\sum_{\nu=1}^{n}\,A^{-1}_{\nu^{\prime}\nu}\,\boldsymbol{e}^{k}\cdot\left(\boldsymbol{n}_{(s)}\left(\boldsymbol{\omega}\right)\times\boldsymbol{r}^{\prime\prime}_{\nu}\right)\\
&\phantom{=} + i\hbar\,\sum_{\mu=1}^{N}\sum_{\nu=1}^{n}\,\frac{\sqrt{M_{\mu}}}{\mathcal{M}}\,A^{-1}_{\nu^{\prime}\nu}\left(\boldsymbol{X}_{\mu\alpha}\cdot\boldsymbol{e}^{k}\right)\,\mathbb{1}\nonumber\\
&= -\,\left[\ P_{\alpha},\ \boldsymbol{e}^{s}\cdot\boldsymbol{\omega}\ \right]\,\sum_{\nu=1}^{n}\,A^{-1}_{\nu^{\prime}\nu}\,\boldsymbol{e}^{k}\cdot\left(\boldsymbol{n}_{(s)}\left(\boldsymbol{\omega}\right)\times\boldsymbol{r}^{\prime\prime}_{\nu}\right) = 0\,.\nonumber
\end{align}
The definitions~\eqref{3.26} and~\eqref{A.5.9} and the commutation relations~\eqref{A.4.11bis} and~\eqref{A.6.1} yield the third commutation relation~\eqref{3.31.1}, i.e.
\begin{align}\label{A.6.9}
&\left[\ P_{\alpha},\ \boldsymbol{e}_{k}\cdot\boldsymbol{p}_{(\nu^{\prime})}\ \right]\nonumber\\
&= \sum_{\mu=1}^{N}\,\frac{1}{\sqrt{M_{\mu}}}\,\left[\ \boldsymbol{X}_{\mu\alpha}\cdot\boldsymbol{P}^{\prime\prime}_{\mu},\ \boldsymbol{e}^{s}\cdot\boldsymbol{\omega}\ \right]\\
&\phantom{=}\cdot\sum_{\nu=1}^{n}\,A^{-1}_{\nu^{\prime}\nu}\,\boldsymbol{e}_{k}\cdot\left(\boldsymbol{n}_{(s)}\left(\boldsymbol{\omega}\right)\times\boldsymbol{p}^{\prime\prime}_{\nu}\right)\nonumber\\
&= \left[\ P_{\alpha},\ \boldsymbol{e}^{s}\cdot\boldsymbol{\omega}\ \right]\,\sum_{\nu=1}^{n}\,A^{-1}_{\nu^{\prime}\nu}\,\boldsymbol{e}_{k}\cdot\left(\boldsymbol{n}_{(s)}\left(\boldsymbol{\omega}\right)\times\boldsymbol{p}^{\prime\prime}_{\nu}\right) = 0\,.\nonumber
\end{align}
The definitions~\eqref{A.5.4} and~\eqref{A.5.9}, the constraint~\eqref{3.17} and the commutation relation~\eqref{A.4.12} yield the first canonical commutation relation~\eqref{3.32.1}, i.e.
\begin{align}\label{A.6.11}
&\left[\ P_{\alpha},\ Q^{\beta}\ \right]\nonumber\\
&=-\,\sum_{\mu=1}^{N}\,\frac{1}{\sqrt{M_{\mu}}}\,\left[\ \boldsymbol{X}_{\mu\alpha}\cdot\boldsymbol{P}^{\prime\prime}_{\mu},\ \boldsymbol{e}^{s}\cdot\boldsymbol{\omega}\ \right]\nonumber\\
&\phantom{=-}\cdot\sum_{\nu=1}^{N}\,\sqrt{M_{\nu}}\,\boldsymbol{X}^{\beta}_{\nu}\cdot\left(\boldsymbol{n}_{(s)}\left(\boldsymbol{\omega}\right)\times\boldsymbol{R}^{\prime\prime}_{\nu}\right)\\
&\phantom{=} -\,i\hbar\,\sum_{\mu,\,\nu=1}^{N}\,\frac{\sqrt{M_{\nu}}}{\sqrt{M_{\mu}}}\,\left(\boldsymbol{X}_{\mu\alpha}\cdot\boldsymbol{X}^{\beta}_{\nu}\right)\,\left(\delta_{\mu\nu} -\,\frac{M_{\mu}}{\mathcal{M}}\right)\,\mathbb{1}\nonumber\\
&= -\,i\hbar\,\sum_{\mu=1}^{N}\,\left(\boldsymbol{X}_{\mu\alpha}\cdot\boldsymbol{X}^{\beta}_{\mu}\right)\,\mathbb{1} = -\,i\hbar\,\delta_{\alpha}^{\beta}\,.\nonumber
\end{align}
The definitions~\eqref{3.15} and~\eqref{A.5.1}, the properties~\eqref{3.16} and~\eqref{A.5.12}, the commutation relations~\eqref{A.4.10bis} and~\eqref{A.6.2bis} and the property~\eqref{A.2.8} yield the third commutation relation~\eqref{3.33.1}, i.e.
\begin{align}\label{A.6.6bis}
&\left[\ \boldsymbol{e}^{k}\cdot\boldsymbol{\Omega},\ \boldsymbol{e}^{j}\cdot\boldsymbol{q}_{(\nu^{\prime})}\ \right]\nonumber\\
&=-\,\left[\ \sum_{\mu=1}^{N}\,\frac{\boldsymbol{e}_{k}\times\boldsymbol{R}^{(0)}_{\mu}}{\boldsymbol{e}_{k}\cdot\mathsf{I}_{0}\cdot\boldsymbol{e}_{k}}\cdot\boldsymbol{P}^{\prime\prime}_{\mu},\ \boldsymbol{e}^{s}\cdot\boldsymbol{\omega}\ \right]\nonumber\\
&\phantom{=-}\cdot\boldsymbol{e}^{j}\cdot\left(\boldsymbol{n}_{(s)}\left(\boldsymbol{\omega}\right)\times\sum_{\nu=1}^{n}A^{-1}_{\nu^{\prime}\nu}\,\boldsymbol{r}^{\prime\prime}_{\nu}\right)\\
&\phantom{=}+ i\hbar\,\sum_{\mu=1}^{N}\,\left( \frac{\boldsymbol{e}_{k}\times\left(M_{\mu}\,\boldsymbol{R}^{(0)}_{\mu}\right)}{\mathcal{M}\,\left(\boldsymbol{e}_{k}\cdot\mathsf{I}_{0}\cdot\boldsymbol{e}_{k}\right)}\right)\cdot\left(\boldsymbol{e}^{j}\,\sum_{\nu=1}^{n}\,A^{-1}_{\nu^{\prime}\nu}\,\mathbb{1}\right)\nonumber\\
&= \left[\ \boldsymbol{e}^{k}\cdot\boldsymbol{\Omega},\ \boldsymbol{e}^{s}\cdot\boldsymbol{\omega}\ \right] \boldsymbol{n}_{(s)}\left(\boldsymbol{\omega}\right)\cdot \left(\boldsymbol{e}^{j}\times\boldsymbol{q}_{(\nu^{\prime})}\right)\nonumber\\
&= -\,i\hbar\left(\boldsymbol{e}^{k}\cdot\mathsf{I}\left(Q^{\,\boldsymbol{.}}\right)^{-1}\!\cdot\boldsymbol{m}^{(s)}\left(\boldsymbol{\omega}\right)\right)\left(\boldsymbol{n}_{(s)}\left(\boldsymbol{\omega}\right)\cdot\left(\boldsymbol{e}^{j}\!\times\!\boldsymbol{q}_{(\nu^{\prime})}\right)\right)\nonumber\\
& =-\,i\hbar\left(\boldsymbol{e}^{k}\cdot\mathsf{I}\left(Q^{\,\boldsymbol{.}}\right)^{-1}\cdot\boldsymbol{e}^{\ell}\right)\,\left(\boldsymbol{e}_{\ell}\times\boldsymbol{e}^{j}\right)\cdot\boldsymbol{q}_{(\nu^{\prime})}\,.\nonumber
\end{align}
Similarly, the definition~\eqref{3.26}, the properties~\eqref{A.2.8} and~\eqref{A.5.12}, and the commutation relations~\eqref{A.4.11bis} and~\eqref{A.6.2bis} yield the fourth commutation relation~\eqref{3.33.1}, i.e.
\begin{align}\label{A.6.10}
&\left[\ \boldsymbol{e}^{k}\cdot\boldsymbol{\Omega},\ \boldsymbol{e}_{j}\cdot\boldsymbol{p}_{(\nu^{\prime})}\ \right]\nonumber\\
&=-\,\left[\ \sum_{\mu=1}^{N}\,\frac{\boldsymbol{e}_{k}\times\boldsymbol{R}^{(0)}_{\mu}}{\boldsymbol{e}_{k}\cdot\mathsf{I}_{0}\cdot\boldsymbol{e}_{k}}\cdot\boldsymbol{P}^{\prime\prime}_{\mu},\ \boldsymbol{e}^{s}\cdot\boldsymbol{\omega}\ \right]\nonumber\\
&\phantom{=-}\cdot\boldsymbol{e}_{j}\cdot\left(\boldsymbol{n}_{(s)}\left(\boldsymbol{\omega}\right)\times\sum_{\nu=1}^{n}A^{-1}_{\nu^{\prime}\nu}\,\boldsymbol{p}^{\prime\prime}_{\nu}\right)\\
&=\left[\ \boldsymbol{e}^{k}\cdot\boldsymbol{\Omega},\ \boldsymbol{e}^{s}\cdot\boldsymbol{\omega}\ \right] \boldsymbol{n}_{(s)}\left(\boldsymbol{\omega}\right)\cdot \left(\boldsymbol{e}_{j}\times\boldsymbol{p}_{(\nu^{\prime})}\right)\nonumber\\
& 
=-\,i\hbar\left(\boldsymbol{e}^{k}\cdot\mathsf{I}\left(Q^{\,\boldsymbol{.}}\right)^{-1}\!\cdot\boldsymbol{m}^{(s)}\left(\boldsymbol{\omega}\right)\right)\left(\boldsymbol{n}_{(s)}\left(\boldsymbol{\omega}\right)\cdot\left(\boldsymbol{e}_{j}\!\times\!\boldsymbol{p}_{(\nu^{\prime})}\right)\right)\nonumber\\
& = -\,i\hbar\left(\boldsymbol{e}^{k}\cdot\mathsf{I}\left(Q^{\,\boldsymbol{.}}\right)^{-1}\cdot\boldsymbol{e}^{\ell}\right)\,\left(\boldsymbol{e}_{\ell}\times\boldsymbol{e}_{j}\right)\cdot\boldsymbol{p}_{(\nu^{\prime})}\,.\nonumber
\end{align}
The definition~\eqref{A.5.4}, the properties~\eqref{3.17},~\eqref{3.17bis},~\eqref{A.2.8} and~\eqref{A.5.12}, and the commutation relations~\eqref{A.4.12} and~\eqref{A.6.2bis} yield the first commutation relation~\eqref{3.33.1}, i.e.
\begin{align}\label{A.6.13}
&\left[\ \boldsymbol{e}^{k}\cdot\boldsymbol{\Omega},\ Q^{\alpha}\ \right]\nonumber\\
&= -\,\left[\ \sum_{\mu=1}^{N}\,\frac{\boldsymbol{e}_{k}\times\boldsymbol{R}^{(0)}_{\mu}}{\boldsymbol{e}_{k}\cdot\mathsf{I}_{0}\cdot\boldsymbol{e}_{k}}\cdot\boldsymbol{P}^{\prime\prime}_{\mu},\ \boldsymbol{e}^{s}\cdot\boldsymbol{\omega}\ \right]\nonumber\\
&\phantom{=-}\cdot\sum_{\nu=1}^{N} \,\sqrt{M_{\nu}}\,\boldsymbol{X}^{\alpha}_{\nu}\cdot\left(\boldsymbol{n}_{(s)}\left(\boldsymbol{\omega}\right)\times\boldsymbol{R}^{\prime\prime}_{\nu}\right)\\
&\phantom{=} -\,i\hbar\,\sum_{\mu=1}^{N}\sum_{\nu=1}^{N}\, \sqrt{M_{\nu}}\left(\delta_{\mu\nu}-\frac{M_{\mu}}{\mathcal{M}}\right)\,\left(\frac{\boldsymbol{e}_{k}\times\boldsymbol{R}^{(0)}_{\mu}}{\boldsymbol{e}_k\cdot\mathsf{I}_{0}\cdot\boldsymbol{e}_k}\right)\cdot\boldsymbol{X}^{\alpha}_{\nu}\,\mathbb{1}\nonumber\\
&= -\,i\hbar\,\left[\ \boldsymbol{e}^{k}\cdot\boldsymbol{\Omega},\ \boldsymbol{e}^{s}\cdot\boldsymbol{\omega}\ \right]\,\sum_{\nu=1}^{N} \,\sqrt{M_{\nu}}\,\left(\boldsymbol{n}_{(s)}\left(\boldsymbol{\omega}\right)\times\boldsymbol{R}^{\prime\prime}_{\nu}\right)\cdot\boldsymbol{X}^{\alpha}_{\nu}\nonumber\\
&= i\hbar\,\left(\boldsymbol{e}^{k}\cdot\mathsf{I}\left(Q^{\,\boldsymbol{.}}\right)^{-1}\cdot\boldsymbol{m}^{(s)}\left(\boldsymbol{\omega}\right)\right)\nonumber\\
&\phantom{=}\cdot\sum_{\nu=1}^{N} \,\sqrt{M_{\nu}}\,\left(\boldsymbol{R}^{\prime\prime}_{\nu}\times\boldsymbol{X}^{\alpha}_{\nu}\right)\cdot\boldsymbol{n}_{(s)}\left(\boldsymbol{\omega}\right)\nonumber\\
&= i\hbar\,Q^{\beta}\left(\boldsymbol{e}^{k}\cdot\mathsf{I}\left(Q^{\,\boldsymbol{.}}\right)^{-1}\cdot\boldsymbol{e}^{\ell}\right)\,\sum_{\nu=1}^{N}\,\left(\boldsymbol{X}_{\nu\beta}\times\boldsymbol{X}^{\alpha}_{\nu}\right)\cdot\boldsymbol{e}_{\ell}\nonumber\\
&=-\,i\hbar\sum_{\mu=1}^{N}\,\left(\boldsymbol{e}^{k}\cdot\mathsf{I}\left(Q^{\,\boldsymbol{.}}\right)^{-1}\cdot\left(\boldsymbol{X}^{\alpha}_{\mu}\times\boldsymbol{X}_{\mu\beta}\right)\,Q^{\beta}\right)\,.\nonumber
\end{align}
The definition~\eqref{A.5.9} and the commutation relations~\eqref{A.4.14} and~\eqref{A.6.1} yield the first commutation relation~\eqref{3.31.1}, i.e.
\begin{align}\label{A.6.14}
&\left[\ P_{\alpha},\ P_{\beta}\ \right]\nonumber\\
&= \sum_{\mu,\,\nu=1}^{N}\,\frac{1}{\sqrt{M_{\mu}}}\,\frac{1}{\sqrt{M_{\nu}}}\,
\left[\ \boldsymbol{X}_{\mu\alpha}\cdot\boldsymbol{P}^{\prime\prime}_{\mu},\ \boldsymbol{X}_{\nu\beta}\cdot\boldsymbol{P}^{\prime\prime}_{\nu}\ \right]\nonumber\\
&= \frac{1}{2}\,\sum_{\nu=1}^{N}\,\frac{1}{\sqrt{M_{\nu}}}\,\Big\{\ \boldsymbol{e}_{\ell}\cdot\boldsymbol{P}^{\prime\prime}_{\nu},\ \left[\ P_{\alpha},\ \boldsymbol{e}^{m}\cdot\boldsymbol{\omega}\ \right]\nonumber\\
&\phantom{= \frac{1}{2}\,\sum_{\nu=1}^{N}\,\frac{1}{\sqrt{M_{\nu}}}\,\Big\{\ }\left(\boldsymbol{n}_{(m)}\left(\boldsymbol{\omega}\right)\times\boldsymbol{X}_{\nu\beta}\right)\cdot\boldsymbol{e}^{\ell}\ \Big\}\\
&-\,\frac{1}{2}\,\sum_{\mu=1}^{N}\,\frac{1}{\sqrt{M_{\mu}}}\,\Big\{\ \boldsymbol{e}_{\ell}\cdot\boldsymbol{P}^{\prime\prime}_{\nu},\ 
\left[\ P_{\beta},\ \boldsymbol{e}^{m}\cdot\boldsymbol{\omega}\ \right]\nonumber\\
&\phantom{= \frac{1}{2}\,\sum_{\mu=1}^{N}\,\frac{1}{\sqrt{M_{\mu}}}\,\Big\{\ }\left(\boldsymbol{n}_{(m)}\left(\boldsymbol{\omega}\right)\times\boldsymbol{X}_{\mu\alpha}\right)\cdot\boldsymbol{e}^{\ell}\ \Big\} = 0\,.\nonumber
\end{align}
The definition~\eqref{A.5.9}, the commutation relation~\eqref{A.4.14},~\eqref{A.6.1} and~\eqref{A.6.2bis}, and the properties~\eqref{3.18} and~\eqref{A.2.8}, the expressions~\eqref{A.5.10ter}-\eqref{A.5.12} yield the second commutation relation~\eqref{3.31.1}, i.e.
\begin{align}\label{A.6.15}
&\left[\ \boldsymbol{e}^{k}\cdot\boldsymbol{\Omega},\ P_{\alpha}\ \right]\nonumber\\
&= \frac{1}{2}\,\sum_{\nu=1}^{N}\,\frac{1}{\sqrt{M_{\nu}}}\,\left\{\ \boldsymbol{e}_{\ell}\cdot\boldsymbol{P}^{\prime\prime}_{\nu},\ \left[\ \boldsymbol{e}^{k}\cdot\boldsymbol{\Omega},\ \boldsymbol{e}^{m}\cdot\boldsymbol{\omega}\ \right]\right.\nonumber\\
&\phantom{= \frac{1}{2}\,\sum_{\nu=1}^{N}\,\frac{1}{\sqrt{M_{\nu}}}\,,\left\{\right.}\left.\cdot\left(\boldsymbol{n}_{(m)}\left(\boldsymbol{\omega}\right)\times\boldsymbol{X}_{\nu\alpha}\right)\cdot\boldsymbol{e}^{\ell}\ \right\}\nonumber\\
&= -\,\frac{1}{2}\,i\hbar\,\sum_{\nu=1}^{N}\,\frac{1}{\sqrt{M_{\nu}}}\,\left\{\,\boldsymbol{e}_{\ell}\!\cdot\boldsymbol{P}^{\prime\prime}_{\nu},\,\left(\boldsymbol{e}^{k}\cdot\mathsf{I}\left(Q^{\,\boldsymbol{.}}\right)^{-1}\!\cdot\boldsymbol{m}^{(m)}\left(\boldsymbol{\omega}\right)\right)\right.\nonumber\\
&\phantom{= -\,\frac{1}{2}\,i\hbar\,\sum_{\nu=1}^{N}\,\frac{1}{\sqrt{M_{\nu}}}\,,\left\{\right.}\left.\cdot\left(\boldsymbol{n}_{(m)}\left(\boldsymbol{\omega}\right)\times\boldsymbol{X}_{\nu\alpha}\right)\cdot\boldsymbol{e}^{\ell}\ \right\}\nonumber\\
&= -\,\frac{1}{2}\,i\hbar\,\sum_{\nu=1}^{N}\,\frac{1}{\sqrt{M_{\nu}}}\,\left\{\ \left(\boldsymbol{e}_{\ell}\cdot\boldsymbol{P}^{\prime\prime}_{\nu}\right)\,\left(\boldsymbol{e}_{m}\times\boldsymbol{X}_{\nu\alpha}\right)\cdot\boldsymbol{e}^{\ell},\right.\nonumber\\
&\phantom{= -\,\frac{1}{2}\,i\hbar\,\sum_{\nu=1}^{N}\,\frac{1}{\sqrt{M_{\nu}}}\,,\left\{\right.}\left.\cdot\,\boldsymbol{e}^{k}\cdot\mathsf{I}\left(Q^{\,\boldsymbol{.}}\right)^{-1}\cdot\boldsymbol{e}^{m}\,\ \right\}\nonumber\\
&= -\,\frac{1}{2}\,i\hbar\!\sum_{\mu,\nu=1}^{N}\!\frac{1}{\sqrt{M_{\nu}}}\,\left\{\ \left(\boldsymbol{X}_{\mu\beta}\cdot\boldsymbol{P}^{\prime\prime}_{\nu}\right)\,\left(\boldsymbol{X}_{\nu\alpha}\times\boldsymbol{X}^{\beta}_{\mu}\right)\cdot\boldsymbol{e}_{m},\right.\nonumber\\
&\phantom{= -\,\frac{1}{2}\,i\hbar\!\sum_{\mu,\nu=1}^{N}\!\frac{1}{\sqrt{M_{\nu}}}\,,\left\{\right.}\left.\cdot\,\boldsymbol{e}^{k}\cdot\mathsf{I}\left(Q^{\,\boldsymbol{.}}\right)^{-1}\cdot\boldsymbol{e}^{m}\,\ \right\}\nonumber\\
&=-\,\frac{1}{2}\,i\hbar\sum_{\mu=1}^{N}\,\left[\ \boldsymbol{e}^{k}\cdot\mathsf{I}\left(Q^{\,\boldsymbol{.}}\right)^{-1}\cdot\boldsymbol{X}_{\mu\alpha},\ P_{\beta}\,\boldsymbol{X}^{\beta}_{\mu}\ \right]_{\times}\,.
\end{align}

\section{Orbital angular momentum}
\label{A.7}

In this appendix, we establish an explicit expression of the orbital angular momentum in terms of the internal observables. 

According to the first and last relations~\eqref{3.13}, the commutation relations between the operators $\mathbf{e}^{j}\cdot\boldsymbol{R}^{\prime}_{\mu}$ and $\mathbf{e}_{k}\cdot\boldsymbol{P}^{\prime}_{\mu}$ and the operators $\mathbf{e}^{j}\cdot\boldsymbol{r}^{\prime}_{\nu}$ and $\mathbf{e}_{k}\cdot\boldsymbol{p}^{\prime}_{\nu}$ are symmetric with respect to the permutation of the basis vectors $\mathbf{e}^{j}$ and $\mathbf{e}_{k}$, which implies that the vector product of these operators satisfies the identities,
\begin{equation}\label{A.7.2}
\begin{split}
&\boldsymbol{R}^{\prime}_{\mu}\times\boldsymbol{P}^{\prime}_{\mu} = -\,\boldsymbol{P}^{\prime}_{\mu}\times\boldsymbol{R}^{\prime}_{\mu}\,,\\
&\boldsymbol{r}^{\prime}_{\nu}\times\boldsymbol{p}^{\prime}_{\nu} = -\,\boldsymbol{p}^{\prime}_{\nu}\times\boldsymbol{r}^{\prime}_{\nu}\,.
\end{split}
\end{equation}
Thus, using the relation~\eqref{A.7.2} the angular momentum~\eqref{3.17pet} is recast as,
\begin{equation}\label{A.7.2bis}
\boldsymbol{L}^{\prime} = \frac{1}{2}\,\sum_{\mu=1}^{N}\,\left[\ \boldsymbol{R}^{\prime}_{\mu},\ \boldsymbol{P}^{\prime}_{\mu}\ \right]_{\boldsymbol{\times}} + \frac{1}{2}\,\sum_{\nu=1}^{n}\,\left[\ \boldsymbol{r}^{\prime}_{\nu},\ \boldsymbol{p}^{\prime}_{\nu}\ \right]_{\boldsymbol{\times}}\,.
\end{equation}
The angular momentum $\boldsymbol{L}$ is a pseudo-vectorial operator that is related to the angular momentum $\boldsymbol{L}^{\prime}$ by,
\begin{equation}\label{A.7.3}
\boldsymbol{e}_{\ell}\cdot\boldsymbol{L} = \frac{1}{2}\Big(\left(\mathsf{R}\left(\boldsymbol{\omega}\right)\cdot\boldsymbol{e}_{\ell}\right)\cdot\boldsymbol{L}^{\prime} + \boldsymbol{L}^{\prime}\cdot\left(\mathsf{R}\left(\boldsymbol{\omega}\right)\cdot\boldsymbol{e}_{\ell}\right)\Big)\,.
\end{equation}
Using the relation~\eqref{A.7.3} and the definition~\eqref{A.7.2bis} of the orbital angular momentum $\boldsymbol{L}^{\prime}$ yields the expression~\eqref{3.26bis} for the orbital angular momentum $\boldsymbol{L}$, i.e.
\begin{equation}\label{A.7.6}
\boldsymbol{L} = \frac{1}{2}\,\sum_{\mu=1}^{N}\,\left[\ \boldsymbol{R}^{\prime\prime}_{\mu},\ \boldsymbol{P}^{\prime\prime}_{\mu}\ \right]_{\boldsymbol{\times}} + \frac{1}{2}\,\sum_{\nu=1}^{n}\,\left[\ \boldsymbol{r}^{\prime\prime}_{\nu},\ \boldsymbol{p}^{\prime\prime}_{\nu}\ \right]_{\boldsymbol{\times}}\,.
\end{equation}
Using the relations~\eqref{3.14}-\eqref{3.26} the expression~\eqref{A.7.6} is recast as,
\begin{align}\label{A.7.6bis}
&\boldsymbol{L} = \frac{1}{2}\sum_{\mu=1}^{N} 
\left(\!\left(\boldsymbol{R}^{(0)}_{\mu}\,\mathbb{1}+\frac{1}{\sqrt{M_{\mu}}}\,Q^{\alpha}\boldsymbol{X}_{\mu\alpha} -\frac{m}{M}\!\!\sum_{\nu,\,\overline{\nu} = 1}^{n}\!A_{\nu\overline{\nu}}\,\boldsymbol{q}_{(\overline{\nu})}\!\right)\right.\nonumber\\
&\times\!\left(\boldsymbol{\Omega}\!\times\!\left(M_{\mu}\,\boldsymbol{R}^{(0)}_{\mu}\right) + \sqrt{M_{\mu}}\,P_{\beta}\boldsymbol{X}^{\beta}_{\mu}
-\frac{M_{\mu}}{M}\!\!\!\sum_{\nu^{\prime},\,\overline{\nu}^{\prime} = 1}^{n}\!\!A_{\nu^{\prime}\overline{\nu}^{\prime}}\,\boldsymbol{p}_{(\overline{\nu}^{\prime})}\!\right)\nonumber\\
&-\left(\boldsymbol{\Omega}\!\times\!\left(M_{\mu}\,\boldsymbol{R}^{(0)}_{\mu}\right) + \sqrt{M_{\mu}}\,P_{\beta}\boldsymbol{X}^{\beta}_{\mu}
-\frac{M_{\mu}}{M}\!\!\!\sum_{\nu^{\prime},\,\overline{\nu}^{\prime} = 1}^{n}\!\!A_{\nu^{\prime}\overline{\nu}^{\prime}}\,\boldsymbol{p}_{(\overline{\nu}^{\prime})}\!\right)\nonumber\\
&\left.\times\!\left(\boldsymbol{R}^{(0)}_{\mu}\,\mathbb{1}+\frac{1}{\sqrt{M_{\mu}}}\,Q^{\alpha}\boldsymbol{X}_{\mu\alpha} -\frac{m}{M}\!\!\sum_{\nu,\,\overline{\nu} = 1}^{n}\!A_{\nu\overline{\nu}}\,\boldsymbol{q}_{(\overline{\nu})}\!\right)\!\right)\nonumber\\
&+ \frac{1}{2}\sum_{\nu=1}^{n}\,\left(\left(\sum_{\overline{\nu} = 1}^{n}A_{\nu\overline{\nu}}\,\boldsymbol{q}_{(\overline{\nu})}\right)\times\left(\sum_{\overline{\nu}^{\prime} = 1}^{n}A_{\nu\overline{\nu}^{\prime}}\,\boldsymbol{p}_{(\overline{\nu}^{\prime})}\right)\right)\nonumber\\ 
&-\,\frac{1}{2}\sum_{\nu=1}^{n}\,\left(\left(\sum_{\overline{\nu}^{\prime} = 1}^{n}A_{\nu\overline{\nu}^{\prime}}\,\boldsymbol{p}_{(\overline{\nu}^{\prime})}\right)\times\left(\sum_{\overline{\nu} = 1}^{n}A_{\nu\overline{\nu}}\,\boldsymbol{q}_{(\overline{\nu})}\right)\right)\,.
\end{align}
Using the definitions~\eqref{3.32bis}-\eqref{3.33} and the properties~\eqref{3.20pre},~\eqref{3.16}-\eqref{3.17bis} and~\eqref{A.6.0five}, $\boldsymbol{L}$ is recast as,
\begin{align}\label{A.7.6ter}
&\boldsymbol{L} = \boldsymbol{e}^{\ell}\,\left(\boldsymbol{e}_{\ell}\cdot\mathsf{I}_{0}\cdot\boldsymbol{e}_{k}\right)\,\left(\boldsymbol{e}^{k}\cdot\boldsymbol{\Omega}\right)\nonumber\\
&+\frac{1}{2}\,\boldsymbol{e}^{\ell}\,\left(\boldsymbol{e}_{\ell}\cdot\mathsf{I}_{\alpha}\cdot\boldsymbol{e}_{k}\right)\,\left\{\ Q^{\alpha},\ \boldsymbol{e}^{k}\cdot\boldsymbol{\Omega}\ \right\}\nonumber\\
&+ \frac{1}{2}\,\sum_{\mu = 1}^{N}\,\left[\ Q^{\alpha}\,\boldsymbol{X}_{\mu\alpha},\ P_{\beta}\,\boldsymbol{X}^{\beta}_{\mu}\ \right]_{\times}\\
&+ \frac{1}{2}\sum_{\overline{\nu},\,\overline{\nu}^{\prime} = 1}^{n}\!\left(\sum_{\nu,\,\nu^{\prime} = 1}^{n}\!A_{\nu\overline{\nu}}\left(\delta_{\nu\nu^{\prime}} + \frac{m}{M}\right)A_{\nu^{\prime}\overline{\nu}^{\prime}}\!\right)\!\left[\,\boldsymbol{q}_{(\overline{\nu})},\,\boldsymbol{p}_{(\overline{\nu}^{\prime})}\,\right]_{\times}\nonumber
\end{align}
The definitions~\eqref{3.4} and~\eqref{3.15.A} yield the identity
\begin{align}\label{A.7.6quad}
&\sum_{\nu,\,\nu^{\prime}=1}^{n}\,A_{\nu\overline{\nu}}\,\left(\delta_{\nu\nu^{\prime}} +\frac{m}{M}\right)\,A_{\nu^{\prime}\overline{\nu}^{\prime}}\nonumber\\
&= \sum_{\nu,\,\nu^{\prime}=1}^{n}\,\left(\delta_{\nu\overline{\nu}} + \frac{1}{n}\,\left(\sqrt{\frac{M}{\mathcal{M}}}-\,1\right)\right)\,\left(\delta_{\nu\nu^{\prime}} + \frac{m}{M}\right)\nonumber\\
&\phantom{= \sum_{\nu,\,\nu^{\prime}=1}^{n}}\cdot\left(\delta_{\nu^{\prime}\overline{\nu}^{\prime}} + \frac{1}{n}\,\left(\sqrt{\frac{M}{\mathcal{M}}}-\,1\right)\right)\\
&= \left(\frac{1}{n}+\frac{m}{M}\right)\,\left(2\,\left(\sqrt{\frac{M}{\mathcal{M}}}-\,1\right) + \left(\sqrt{\frac{M}{\mathcal{M}}}-\,1\right)^2\right)\nonumber\\
&\phantom{=} + \frac{m}{M} + \delta_{\overline{\nu}\overline{\nu}^{\prime}} = \delta_{\overline{\nu}\overline{\nu}^{\prime}}\,.\nonumber
\end{align}
Using the identity~\eqref{A.7.6quad} and the relation~\eqref{3.31}, the expression~\eqref{A.7.6ter} yields the expression~\eqref{3.27} for the orbital angular momentum, i.e.
\begin{align}\label{A.7.7}
&\boldsymbol{L} =  \frac{1}{2}\,\Big\{\ \mathsf{I}\left(Q^{\,\boldsymbol{.}}\right),\ \boldsymbol{\Omega}\ \Big\}_{\mathsmaller{\bullet}} + \frac{1}{2}\,\sum_{\mu = 1}^{N}\,\left[\ Q^{\alpha}\,\boldsymbol{X}_{\mu\alpha},\ P_{\beta}\,\boldsymbol{X}^{\beta}_{\mu}\ \right]_{\times}\nonumber\\
&\phantom{\boldsymbol{L} =} + \frac{1}{2}\,\sum_{\nu=1}^{n}\,\left[\ \boldsymbol{q}_{(\nu)},\ \boldsymbol{p}_{(\nu)}\ \right]_{\times}\,.
\end{align}

\section{Commutation relations of orbital angular momentum and the internal observables}
\label{A.8}

In this appendix, we determine explicitly the commutation relations between the orbital angular momentum $\boldsymbol{L}$ and the internal observables $Q^{\alpha}$, $P_{\alpha}$, $\boldsymbol{q}_{(\nu)}$, $\boldsymbol{p}_{(\nu)}$, $\boldsymbol{\omega}$ and $\boldsymbol{\Omega}$.

Since the operator $\boldsymbol{\Omega}$ commutes with the operators $Q^{\alpha}$, $P_{\beta}$, $\boldsymbol{q}_{(\nu)}$ and $\boldsymbol{p}_{(\nu)}$, the commutation relations~\eqref{3.30.1}-\eqref{3.33} and the expression~\eqref{A.7.7} yield the first commutation relation~\eqref{3.35.1}, i.e.
\begin{align}\label{A.8.0}
&\left[\ \boldsymbol{L},\ \boldsymbol{e}^{\ell}\cdot{\boldsymbol{\omega}}\ \right] =  \frac{1}{2}\,\left\{\ \mathsf{I}\left(Q^{\,\boldsymbol{.}}\right)\cdot\boldsymbol{e}_{k},\ \left[\ \boldsymbol{e}^{k}\cdot\boldsymbol{\Omega},\ \boldsymbol{e}^{\ell}\cdot{\boldsymbol{\omega}}\ \right]\ \right\}\nonumber\\ 
&= \frac{1}{2}\,\left\{\ \mathsf{I}\left(Q^{\,\boldsymbol{.}}\right)\cdot\boldsymbol{e}_{k},\ -\,i\hbar\,\left(\boldsymbol{e}^{k}\cdot\mathsf{I}\left(Q^{\,\boldsymbol{.}}\right)^{-1}\cdot\boldsymbol{m}^{(\ell)}\left(\boldsymbol{\omega}\right)\right)\ \right\}\nonumber\\
&= -\,i\hbar\ \boldsymbol{m}^{(\ell)}\left(\boldsymbol{\omega}\right)\,.
\end{align}
The commutation relation~\eqref{A.8.0} and the property~\eqref{A.2.8} yield the canonical commutation relation in rotation~\eqref{3.36.1}, i.e.
\begin{equation}\label{A.8.0 prime}
\left[\ \boldsymbol{n}_{(k)}\left(\boldsymbol{\omega}\right)\cdot\boldsymbol{L},\ \boldsymbol{e}^{\ell}\cdot{\boldsymbol{\omega}}\ \right] = -\,i\hbar\,\left(\boldsymbol{e}_{k}\cdot\boldsymbol{e}^{\ell}\right)\,, 
\end{equation}
which means that $\boldsymbol{n}_{(k)}\left(\boldsymbol{\omega}\right)\cdot\boldsymbol{L}$ and $\boldsymbol{e}^{\ell}\cdot{\boldsymbol{\omega}}$ are canonically conjugated operators. 

Since the operator $Q^{\alpha}$ commutes with the operators $Q^{\gamma}$, $\boldsymbol{q}_{(\nu)}$ and $\boldsymbol{p}_{(\nu)}$, the commutation relations~\eqref{3.30.1}-\eqref{3.33} and the expression~\eqref{A.7.7} yield the first commutation relation~\eqref{3.34.1}, i.e.
\begin{align}\label{A.8.0 bis}
&\left[\ \boldsymbol{L},\ Q^{\alpha}\ \right] =  \frac{1}{2}\,\left\{\ \mathsf{I}\left(Q^{\,\boldsymbol{.}}\right)\cdot\boldsymbol{e}_{k},\ \left[\ \boldsymbol{e}^{k}\cdot\boldsymbol{\Omega},\ Q^{\alpha}\ \right]\ \right\}\nonumber\\
&\phantom{=}+ \frac{1}{2}\,\sum_{\mu=1}^{N}\,\left(\boldsymbol{X}_{\gamma\mu}\times\boldsymbol{X}^{\beta}_{\mu}\right)\,\left\{\ Q^{\gamma},\ \left[\ P_{\beta},\ Q^{\alpha}\ \right]\ \right\}\\ 
&= \frac{1}{2}\,\left\{\ \mathsf{I}\left(Q^{\,\boldsymbol{.}}\right),\ -\,i\hbar\ \mathsf{I}\left(Q^{\,\boldsymbol{.}}\right)^{-1}\cdot\sum_{\mu=1}^{N}\left(\boldsymbol{X}^{\alpha}_{\mu}\times\boldsymbol{X}_{\mu\beta}\right)\,Q^{\beta}\ \right\}_{\mathsmaller{\bullet}}
\nonumber\\
&\phantom{=}-\,i\hbar\,\sum_{\mu=1}^{N}\left(\boldsymbol{X}_{\mu\beta}\times\boldsymbol{X}^{\alpha}_{\mu}\right)\,Q^{\beta}\nonumber\\
&= -\,i\hbar\,\sum_{\mu=1}^{N}\left(\boldsymbol{X}^{\alpha}_{\mu}\times\boldsymbol{X}_{\mu\beta} + \boldsymbol{X}_{\mu\beta}\times\boldsymbol{X}^{\alpha}_{\mu}\right)\,Q^{\beta} = \boldsymbol{0}\,.\nonumber
\end{align}
Since the operator $\boldsymbol{q}_{(\nu)}$ commutes with the operators $Q^{\alpha}$ and $P_{\beta}$, the commutation relations~\eqref{3.30.1}-\eqref{3.33} and the expression~\eqref{A.7.7} yield the third commutation relation~\eqref{3.34.1}, i.e.
\begin{align}\label{A.8.0 quad}
&\left[\ \boldsymbol{L},\ \boldsymbol{e}^{j}\cdot\boldsymbol{q}_{(\nu^{\prime})}\ \right] = \frac{1}{2}\left\{\, \mathsf{I}\left(Q^{\,\boldsymbol{.}}\right)\cdot\boldsymbol{e}_{k},\ \left[\ \boldsymbol{e}^{k}\cdot\boldsymbol{\Omega},\ \boldsymbol{e}^{j}\cdot\boldsymbol{q}_{(\nu^{\prime})}\ \right]\,\right\}\nonumber\\
&\phantom{=}+ \frac{1}{2}\,\sum_{\nu=1}^{n}\,\left[\ \boldsymbol{q}_{(\nu)},\ \left[\ \boldsymbol{p}_{(\nu)},\ \boldsymbol{e}^{j}\cdot\boldsymbol{q}_{(\nu^{\prime})}\ \right]\ \right]_{\times}\\
&= \frac{1}{2}\,\left\{\ \mathsf{I}\left(Q^{\,\boldsymbol{.}}\right)\cdot\boldsymbol{e}_{k},\ -\,i\hbar\,\left(\boldsymbol{e}^{k}\cdot\mathsf{I}\left(Q^{\,\boldsymbol{.}}\right)^{-1}\cdot\left(\boldsymbol{e}^{j}\times\boldsymbol{q}_{(\nu^{\prime})}\right)\right)\ \right\}\nonumber\\
&\phantom{=}+ \frac{1}{2}\,\sum_{\nu=1}^{n}\,\left[\ \boldsymbol{q}_{(\nu)},\ -\,i\hbar\,\delta_{\nu\nu^{\prime}}\, \boldsymbol{e}^{j}\ \right]_{\times}\nonumber\\
& = 
-\,i\hbar\left(\boldsymbol{e}^{j}\times\boldsymbol{q}_{(\nu^{\prime})} + \boldsymbol{q}_{(\nu^{\prime})}\times\boldsymbol{e}^{j}\right) = \boldsymbol{0}\,.\nonumber
\end{align}
Since the operator $\boldsymbol{p}_{(\nu)}$ commutes with the operators $Q^{\alpha}$ and $P_{\beta}$, the commutation relations~\eqref{3.31.1}-\eqref{3.33} and the expression~\eqref{A.7.7} yield the fourth commutation relation~\eqref{3.34.1}, i.e.
\begin{align}\label{A.8.1 bis} 
&\left[\ \boldsymbol{L},\ \boldsymbol{e}_{j}\cdot\boldsymbol{p}_{(\nu^{\prime})}\ \right] =  \frac{1}{2}\left\{\, \mathsf{I}\left(Q^{\,\boldsymbol{.}}\right)\cdot\boldsymbol{e}_{k},\ \left[\ \boldsymbol{e}^{k}\cdot\boldsymbol{\Omega},\ \boldsymbol{e}_{j}\cdot\boldsymbol{p}_{(\nu^{\prime})}\ \right]\,\right\}\nonumber\\
&\phantom{=}+ \frac{1}{2}\,\sum_{\nu=1}^{n}\,\left[\ \left[\ \boldsymbol{q}_{(\nu)},\ \boldsymbol{e}_{j}\cdot\boldsymbol{p}_{(\nu^{\prime})}\ \right],\ \boldsymbol{p}_{(\nu)}\ \right]_{\times}\\
&= \frac{1}{2}\,\left\{\ \mathsf{I}\left(Q^{\,\boldsymbol{.}}\right)\cdot\boldsymbol{e}_{k},\ -\,i\hbar\,\left(\boldsymbol{e}^{k}\cdot\mathsf{I}\left(Q^{\,\boldsymbol{.}}\right)^{-1}\cdot\left(\boldsymbol{e}^{j}\times\boldsymbol{p}_{(\nu^{\prime})}\right)\right)\ \right\}\nonumber\\
&\phantom{=}+ \frac{1}{2}\,\sum_{\nu=1}^{n}\,\left[\ i\hbar\,\delta_{\nu\nu^{\prime}}\,\boldsymbol{e}_{j},\ \boldsymbol{p}_{(\nu)}\ \right]_{\times}\nonumber\\
&= 
i\hbar\left(\boldsymbol{p}_{(\nu^{\prime})}\times\boldsymbol{e}_{j} + \boldsymbol{e}_{j}\times\boldsymbol{p}_{(\nu^{\prime})}\right) = \boldsymbol{0}\,.\nonumber
\end{align}
Since the operator $P_{\alpha}$ commutes with the operators $Q^{\alpha}$ and $\boldsymbol{p}_{(\nu)}$, the commutation relations~\eqref{3.31.1}-\eqref{3.33} and the expression~\eqref{A.7.7} imply that
\begin{align}\label{A.8.1 ter}
&\left[\ \boldsymbol{L},\ P_{\alpha}\ \right] = \frac{1}{2}\left\{\ \mathsf{I}\left(Q^{\,\boldsymbol{.}}\right)\cdot\boldsymbol{e}_{k},\ \left[\ \boldsymbol{e}^{k}\cdot\boldsymbol{\Omega},\ P_{\alpha}\ \right]\ \right\}\nonumber\\
&\phantom{=}+\frac{1}{2}\,\sum_{\mu=1}^{N}\,\left(\boldsymbol{X}_{\gamma\mu}\times\boldsymbol{X}^{\beta}_{\mu}\right)\,\left\{\ \left[\,Q^{\gamma},\ P_{\alpha}\ \right],\ P_{\beta}\ \right\}\\ 
&= \frac{1}{2}\,\left\{\,\mathsf{I}\left(Q^{\,\boldsymbol{.}}\right),\ -\,\frac{1}{2}\,i\hbar\,\sum_{\mu=1}^{N}\,\left[\ \mathsf{I}\left(Q^{\,\boldsymbol{.}}\right)^{-1}\cdot\boldsymbol{X}_{\mu\alpha},\ P_{\beta}\,\boldsymbol{X}^{\beta}_{\mu}\ \right]_{\times}\,\right\}_{\mathsmaller{\bullet}}\nonumber\\
&\phantom{=}+ i\hbar\,\sum_{\mu=1}^{N}\left(\boldsymbol{X}_{\mu\alpha}\times\boldsymbol{X}^{\beta}_{\mu}\right)\,P_{\beta}\nonumber\\
&= -\,\frac{1}{4}\,i\hbar\,\left[\ \mathsf{I}\left(Q^{\,\boldsymbol{.}}\right),\ P_{\beta}\ \right]\cdot\left(\mathsf{I}\left(Q^{\,\boldsymbol{.}}\right)^{-1}\cdot\sum_{\mu=1}^{N}\,\left(\boldsymbol{X}_{\mu\alpha}\times\boldsymbol{X}^{\beta}_{\mu}\right)\right)\nonumber\\
&\phantom{=}-\,\frac{1}{4}\,i\hbar\, \left(\mathsf{I}\left(Q^{\,\boldsymbol{.}}\right)^{-1}\cdot\sum_{\mu=1}^{N}\,\left(\boldsymbol{X}_{\mu\alpha}\times\boldsymbol{X}^{\beta}_{\mu}\right)\right)\cdot\left[\ P_{\beta},\ \mathsf{I}\left(Q^{\,\boldsymbol{.}}\right)\ \right].\nonumber
\end{align}
Now, the definition~\eqref{3.31} implies that,
\begin{equation}\label{A.8.1 quad}
\left[\ \mathsf{I}\left(Q^{\,\boldsymbol{.}}\right),\ P_{\beta}\ \right] = \mathsf{I}_{\gamma}\,\left[\  Q^{\gamma},\ P_{\beta}\ \right] = i\hbar\,\mathsf{I}_{\beta}\,.
\end{equation}
Substituting the commutation relation~\eqref{A.8.1 quad} into the commutation relation~\eqref{A.8.1 ter}, the latter yields the second commutation relation~\eqref{3.34.1}, i.e.
\begin{equation}\label{A.8.1 pet}
\left[\ \boldsymbol{L},\ P_{\alpha}\ \right] = \boldsymbol{0}\,.
\end{equation}
The commutation relation~\eqref{A.3.2quart} applies for the operator $\boldsymbol{L}$ as well as for the operator $\boldsymbol{P}^{\prime\prime}_{\nu}$. The commutation relations~\eqref{A.3.2quart} and~\eqref{A.8.0} and the property~\eqref{A.2.8} imply that,
\begin{equation}\label{A.8.1 pet 1}
\begin{split}
&\mathsf{R}\left(\boldsymbol{\omega}\right)^{-1}\,\left[\ \boldsymbol{L},\ \mathsf{R}\left(\boldsymbol{\omega}\right)\ \right] = \left[\ \boldsymbol{L},\ \boldsymbol{e}^{j}\cdot\boldsymbol{\omega}\ \right]\,\boldsymbol{n}_{(j)}\left(\boldsymbol{\omega}\right)\cdot\boldsymbol{\mathsf{G}}\\
&= -\,i\hbar\,\left(\boldsymbol{m}^{(j)}\left(\boldsymbol{\omega}\right)\cdot\boldsymbol{n}_{(j)}\left(\boldsymbol{\omega}\right)\right)\,\boldsymbol{\mathsf{G}} = -\,i\hbar\,\boldsymbol{\mathsf{G}}\,.
\end{split}
\end{equation}
The commutation relation~\eqref{A.8.1 pet 1} implies that,
\begin{equation}\label{A.8.1 pet 2}
\left[\ \boldsymbol{e}_{j}\cdot\boldsymbol{L},\ \mathsf{R}\left(\boldsymbol{\omega}\right)\ \right] = -\,i\hbar\,\mathsf{R}\left(\boldsymbol{\omega}\right)\,\left(\boldsymbol{e}_{j}\cdot\boldsymbol{\mathsf{G}}\right)\,.
\end{equation}
Now, the commutation relations~\eqref{3.23} and~\eqref{A.8.1 pet 2} imply that,
\begin{align}\label{A.8.9}
&\left[\ \left[\ \boldsymbol{e}_{j}\cdot\boldsymbol{L},\ \boldsymbol{e}_{k}\cdot\boldsymbol{L}\ \right],\ \mathsf{R}\left(\boldsymbol{\omega}\right)\ \right]\nonumber\vphantom{\Big(\Big)}\\
&=\left[\,\boldsymbol{e}_{j}\cdot\boldsymbol{L},\,\left[\,\boldsymbol{e}_{k}\cdot\boldsymbol{L},\,\mathsf{R}\left(\boldsymbol{\omega}\right)\,\right]\,\right] -\left[\,\boldsymbol{e}_{k}\cdot\boldsymbol{L},\,\left[\,\boldsymbol{e}_{j}\cdot\boldsymbol{L},\,\mathsf{R}\left(\boldsymbol{\omega}\right)\,\right]\,\right]\nonumber\\
&= -\,i\hbar\Big(\!\left(\boldsymbol{e}_{k}\!\cdot\!\boldsymbol{\mathsf{G}}\right)\left[\,\boldsymbol{e}_{j}\cdot\boldsymbol{L},\,\mathsf{R}\left(\boldsymbol{\omega}\right)\,\right] -\left(\boldsymbol{e}_{j}\!\cdot\!\boldsymbol{\mathsf{G}}\right)\left[\,\boldsymbol{e}_{k}\cdot\boldsymbol{L},\,\mathsf{R}\left(\boldsymbol{\omega}\right)\,\right]\!\Big)\nonumber\\
&= \hbar^2\,\mathsf{R}\left(\boldsymbol{\omega}\right)\,\left[\ \boldsymbol{e}_{j}\cdot\boldsymbol{\mathsf{G}},\ \boldsymbol{e}_{k}\cdot\boldsymbol{\mathsf{G}}\ \right]\\
&= i\hbar\,\Big(\!-\,i\hbar\,\mathsf{R}\left(\boldsymbol{\omega}\right)\,\left(\boldsymbol{e}_{j}\times\boldsymbol{e}_{k}\right)\cdot\boldsymbol{\mathsf{G}}\Big)\nonumber\\
&= \left[\ i\hbar\,\left(\boldsymbol{e}_{j}\times\boldsymbol{e}_{k}\right)\cdot\boldsymbol{L},\ \mathsf{R}\left(\boldsymbol{\omega}\right)\ \right]\,.\nonumber
\end{align}
Identifying the terms on the LHS of the commutation relation~\eqref{A.8.9} yields the second commutation relation~\eqref{3.35.1}, i.e.
\begin{equation}\label{A.8.10}
\left[\ \boldsymbol{e}_{j}\cdot\boldsymbol{L},\ \boldsymbol{e}_{k}\cdot\boldsymbol{L}\ \right] = i\hbar\,\left(\boldsymbol{e}_{j}\times\boldsymbol{e}_{k}\right)\cdot\boldsymbol{L}\,,
\end{equation}
which is recast as,
\begin{equation}\label{A.8.11}
\left[\ \boldsymbol{L},\ \boldsymbol{e}_{k}\cdot\boldsymbol{L}\ \right] = i\hbar\,\left(\boldsymbol{e}_{k}\times\boldsymbol{L}\right)\,.
\end{equation}
Since the operator $\boldsymbol{L}$ commutes with the operators $Q^{\alpha}$, $P_{\alpha}$, $q^{\alpha}$, $p_{\alpha}$,  the commutation relations~\eqref{A.8.0 bis}-\eqref{A.8.1 bis} and \eqref{A.8.1 pet}, and the expression~\eqref{A.7.7} imply that
\begin{align}\label{A.8.12}
&\left[\ \boldsymbol{L},\ \boldsymbol{e}_{k}\cdot\boldsymbol{L}\ \right] = \frac{1}{2}\,\left[\ \left\{\ \mathsf{I}\left(Q^{\,\boldsymbol{.}}\right),\ \boldsymbol{\Omega}\ \right\}_{\mathsmaller{\bullet}},\ \boldsymbol{e}_{k}\cdot\boldsymbol{L}\ \right]\nonumber\\
&= \frac{1}{2}\,\left[\ \left\{\ \mathsf{I}\left(Q^{\,\boldsymbol{.}}\right)\cdot\boldsymbol{e}_{\ell},\ \boldsymbol{e}^{\ell}\cdot\boldsymbol{\Omega}\ \right\}_{\mathsmaller{\bullet}},\ \boldsymbol{e}_{k}\cdot\boldsymbol{L}\ \right]\nonumber\\
&= \frac{1}{2}\,\left\{\ \mathsf{I}\left(Q^{\,\boldsymbol{.}}\right)\cdot\boldsymbol{e}_{\ell},\ \left[\ \boldsymbol{e}^{\ell}\cdot\boldsymbol{\Omega},\ \boldsymbol{e}_{k}\cdot\boldsymbol{L}\ \right]\ \right\}\\
&\phantom{=}+ \frac{1}{2}\,\left\{\ \boldsymbol{e}^{\ell}\cdot\boldsymbol{\Omega},\ \left[\ \mathsf{I}\left(Q^{\,\boldsymbol{.}}\right)\cdot\boldsymbol{e}_{\ell},\ \boldsymbol{e}_{k}\cdot\boldsymbol{L}\ \right]\ \right\}\,.\nonumber
\end{align}
The definition~\eqref{3.31} and the commutation relation~\eqref{A.8.0 bis} imply that,
\begin{equation}\label{A.8.13}
\begin{split}
&\left[\ \mathsf{I}\left(Q^{\,\boldsymbol{.}}\right)\cdot\boldsymbol{e}_{\ell},\ \boldsymbol{e}_{k}\cdot\boldsymbol{L}\ \right] = 
\left[\ \mathsf{I}_0\cdot\boldsymbol{e}_{\ell} + Q^{\alpha}\,\left(\mathsf{I}_{\alpha}\cdot\boldsymbol{e}_{\ell}\right),\ \boldsymbol{e}_{k}\cdot\boldsymbol{L}\ \right]\\
&= \mathsf{I}_{\alpha}\cdot\boldsymbol{e}_{\ell}\,\left[\ Q^{\alpha},\ \boldsymbol{e}_{k}\cdot\boldsymbol{L}\ \right] = \boldsymbol{0}\,,
\end{split}
\end{equation}
which implies in turn that,
\begin{equation}\label{A.8.14}
\left[\ \boldsymbol{e}^{\ell}\cdot\mathsf{I}\left(Q^{\,\boldsymbol{.}}\right)^{-1},\ \boldsymbol{e}_{k}\cdot\boldsymbol{L}\ \right] = \boldsymbol{0}\,.
\end{equation}
The commutation relation~\eqref{A.8.13} implies that the commutation relation~\eqref{A.8.12} is recast as,
\begin{equation}\label{A.8.15}
\left[\ \boldsymbol{L},\ \boldsymbol{e}_{k}\cdot\boldsymbol{L}\ \right] = \frac{1}{2}\,\left\{\ \mathsf{I}\left(Q^{\,\boldsymbol{.}}\right)\cdot\boldsymbol{e}_{\ell},\ \left[\ \boldsymbol{e}^{\ell}\cdot\boldsymbol{\Omega},\ \boldsymbol{e}_{k}\cdot\boldsymbol{L}\ \right]\ \right\}\,.
\end{equation}
The commutation relations~\eqref{A.8.11} and~\eqref{A.8.14}-\eqref{A.8.15} imply that,
\begin{align}\label{A.8.16}
&\left[\ \boldsymbol{e}_{k}\cdot\boldsymbol{L},\ \boldsymbol{e}^{\ell}\cdot\boldsymbol{\Omega}\ \right] = -\,\left(\boldsymbol{e}^{\ell}\cdot\mathsf{I}\left(Q^{\,\boldsymbol{.}}\right)^{-1}\right)\cdot\left[\ \boldsymbol{L},\ \boldsymbol{e}_{k}\cdot\boldsymbol{L}\ \right]\nonumber\\
&= -\,i\hbar\,\left(\boldsymbol{e}^{\ell}\cdot\mathsf{I}\left(Q^{\,\boldsymbol{.}}\right)^{-1}\cdot\boldsymbol{e}^{j}\right)\left(\boldsymbol{e}_{j}\cdot\left(\boldsymbol{e}_{k}\times\boldsymbol{L}\right)\right)\\
&= i\hbar\,\left(\boldsymbol{e}^{\ell}\cdot\mathsf{I}\left(Q^{\,\boldsymbol{.}}\right)^{-1}\cdot\boldsymbol{e}^{j}\right)\left(\boldsymbol{e}_{k}\cdot\left(\boldsymbol{e}_{j}\times\boldsymbol{L}\right)\right)\,,\nonumber
\end{align}
which yields the second commutation relation~\eqref{3.35.1}, i.e.
\begin{equation}\label{A.8.17}
\left[\ \boldsymbol{L},\ \boldsymbol{e}^{\ell}\cdot\boldsymbol{\Omega}\ \right] = i\hbar\,\left(\boldsymbol{e}^{\ell}\cdot\mathsf{I}\left(Q^{\,\boldsymbol{.}}\right)^{-1}\cdot\boldsymbol{e}^{j}\right)\left(\boldsymbol{e}_{j}\times\boldsymbol{L}\right)\,.
\end{equation}

\section{Commutation relations of spin and the internal observables}
\label{A.9}

In this appendix, we determine explicitly the commutation relations between the nuclear spin $\boldsymbol{S}_{(\mu)}$, the electronic spin $\boldsymbol{s}_{(\mu)}$ and the internal observables $P_{\alpha}$, $\boldsymbol{p}_{(\nu)}$ and $\boldsymbol{L}$.

The definitions~\eqref{3.36bis} and the commutations relations~\eqref{2.3 bis} and~\eqref{2.7} imply that
\begin{align}
\label{A.9.1}
&\left[\ \boldsymbol{e}_{i}\cdot\boldsymbol{S}_{(\mu)},\ \boldsymbol{e}_{j}\cdot\boldsymbol{S}_{(\nu)}\ \right]\\
& = \left(\boldsymbol{e}^{k}\cdot\mathsf{R}\left(\boldsymbol{\omega}\right)\cdot\boldsymbol{e}_{i}\right)\,\left(\boldsymbol{e}^{\ell}\cdot\mathsf{R}\left(\boldsymbol{\omega}\right)\cdot\boldsymbol{e}_{j}\right)\,\left[\ \boldsymbol{e}_{k}\cdot\boldsymbol{S}_{\mu},\ \boldsymbol{e}_{\ell}\cdot\boldsymbol{S}_{\nu}\ \right]\nonumber\\ 
& = i\hbar\ \delta_{\mu\nu} \left(\boldsymbol{e}^{k}\cdot\mathsf{R}\left(\boldsymbol{\omega}\right)\cdot\boldsymbol{e}_{i}\right)\,\left(\boldsymbol{e}^{\ell}\cdot\mathsf{R}\left(\boldsymbol{\omega}\right)\cdot\boldsymbol{e}_{j}\right)\nonumber\\
&\phantom{=}\cdot\left(\boldsymbol{e}_k\times\boldsymbol{e}_{\ell}\right)\cdot\boldsymbol{e}^m\,\left(\boldsymbol{e}_m\cdot\boldsymbol{S}_{\mu}\right)\nonumber\\
&\nonumber\\
\label{A.9.2}
&\left[\ \boldsymbol{e}_{i}\cdot\boldsymbol{s}_{(\mu)},\ \boldsymbol{e}_{j}\cdot\boldsymbol{s}_{(\nu)}\ \right]\\
& = \left(\boldsymbol{e}^{k}\cdot\mathsf{R}\left(\boldsymbol{\omega}\right)\cdot\boldsymbol{e}_{i}\right)\,\left(\boldsymbol{e}^{\ell}\cdot\mathsf{R}\left(\boldsymbol{\omega}\right)\cdot\boldsymbol{e}_{j}\right)\,\left[\ \boldsymbol{e}_{k}\cdot\boldsymbol{s}_{\mu},\ \boldsymbol{e}_{\ell}\cdot\boldsymbol{s}_{\nu}\ \right]\nonumber\\ 
& = i\hbar\ \delta_{\mu\nu} \left(\boldsymbol{e}^{k}\cdot\mathsf{R}\left(\boldsymbol{\omega}\right)\cdot\boldsymbol{e}_{i}\right)\,\left(\boldsymbol{e}^{\ell}\cdot\mathsf{R}\left(\boldsymbol{\omega}\right)\cdot\boldsymbol{e}_{j}\right)\nonumber\\
&\phantom{=}\cdot\left(\boldsymbol{e}_k\times\boldsymbol{e}_{\ell}\right)\cdot\boldsymbol{e}^m\,\left(\boldsymbol{e}_m\cdot\boldsymbol{s}_{\mu}\right)\nonumber
\end{align}
The triple product $\left(\boldsymbol{e}_k\times\boldsymbol{e}_{\ell}\right)\cdot\boldsymbol{e}^m$ is invariant under rotation, which implies that,
\begin{equation}\label{A.9.3}
\begin{split}
&\left(\boldsymbol{e}^{k}\cdot\mathsf{R}\left(\boldsymbol{\omega}\right)\cdot\boldsymbol{e}_{i}\right)\,\left(\boldsymbol{e}^{\ell}\cdot\mathsf{R}\left(\boldsymbol{\omega}\right)\cdot\boldsymbol{e}_{j}\right)\,\left(\boldsymbol{e}_k\times\boldsymbol{e}_{\ell}\right)\cdot\boldsymbol{e}^m\\
&= \left(\boldsymbol{e}_i\times\boldsymbol{e}_j\right)\cdot\boldsymbol{e}^{n}\,\left(\boldsymbol{e}^m\cdot\mathsf{R}\left(\boldsymbol{\omega}\right)\cdot\boldsymbol{e}_n\right)
\end{split}
\end{equation}
Using the vectorial identity~\eqref{A.9.3} and the definitions~\eqref{3.36bis}, the commutation relations~\eqref{A.9.1} and~\eqref{A.9.2} yield the commutation relations~\eqref{3.36ter}, i.e.
\begin{align}
\label{A.9.4}
&\left[\ \boldsymbol{e}_{i}\cdot\boldsymbol{S}_{(\mu)},\ \boldsymbol{e}_{j}\cdot\boldsymbol{S}_{(\nu)}\ \right]\nonumber\\*
& = i\hbar\ \delta_{\mu\nu} \left(\boldsymbol{e}_i\times\boldsymbol{e}_j\right)\cdot\boldsymbol{e}^{n}\,\left(\boldsymbol{e}^m\cdot\mathsf{R}\left(\boldsymbol{\omega}\right)\cdot\boldsymbol{e}_n\right)\,\left(\boldsymbol{e}_m\cdot\boldsymbol{S}_{\mu}\right)\\*
& = i\hbar\ \delta_{\mu\nu} \left(\boldsymbol{e}_i\times\boldsymbol{e}_j\right)\cdot\boldsymbol{S}_{(\mu)}\nonumber\\*
&\nonumber\\*
\label{A.9.5}
&\left[\ \boldsymbol{e}_{i}\cdot\boldsymbol{s}_{(\mu)},\ \boldsymbol{e}_{j}\cdot\boldsymbol{s}_{(\nu)}\ \right]\nonumber\\*
& = i\hbar\ \delta_{\mu\nu} \left(\boldsymbol{e}_i\times\boldsymbol{e}_j\right)\cdot\boldsymbol{e}^{n}\,\left(\boldsymbol{e}^m\cdot\mathsf{R}\left(\boldsymbol{\omega}\right)\cdot\boldsymbol{e}_n\right)\,\left(\boldsymbol{e}_m\cdot\boldsymbol{s}_{\mu}\right)\\*
& = i\hbar\ \delta_{\mu\nu} \left(\boldsymbol{e}_i\times\boldsymbol{e}_j\right)\cdot\boldsymbol{s}_{(\mu)}\nonumber
\end{align}
Using the group action~\eqref{3.23bis}, the identity~\eqref{A.2.8}, the definitions~\eqref{3.36bis}, the commutation relation~\eqref{A.3.2quart}  where $\boldsymbol{e}_{\ell}\cdot\boldsymbol{P}^{\prime\prime}_{\nu}$ is replaced by $\boldsymbol{e}^{j}\cdot\boldsymbol{L}$ and the commutation relation~\eqref{A.8.0}, we compute the commutation relations,
\begin{align}
\label{A.9.6}
&\left[\ \boldsymbol{e}_{j}\cdot\boldsymbol{L},\ \boldsymbol{S}_{(\mu)}\ \right]\nonumber\\
& = \left[\ \boldsymbol{e}_{j}\cdot\boldsymbol{L},\ \boldsymbol{e}^k\cdot\mathsf{R}\left(\boldsymbol{\omega}\right)\ \right]\,\left(\boldsymbol{e}_k\cdot\boldsymbol{S}_{\mu}\right)\nonumber\\
& = \left[\ \boldsymbol{e}_{j}\cdot\boldsymbol{L},\ \boldsymbol{e}^{\ell}\cdot\boldsymbol{\omega}\ \right]\,\left(\boldsymbol{n}_{(\ell)}\left(\boldsymbol{\omega}\right)\cdot\boldsymbol{\mathsf{G}}\right)\,\boldsymbol{S}_{(\mu)}\\
& = -\,i\hbar\,\left(\boldsymbol{e}_{j}\cdot\boldsymbol{m}^{(\ell)}\left(\boldsymbol{\omega}\right)\right)\,\left(\boldsymbol{n}_{(\ell)}\left(\boldsymbol{\omega}\right)\cdot\boldsymbol{\mathsf{G}}\right)\,\boldsymbol{S}_{(\mu)}\nonumber\\
& = -\,i\hbar\,\left(\boldsymbol{e}_{j}\cdot\boldsymbol{\mathsf{G}}\right)\,\boldsymbol{S}_{(\mu)} = -\,i\hbar\,\left(\boldsymbol{e}_{j}\times\boldsymbol{S}_{(\mu)}\right) \nonumber\\
&\nonumber\\
\label{A.9.7}
&\left[\ \boldsymbol{e}_{j}\cdot\boldsymbol{L},\ \boldsymbol{s}_{(\mu)}\ \right]\nonumber\\
& = \left[\ \boldsymbol{e}_{j}\cdot\boldsymbol{L},\ \boldsymbol{e}^k\cdot\mathsf{R}\left(\boldsymbol{\omega}\right)\ \right]\,\left(\boldsymbol{e}_k\cdot\boldsymbol{s}_{\mu}\right)\nonumber\\
& = \left[\ \boldsymbol{e}_{j}\cdot\boldsymbol{L},\ \boldsymbol{e}^{\ell}\cdot\boldsymbol{\omega}\ \right]\,\left(\boldsymbol{n}_{(\ell)}\left(\boldsymbol{\omega}\right)\cdot\boldsymbol{\mathsf{G}}\right)\,\boldsymbol{s}_{(\mu)}\\
& = -\,i\hbar\,\left(\boldsymbol{e}_{j}\cdot\boldsymbol{m}^{(\ell)}\left(\boldsymbol{\omega}\right)\right)\,\left(\boldsymbol{n}_{(\ell)}\left(\boldsymbol{\omega}\right)\cdot\boldsymbol{\mathsf{G}}\right)\,\boldsymbol{s}_{(\mu)}\nonumber\\
& = -\,i\hbar\,\left(\boldsymbol{e}_{j}\cdot\boldsymbol{\mathsf{G}}\right)\,\boldsymbol{s}_{(\mu)} = -\,i\hbar\,\left(\boldsymbol{e}_{j}\times\boldsymbol{s}_{(\mu)}\right)\nonumber
\end{align}
Using the definitions~\eqref{3.36bis} and~\eqref{A.5.9}, the commutation relations~\eqref{A.3.2quart} and~\eqref{A.6.1}, we compute the commutation relations,
\begin{align}
\label{A.9.8}
&\left[\ P_{\alpha},\ \boldsymbol{S}_{(\mu)}\ \right]\nonumber\\
& = \sum_{\nu=1}^{N}\,\frac{1}{\sqrt{M_{\nu}}}\,\left(\boldsymbol{e}^j\cdot\boldsymbol{X}_{\alpha\nu}\right)\left[\ \boldsymbol{e}_{j}\cdot\boldsymbol{P}^{\prime\prime}_{\nu},\ \boldsymbol{e}^k\cdot\mathsf{R}\left(\boldsymbol{\omega}\right)\ \right]\,\left(\boldsymbol{e}_k\cdot\boldsymbol{S}_{\mu}\right)\nonumber\\
& = \sum_{\nu=1}^{N}\,\frac{1}{\sqrt{M_{\nu}}}\,\left(\boldsymbol{e}^j\cdot\boldsymbol{X}_{\alpha\nu}\right)\left[\ \boldsymbol{e}_{j}\cdot\boldsymbol{P}^{\prime\prime}_{\nu},\ \boldsymbol{e}^{\ell}\cdot\boldsymbol{\omega}\ \right]\\
&\phantom{=\sum_{\nu=1}^{N}\,\frac{1}{\sqrt{M_{\nu}}}}\cdot\left(\boldsymbol{n}_{(\ell)}\left(\boldsymbol{\omega}\right)\cdot\boldsymbol{\mathsf{G}}\right)\,\boldsymbol{S}_{(\mu)}\nonumber\\
& = \left[\ P_{\alpha},\ \boldsymbol{e}^{\ell}\cdot\boldsymbol{\omega}\ \right]\,\left(\boldsymbol{n}_{(\ell)}\left(\boldsymbol{\omega}\right)\cdot\boldsymbol{\mathsf{G}}\right)\,\boldsymbol{S}_{(\mu)} = \boldsymbol{0}\nonumber\\
&\nonumber\\
\label{A.9.9}
&\left[\ P_{\alpha},\ \boldsymbol{s}_{(\mu)}\ \right]\nonumber\\*
& = \sum_{\nu=1}^{N}\,\frac{1}{\sqrt{M_{\nu}}}\,\left(\boldsymbol{e}^j\cdot\boldsymbol{X}_{\alpha\nu}\right)\left[\ \boldsymbol{e}_{j}\cdot\boldsymbol{P}^{\prime\prime}_{\nu},\ \boldsymbol{e}^k\cdot\mathsf{R}\left(\boldsymbol{\omega}\right)\ \right]\,\left(\boldsymbol{e}_k\cdot\boldsymbol{s}_{\mu}\right)\nonumber\\*
& = \sum_{\nu=1}^{N}\,\frac{1}{\sqrt{M_{\nu}}}\,\left(\boldsymbol{e}^j\cdot\boldsymbol{X}_{\alpha\nu}\right)\left[\ \boldsymbol{e}_{j}\cdot\boldsymbol{P}^{\prime\prime}_{\nu},\ \boldsymbol{e}^{\ell}\cdot\boldsymbol{\omega}\ \right]\\
&\phantom{=\sum_{\nu=1}^{N}\,\frac{1}{\sqrt{M_{\nu}}}}\cdot\left(\boldsymbol{n}_{(\ell)}\left(\boldsymbol{\omega}\right)\cdot\boldsymbol{\mathsf{G}}\right)\,\boldsymbol{s}_{(\mu)}\nonumber\\
& = \left[\ P_{\alpha},\ \boldsymbol{e}^{\ell}\cdot\boldsymbol{\omega}\ \right]\,\left(\boldsymbol{n}_{(\ell)}\left(\boldsymbol{\omega}\right)\cdot\boldsymbol{\mathsf{G}}\right)\,\boldsymbol{s}_{(\mu)} = \boldsymbol{0}\nonumber
\end{align}
Using the definitions~\eqref{3.36bis}, the commutation relations~\eqref{A.3.2quart} and~\eqref{A.6.0before}, we compute the commutation relations,
\begin{align}
\label{A.9.10}
&\left[\ \boldsymbol{e}_j\cdot\boldsymbol{p}_{(\nu)},\ \boldsymbol{S}_{(\mu)}\ \right]\nonumber\\
& = \left[\ \boldsymbol{e}_j\cdot\boldsymbol{p}_{(\nu)},\ \boldsymbol{e}^k\cdot\mathsf{R}\left(\boldsymbol{\omega}\right)\ \right]\,\left(\boldsymbol{e}_k\cdot\boldsymbol{S}_{\mu}\right)\\
& = \left[\ \boldsymbol{e}_j\cdot\boldsymbol{p}_{(\nu)},\ \boldsymbol{e}^{\ell}\cdot\boldsymbol{\omega}\ \right]\,\left(\boldsymbol{n}_{(\ell)}\left(\boldsymbol{\omega}\right)\cdot\boldsymbol{\mathsf{G}}\right)\,\boldsymbol{S}_{(\mu)} = \boldsymbol{0}\nonumber\\
&\nonumber\\
\label{A.9.11}
&\left[\ \boldsymbol{e}_j\cdot\boldsymbol{p}_{(\nu)},\ \boldsymbol{s}_{(\mu)}\ \right]\nonumber\\*
& = \left[\ \boldsymbol{e}_j\cdot\boldsymbol{p}_{(\nu)},\ \boldsymbol{e}^k\cdot\mathsf{R}\left(\boldsymbol{\omega}\right)\ \right]\,\left(\boldsymbol{e}_k\cdot\boldsymbol{s}_{\mu}\right)\\
& = \left[\ \boldsymbol{e}_j\cdot\boldsymbol{p}_{(\nu)},\ \boldsymbol{e}^{\ell}\cdot\boldsymbol{\omega}\ \right]\,\left(\boldsymbol{n}_{(\ell)}\left(\boldsymbol{\omega}\right)\cdot\boldsymbol{\mathsf{G}}\right)\,\boldsymbol{s}_{(\mu)} = \boldsymbol{0}\nonumber
\end{align}

\section{Kinetic energy operator}
\label{B.1}

In this appendix, we determine the expression of the kinetic energy operator $T$ in terms of the internal observables.

Using the relation~\eqref{3.20pre}and the definition~\eqref{3.10}, the expression~\eqref{4.0.0} of the kinetic energy operator $T$ is recast as,
\begin{align}\label{B.1.0 bis}
&T = \sum_{\mu=1}^{N}\frac{1}{2\,M_{\mu}}\,\left(\boldsymbol{P}^{\prime}_{\mu} + \frac{M_{\mu}}{\mathcal{M}}\,\boldsymbol{\mathcal{P}}\right)^{2} +\sum_{\nu=1}^{n}\frac{1}{2\,m}\,\left(\boldsymbol{p}^{\prime}_{\nu}+\frac{m}{\mathcal{M}}\boldsymbol{\mathcal{P}}\right)^{2}\nonumber\\
&\phantom{T} = \left(\sum_{\mu=1}^{N}\,M_{\mu} + \sum_{\nu=1}^{n}\,m\right)\,\frac{\boldsymbol{\mathcal{P}}^{2}}{2\,\mathcal{M}^2} + \sum_{\mu=1}^{N}\,\frac{\boldsymbol{P}^{\prime\,2}_{\mu}}{2\,M_{\mu}} + \sum_{\nu=1}^{n}\,\frac{\boldsymbol{p}^{\prime\,2}_{\nu}}{2\,m}\nonumber\\ 
&\phantom{T =}+ \left(\sum_{\mu=1}^{N}\,\boldsymbol{P}^{\prime}_{\mu} + \sum_{\nu=1}^{n}\,\boldsymbol{p}^{\prime}_{\nu}\right)\cdot\frac{\boldsymbol{\mathcal{P}}}{\mathcal{M}}\,.
\end{align}
Using the relations~\eqref{3.4}-\eqref{3.20pre} and~\eqref{3.12}, the kinetic energy operator~\eqref{B.1.0 bis} reduces to,
\begin{equation}\label{B.1.1}
T = \frac{\boldsymbol{\mathcal{P}}^{2}}{2\,\mathcal{M}} + \sum_{\mu=1}^{N}\,\frac{\boldsymbol{P}^{\prime\,2}_{\mu}}{2\,M_{\mu}} + \sum_{\nu=1}^{n}\,\frac{\boldsymbol{p}^{\prime\,2}_{\nu}}{2\,m}\,.
\end{equation}
Using the relation~\eqref{3.13ter}, the first term on the RHS of the expression~\eqref{B.1.1} is recast as,
\begin{equation}\label{B.1.3}
\begin{split}
&\sum_{\mu=1}^{N}\frac{\boldsymbol{P}^{\prime\,2}_{\mu}}{2\,M_{\mu}} = \sum_{\mu=1}^{N}\frac{\boldsymbol{P}^{\prime\prime\,2}_{\mu}}{2\ M_{\mu}}
+ \sum_{\mu=1}^{N}\,\frac{\hbar^{2}}{8\,M_{\mu}}\,\boldsymbol{A}_{\mu}\left(Q^{\,\boldsymbol{.}}\right)^{2}\\
&-\,\sum_{\mu=1}^{N}\,\frac{i\hbar}{4\,M_{\mu}}\,\left[\ \boldsymbol{e}_{\ell}\cdot\boldsymbol{P}^{\prime\prime}_{\mu},\ \boldsymbol{e}^{\ell}\cdot\boldsymbol{A}_{\mu}\left(Q^{\,\boldsymbol{.}}\right)\ \right]\,,
\end{split}
\end{equation}
where the operator $\boldsymbol{A}_{\mu}\left(Q^{\,\boldsymbol{.}}\right)$ is defined as,
\begin{align}\label{B.1.2}
&-\,i\hbar\,\left(\boldsymbol{e}^{\ell}\cdot\boldsymbol{A}_{\mu}\left(Q^{\,\boldsymbol{.}}\right)\right)\\
&=\left[\ \boldsymbol{e}^{k}\cdot\boldsymbol{P}^{\prime\prime}_{\mu},\ \boldsymbol{e}^{j}\cdot\mathsf{R}\left(\boldsymbol{\omega}\right)\cdot\boldsymbol{e}_{k}\ \right]\,\left(\boldsymbol{e}^{\ell}\cdot\mathsf{R}\left(\boldsymbol{\omega}\right)^{-1}\cdot\boldsymbol{e}_{j}\right)\,.\nonumber
\end{align}
Using the commutation relation~\eqref{A.6.2 extra 2}, the operator $\boldsymbol{A}_{\mu}\left(Q^{\,\boldsymbol{.}}\right)$ is recast as,
\begin{equation}\label{B.1.2bis}                              
\begin{split}
&\boldsymbol{A}_{\mu}\left(Q^{\,\boldsymbol{.}}\right) = \left(\boldsymbol{e}^{j}\cdot\mathsf{I}\left(Q^{\,\boldsymbol{.}}\right)^{-1}\cdot\boldsymbol{e}^{k}\right)\left(\boldsymbol{e}_{j}\times\left(\boldsymbol{e}_{k}\times\left(M_{\mu}\boldsymbol{R}^{(0)}_{\mu}\right)\right)\right)\\              
&= \left(\boldsymbol{e}^{j}\cdot\mathsf{I}\left(Q^{\,\boldsymbol{.}}\right)^{-1}\cdot\boldsymbol{e}^{k}\right)\\
&\phantom{=}\cdot\left(M_{\mu}\,\left(\boldsymbol{e}_j\cdot\boldsymbol{R}^{(0)}_{\mu}\right)\,\boldsymbol{e}_{k} -\,\left(\boldsymbol{e}_{j}\cdot\boldsymbol{e}_{k}\right)\,M_{\mu}\,\boldsymbol{R}^{(0)}_{\mu}\right)\\
&= M_{\mu}\,\left(\boldsymbol{R}^{(0)}_{\mu}\cdot\mathsf{I}\left(Q^{\,\boldsymbol{.}}\right)^{-1}\right) -\,M_{\mu}\,\boldsymbol{R}^{(0)}_{\mu}\, \text{Tr}\left(\mathsf{I}\left(Q^{\,\boldsymbol{.}}\right)^{-1}\right)\,. 
\end{split}
\end{equation}
Using the definitions~\eqref{A.2.1bis} and~\eqref{3.13ter}, the third term on the RHS of the relation~\eqref{B.1.1} is recast as,
\begin{equation}\label{B.1.4}
\sum_{\nu=1}^{n}\,\frac{\boldsymbol{p}^{\prime\,2}_{\nu}}{2\,m} = \sum_{\nu=1}^{n}\frac{\boldsymbol{p}^{\prime\prime\,2}_{\nu}}{2\,m}\,.
\end{equation}
The identities~\eqref{B.1.3} and~\eqref{B.1.4} imply that the kinetic energy~\eqref{B.1.1 sec} is recast as,
\begin{equation}\label{B.1.1 sec}
T = \frac{\boldsymbol{\mathcal{P}}^{2}}{2\,\mathcal{M}} + \sum_{\mu=1}^{N}\,\frac{\boldsymbol{P}^{\prime\prime\,2}_{\mu}}{2\,M_{\mu}} + \sum_{\nu=1}^{n}\,\frac{\boldsymbol{p}^{\prime\prime\,2}_{\nu}}{2\,m} + \frac{\hbar^2}{8}\,\Phi_{\text{res}}\left(Q^{\,\boldsymbol{.}}\right)\,,
\end{equation}
where the operator accounting for the residual terms is given by,
\begin{align}\label{B.1.1 res}
&\Phi_{\text{res}}\left(Q^{\,\boldsymbol{.}}\right)\\
&= -\,\frac{2i}{\hbar}\,\sum_{\mu=1}^{N}\,\frac{1}{M_{\mu}}\,\left[\,\boldsymbol{P}^{\prime\prime}_{\mu},\,\boldsymbol{A}_{\mu}\left(Q^{\,\boldsymbol{.}}\right)\,\right]_{\mathsmaller{\bullet}} + \sum_{\mu=1}^{N}\,\frac{1}{M_{\mu}}\,\boldsymbol{A}_{\mu}\left(Q^{\,\boldsymbol{.}}\right)^{2}\,.\nonumber
\end{align}

Now, we recast the kinetic energy operator $T$ in terms of the internal observables. The relation~\eqref{3.25}, the conditions~\eqref{3.16}-\eqref{3.17bis} and the definitions~\eqref{3.20pre} and~\eqref{3.32} imply that the second term on the RHS of the expression~\eqref{B.1.1 sec} is recast as,
\begin{equation}\label{B.1.5}
\begin{split}
&\sum_{\mu=1}^{N}\frac{\boldsymbol{P}^{\prime\prime\,2}_{\mu}}{2\,M_{\mu}} = \frac{1}{2}\,\boldsymbol{\Omega}\cdot\mathsf{I}_{0}\cdot\boldsymbol{\Omega} +\frac{1}{2}\,P_{\alpha}\,P^{\alpha}\\
&+\frac{1}{2\, m}\sum_{\nu,\,\overline{\nu}=1}^{n}\sum_{\nu^{\prime},\,\overline{\nu}^{\prime}=1}^{n}\frac{m}{M}\,A_{\nu\overline{\nu}}\,A_{\nu^{\prime}\overline{\nu}^{\prime}}\,\left(\boldsymbol{p}_{(\overline{\nu})}\cdot\boldsymbol{p}_{(\overline{\nu}^{\prime})}\right)
\,.
\end{split}
\end{equation}
Similarly, the relation~\eqref{3.26} implies that the third term on the RHS of the expression~\eqref{B.1.1 sec} is recast as,
\begin{equation}\label{B.1.5 bis}
\sum_{\nu=1}^{n}\frac{\boldsymbol{p}^{\prime\prime\,2}_{\nu}}{2\,m}
= \frac{1}{2\, m}\sum_{\nu,\,\overline{\nu}=1}^{n}\sum_{\nu^{\prime},\,\overline{\nu}^{\prime}=1}^{n}\,\delta_{\nu\nu^{\prime}}\,A_{\nu\overline{\nu}}\,A_{\nu^{\prime}\overline{\nu}^{\prime}}\!\left(\boldsymbol{p}_{(\overline{\nu})}\cdot\boldsymbol{p}_{(\overline{\nu}^{\prime})}\right)\,.
\end{equation}
Using the identity~\eqref{A.7.6quad}, the sum of the relations~\eqref{B.1.5} and~\eqref{B.1.5 bis} reduces to,
\begin{equation}\label{B.1.5 quad}
\sum_{\mu=1}^{N}\frac{\boldsymbol{P}^{\prime\prime\,2}_{\mu}}{2\,M_{\mu}} + \sum_{\nu=1}^{n}\frac{\boldsymbol{p}^{\prime\prime\,2}_{\nu}}{2\,m}
= \frac{1}{2}\,\boldsymbol{\Omega}\cdot\mathsf{I}_{0}\cdot\boldsymbol{\Omega} +\frac{1}{2}\,P_{\alpha}\,P^{\alpha} + \sum_{\nu=1}^{n}\,\frac{\boldsymbol{p}_{(\nu)}^2}{2\,m}
\end{equation}
which implies that the kinetic energy yields the expression~\eqref{4.1}, i.e.
\begin{equation}\label{B.1.1 tri}
T = \frac{\boldsymbol{\mathcal{P}}^{2}}{2\,\mathcal{M}} + \frac{1}{2}\,\boldsymbol{\Omega}\cdot\mathsf{I}_{0}\cdot\boldsymbol{\Omega} +\frac{1}{2}\,P_{\alpha}\,P^{\alpha} + \sum_{\nu=1}^{n}\,\frac{\boldsymbol{p}_{(\nu)}^2}{2\,m} + \frac{\hbar^2}{8}\,\Phi_{\text{res}}\left(Q^{\,\boldsymbol{.}}\right)
\end{equation}

Now, we determine the expression of the residual terms in the expression~\eqref{B.1.1 tri}. Using the commutation relation~\eqref{3.31.1} and the relation~\eqref{3.25}, the first residual term appearing in the expression~\eqref{B.1.3} is recast as,
\begin{align}\label{B.1.6}
&-\,\frac{2i}{\hbar}\,\sum_{\mu=1}^{N}\,\frac{1}{M_{\mu}}\,\left[\,\boldsymbol{P}^{\prime\prime}_{\mu},\ \boldsymbol{A}_{\mu}\left(Q^{\,\boldsymbol{.}}\right)\ \right]_{\mathsmaller{\bullet}}\nonumber\\
&= -\,\frac{2i}{\hbar}\,\sum_{\mu=1}^{N}\,\left[\ \boldsymbol{e}_{\ell}\cdot\left(\boldsymbol{\Omega}\times\boldsymbol{R}^{(0)}_{\mu}\right),\ \boldsymbol{e}^{\ell}\cdot\boldsymbol{A}_{\mu}\left(Q^{\,\boldsymbol{.}}\right)\ \right]\nonumber\\
&\phantom{=}-\,\frac{2i}{\hbar}\,\sum_{\mu=1}^{N}\,\left[\  \frac{1}{\sqrt{M_{\mu}}}\,P_{\alpha}\,\left(\boldsymbol{e}_{\ell}\cdot\boldsymbol{X}^{\alpha}_{\mu}\right),\ \boldsymbol{e}^{\ell}\cdot\boldsymbol{A}_{\mu}\left(Q^{\,\boldsymbol{.}}\right)\ \right]\nonumber\\
&= -\,\frac{2i}{\hbar}\, 
\left[\ \boldsymbol{e}^{k}\cdot\boldsymbol{\Omega},\ \boldsymbol{e}_{k}\cdot\left(\sum_{\mu=1}^{N}\,\boldsymbol{R}^{(0)}_{\mu}\times\boldsymbol{A}_{\mu}\left(Q^{\,\boldsymbol{.}}\right)\right)\ \right]\nonumber\\
&\phantom{=} -\,\frac{2i}{\hbar}\, 
\left[\ P_{\alpha},\ \sum_{\mu=1}^{N}\,\frac{1}{\sqrt{M_{\mu}}}\,\left(\boldsymbol{X}^{\alpha}_{\mu}\cdot\boldsymbol{A}_{\mu}\left(Q^{\,\boldsymbol{.}}\right)\right)\ \right]
\end{align}
We determine in turn the two terms in the commutation relation~\eqref{B.1.6}. The definition~\eqref{B.1.2bis} implies that,
\begin{equation}\label{B.1.7}
\begin{split}
&\sum_{\mu=1}^{N}\,\boldsymbol{R}^{(0)}_{\mu}\times\boldsymbol{A}_{\mu}(Q^{\,\boldsymbol{.}})\\
&= \sum_{\mu=1}^{N}\,\boldsymbol{R}^{(0)}_{\mu}\times\left(M_{\mu}\,\left(\boldsymbol{R}^{(0)}_{\mu}\cdot\mathsf{I}\left(Q^{\,\boldsymbol{.}}\right)^{-1}\right)\right)\\
&= \left(\boldsymbol{e}^{m}\cdot\mathsf{I}\left(Q^{\,\boldsymbol{.}}\right)^{-1}\cdot\boldsymbol{e}^{j}\right)\\
&\phantom{=}\cdot\sum_{\mu=1}^{N}\,M_{\mu}\,\left(\boldsymbol{e}_{j}\cdot\boldsymbol{R}^{(0)}_{\mu}\right)\,\left(\boldsymbol{e}_{\ell}\cdot\boldsymbol{R}^{(0)}_{\mu}\right)\,\left(\boldsymbol{e}^{\ell}\times\boldsymbol{e}_{m}\right)
\end{split}
\end{equation}
Using the definition~\eqref{4.0.3}, equation~\eqref{B.1.7} is recast as,
\begin{equation}\label{B.1.8}
\sum_{\mu=1}^{N}\,\boldsymbol{R}^{(0)}_{\mu}\times\boldsymbol{A}_{\mu}(Q^{\,\boldsymbol{.}}) = \left(\boldsymbol{e}^{m}\cdot\mathsf{I}\left(Q^{\,\boldsymbol{.}}\right)^{-1}\cdot\mathsf{I}_{00}\cdot\boldsymbol{e}^{\ell}\right)\,\left(\boldsymbol{e}_{\ell}\times\boldsymbol{e}_{m}\right)
\end{equation}
which implies in turn that the first term in the commutation relation~\eqref{B.1.6} becomes
\begin{equation}\label{B.1.8 prime 3}
\begin{split}
&-\,\frac{2i}{\hbar}\,\left[\ \boldsymbol{e}^{k}\cdot\boldsymbol{\Omega},\ \boldsymbol{e}_{k}\cdot\left(\sum_{\mu=1}^{N}\boldsymbol{R}^{(0)}_{\mu}\times\boldsymbol{A}_{\mu}(Q^{\,\boldsymbol{.}})\right)\ \right]\\
&= -\,\frac{2i}{\hbar}\left(\boldsymbol{e}^{m}\cdot\!\left[\ \boldsymbol{e}^{k}\cdot\boldsymbol{\Omega},\ \mathsf{I}\left(Q^{\,\boldsymbol{.}}\right)^{-1}\ \right]\!\cdot\mathsf{I}_{00}\cdot\boldsymbol{e}^{\ell}\right)\,\boldsymbol{e}_{k}\cdot\left(\boldsymbol{e}_{\ell}\times\boldsymbol{e}_{m}\right)
\end{split}
\end{equation}
The definition~\eqref{3.31} and the first commutation relation~\eqref{3.33.1} and imply that,
\begin{equation}\label{B.1.8 prime 4}
\begin{split}
&\left[\ \boldsymbol{e}^{k}\cdot\boldsymbol{\Omega},\ \mathsf{I}\left(Q^{\,\boldsymbol{.}}\right)\ \right] = \mathsf{I}_{\alpha}\,\left[\ \boldsymbol{e}^{k}\cdot\boldsymbol{\Omega},\ Q^{\alpha}\ \right]\\
&= -\,i\hbar\,\mathsf{I}_{\alpha}\,\left(\boldsymbol{e}^{k}\cdot\mathsf{I}\left(Q^{\,\boldsymbol{.}}\right)^{-1}\cdot\sum_{\mu=1}\,\left(\boldsymbol{X}^{\alpha}_{\mu}\times\boldsymbol{X}_{\mu\beta}\right)\,Q^{\beta}\right)
\end{split}
\end{equation}
which implies that,
\begin{align}\label{B.1.8 prime 5}
&\left[\ \boldsymbol{e}^{k}\cdot\boldsymbol{\Omega},\ \mathsf{I}\left(Q^{\,\boldsymbol{.}}\right)^{-1}\ \right]\nonumber\\
&= -\,\mathsf{I}\left(Q^{\,\boldsymbol{.}}\right)^{-1}\cdot\left[\ \boldsymbol{e}^{k}\cdot\boldsymbol{\Omega},\ \mathsf{I}\left(Q^{\,\boldsymbol{.}}\right)\ \right]\cdot\mathsf{I}\left(Q^{\,\boldsymbol{.}}\right)^{-1}\nonumber\\
&= i\hbar\,\left(\mathsf{I}\left(Q^{\,\boldsymbol{.}}\right)^{-1}\cdot\mathsf{I}_{\alpha}\cdot\mathsf{I}\left(Q^{\,\boldsymbol{.}}\right)^{-1}\right)\\
&\phantom{=}\cdot\left(\boldsymbol{e}^{k}\cdot\mathsf{I}\left(Q^{\,\boldsymbol{.}}\right)^{-1}\cdot\sum_{\mu=1}\,\left(\boldsymbol{X}^{\alpha}_{\mu}\times\boldsymbol{X}_{\mu\beta}\right)\,Q^{\beta}\right)\nonumber
\end{align}
Using the commutation relation~\eqref{B.1.8 prime 5} and the definition~\eqref{4.0.3}, the commutation relation~\eqref{B.1.8 prime 3} can be recast as,
\begin{equation}\label{B.1.8 ter}
\begin{split}
&-\,\frac{2i}{\hbar}\,\left[\ \boldsymbol{e}^{k}\cdot\boldsymbol{\Omega},\ \boldsymbol{e}_{k}\cdot\left(\sum_{\mu=1}^{N}\boldsymbol{R}^{(0)}_{\mu}\times\boldsymbol{A}_{\mu}(Q^{\,\boldsymbol{.}})\right)\ \right]\\
&=2\,\boldsymbol{e}^{m}\cdot\left(\mathsf{I}\left(Q^{\,\boldsymbol{.}}\right)^{-1}\cdot\mathsf{I}_{\alpha}\cdot\mathsf{I}\left(Q^{\,\boldsymbol{.}}\right)^{-1}\cdot\mathsf{I}_{00}\right)\cdot\boldsymbol{e}^{\ell}\\
&\phantom{=2\,\boldsymbol{e}^{m}\cdot}\left(\boldsymbol{e}_{\ell}\times\boldsymbol{e}_{m}\right)\cdot\mathsf{I}\left(Q^{\,\boldsymbol{.}}\right)^{-1}\cdot\sum_{\mu=1}\,\left(\boldsymbol{X}^{\alpha}_{\mu}\times\boldsymbol{X}_{\mu\beta}\right)\,Q^{\beta}\\
&=2\,\boldsymbol{e}^{m}\cdot\left(\mathsf{I}\left(Q^{\,\boldsymbol{.}}\right)^{-1}\cdot\mathsf{I}_{\alpha}\cdot\mathsf{I}\left(Q^{\,\boldsymbol{.}}\right)^{-1}\cdot\mathsf{I}^{\alpha}_{0\beta}\,Q^{\beta}\right)\cdot\boldsymbol{e}_{m}\\
&=2\,\text{Tr}\left(\mathsf{I}\left(Q^{\,\boldsymbol{.}}\right)^{-1}\cdot\mathsf{I}_{\alpha}\cdot\mathsf{I}\left(Q^{\,\boldsymbol{.}}\right)^{-1}\cdot\mathsf{I}^{\alpha}_{0\beta}\,Q^{\beta}\right)
\end{split}
\end{equation}
Similarly, using the definition~\eqref{B.1.2bis}, the second term in the commutation relation~\eqref{B.1.6} is recast as,
\begin{align}\label{B.1.8 quad}
&-\,\frac{2i}{\hbar}\,\left[\ P_{\alpha},\ \sum_{\mu=1}^{N}\,\frac{1}{\sqrt{M_{\mu}}}\,\left(\boldsymbol{X}^{\alpha}_{\mu}\cdot\boldsymbol{A}_{\mu}\left(Q^{\,\boldsymbol{.}}\right)\right)\ \right]\nonumber\\
&= -\,\frac{2i}{\hbar}\,\left[\ P_{\alpha},\ \left(\boldsymbol{e}^{j}\cdot\mathsf{I}\left(Q^{\,\boldsymbol{.}}\right)^{-1}\cdot\boldsymbol{e}^{\ell}\right)\right.\\
&\phantom{= -\,\frac{2i}{\hbar}\,\left[\right.}\left.\sum_{\mu=1}^{N} \,\sqrt{M_{\mu}}\,\left(\boldsymbol{X}^{\alpha}_{\mu}\cdot\left(\boldsymbol{e}_{j}\times\left(\boldsymbol{e}_{\ell}\times\boldsymbol{R}^{(0)}_{\mu}\right)\right)\right)\ \right]\,.\nonumber
\end{align}
The definition~\eqref{3.33} implies that,
\begin{equation}\label{B.1.8 pet}
\begin{split}
&\sum_{\mu=1}^{N}\,\sqrt{M_{\mu}}\,\left(\boldsymbol{X}^{\alpha}_{\mu}\cdot\left(\boldsymbol{e}_{j}\times\left(\boldsymbol{e}_{\ell}\times\boldsymbol{R}^{(0)}_{\mu}\right)\right)\right)\\
&=-\,\sum_{\mu=1}^{N}\,\sqrt{M_{\mu}}\,\left(\boldsymbol{e}_{j}\times\boldsymbol{X}^{\alpha}_{\mu}\right)\cdot\left(\boldsymbol{e}_{\ell}\times\boldsymbol{R}^{(0)}_{\mu}\right)\\ &= -\,\left(\boldsymbol{e}_{j}\cdot\mathsf{I}^{\alpha}\cdot\boldsymbol{e}_{\ell}\right)\,.
\end{split}
\end{equation}
Using the relation~\eqref{B.1.8 pet}, the commutation relation~\eqref{B.1.8 quad} is recast as,
\begin{equation}\label{B.1.8 hex}
\begin{split}
&-\,\frac{2i}{\hbar}\,\left[\ P_{\alpha},\ \sum_{\mu=1}^{N}\,\frac{1}{\sqrt{M_{\mu}}}\,\left(\boldsymbol{X}^{\alpha}_{\mu}\cdot\boldsymbol{A}_{\mu}\left(Q^{\,\boldsymbol{.}}\right)\right)\ \right]\\
&= \frac{2i}{\hbar}\,\left(\boldsymbol{e}_{j}\cdot\mathsf{I}^{\alpha}\cdot\boldsymbol{e}_{\ell}\right)\,\left[\ P_{\alpha},\ \boldsymbol{e}^{j}\cdot\mathsf{I}\left(Q^{\,\boldsymbol{.}}\right)^{-1}\cdot\boldsymbol{e}^{\ell}\ \right]\,.
\end{split}
\end{equation}
Using the definition~\eqref{3.31} and the canonical commutation relation~\eqref{3.33.1}, we commute the commutation relation on the RHS of the expression~\eqref{B.1.8 hex}, i.e.
\begin{equation}\label{B.1.8 hep}
\begin{split}
&\left[\ P_{\alpha},\ \boldsymbol{e}^{j}\cdot\mathsf{I}\left(Q^{\,\boldsymbol{.}}\right)^{-1}\cdot\boldsymbol{e}^{\ell}\ \right]\\
&=i\hbar\,\left(\boldsymbol{e}^{j}\cdot\mathsf{I}\left(Q^{\,\boldsymbol{.}}\right)^{-1}\cdot\mathsf{I}_{\alpha}\cdot\mathsf{I}\left(Q^{\,\boldsymbol{.}}\right)^{-1}\cdot\boldsymbol{e}^{\ell}\right)\,.
\end{split}
\end{equation}
Using the commutation relation~\eqref{B.1.8 hep} and the fact that the tensor $\mathsf{I}_{\alpha}$ is symmetric, the commutation relation~\eqref{B.1.8 hep} is recast as,
\begin{equation}\label{B.1.8 oct}
\begin{split}
&-\,\frac{2i}{\hbar}\,\left[\ P_{\alpha},\ \sum_{\mu=1}^{N}\,\frac{1}{\sqrt{M_{\mu}}}\,\left(\boldsymbol{X}^{\alpha}_{\mu}\cdot\boldsymbol{A}_{\mu}\left(Q^{\,\boldsymbol{.}}\right)\right)\ \right]\\
&=-\,2\,\boldsymbol{e}_{\ell}\cdot\left(\mathsf{I}^{\alpha}\cdot\mathsf{I}\left(Q^{\,\boldsymbol{.}}\right)^{-1}\cdot\mathsf{I}_{\alpha}\cdot\mathsf{I}\left(Q^{\,\boldsymbol{.}}\right)^{-1}\right)\cdot\boldsymbol{e}^{\ell}\\
&=-\,2\,\text{Tr}\left(\left(\mathsf{I}_{\alpha}\cdot\mathsf{I}\left(Q^{\,\boldsymbol{.}}\right)^{-1}\right)^2\right)\,.
\end{split}
\end{equation}
The commutation relations~\eqref{B.1.6},~\eqref{B.1.8 ter} and~\eqref{B.1.8 oct} imply that the first term of the operator~\eqref{B.1.1 res} is recast as,
\begin{equation}\label{B.1.9}
\begin{split}
&-\,\frac{2i}{\hbar}\,\sum_{\mu=1}^{N}\,\frac{1}{M_{\mu}}\,\left[\,\boldsymbol{P}^{\prime\prime}_{\mu},\ \boldsymbol{A}_{\mu}\left(Q^{\,\boldsymbol{.}}\right)\ \right]_{\mathsmaller{\bullet}}\\
&= 2\,\text{Tr}\left(\mathsf{I}\left(Q^{\,\boldsymbol{.}}\right)^{-1}\cdot\mathsf{I}_{\alpha}\cdot \mathsf{I}\left(Q^{\,\boldsymbol{.}}\right)^{-1}\cdot\mathsf{I}^{\alpha}_{0\beta}\,Q^{\beta}\right)\\
&\phantom{=}-\,2\,\text{Tr}\left(\left(\mathsf{I}^{\alpha}\cdot\mathsf{I}\left(Q^{\,\boldsymbol{.}}\right)^{-1}\right)^2\right)\,.
\end{split}
\end{equation}
Using the definitions~\eqref{4.0.3} and~\eqref{B.1.2bis} the last term of the the RHS of the expression~\eqref{B.1.1 res} is recast as,
\begin{equation}\label{B.1.10}
\begin{split}
&\sum_{\mu=1}^{N}\,\frac{1}{M_{\mu}}\,\boldsymbol{A}_{\mu}\left(Q^{\,\boldsymbol{.}}\right)^{2}\\
&=\sum_{\mu=1}^{N}\,M_{\mu}\,\left(\left(\boldsymbol{R}^{(0)}_{\mu}\cdot\mathsf{I}\left(Q^{\,\boldsymbol{.}}\right)^{-1}\right) -\,\boldsymbol{R}^{(0)}_{\mu}\,\text{Tr}\left(\mathsf{I}\left(Q^{\,\boldsymbol{.}}\right)^{-1}\right)\right)^2\\
&= \text{Tr}\left(\mathsf{I}_{00}\cdot\mathsf{I}\left(Q^{\,\boldsymbol{.}}\right)^{-2}\right) -\,2\,\text{Tr}\left(\mathsf{I}_{00}\cdot\mathsf{I}\left(Q^{\,\boldsymbol{.}}\right)^{-1}\right)\,\text{Tr}\left(\mathsf{I}\left(Q^{\,\boldsymbol{.}}\right)^{-1}\right)\\
&\phantom{=}+ \text{Tr}\left(\mathsf{I}_{00}\right)\,\text{Tr}\left(\mathsf{I}\left(Q^{\,\boldsymbol{.}}\right)^{-2}\right)\,.
\end{split}
\end{equation}
The relations~\eqref{B.1.9} and~\eqref{B.1.10} imply that the expression~\eqref{B.1.1 res} yields the residual operator~\eqref{4.0.2}, i.e.
\begin{equation}\label{B.1.13}
\begin{split}
&\Phi_{\text{res}}\left(Q^{\,\boldsymbol{.}}\right) = \text{Tr}\left(\mathsf{I}_{00}\right)\,\text{Tr}\left(\mathsf{I}\left(Q^{\,\boldsymbol{.}}\right)^{-2}\right)\\
&-\,2\,\text{Tr}\left(\mathsf{I}_{00}\cdot\mathsf{I}\left(Q^{\,\boldsymbol{.}}\right)^{-1}\right)\,\text{Tr}\left(\mathsf{I}\left(Q^{\,\boldsymbol{.}}\right)^{-1}\right)\\ 
&+ \text{Tr}\left(\mathsf{I}_{00}\cdot\mathsf{I}\left(Q^{\,\boldsymbol{.}}\right)^{-2} -\,2\,\left(\mathsf{I}_{\alpha}\cdot\mathsf{I}\left(Q^{\,\boldsymbol{.}}\right)^{-1}\right)^2\right)\\
&+2\,\text{Tr}\left(\mathsf{I}\left(Q^{\,\boldsymbol{.}}\right)^{-1}\cdot\mathsf{I}_{\alpha}\cdot \mathsf{I}\left(Q^{\,\boldsymbol{.}}\right)^{-1}\cdot\mathsf{I}^{\alpha}_{0\beta}\,Q^{\beta}\right)\,.
\end{split}
\end{equation}

\section{Vibrational modes}
\label{Vibrational modes}

In this appendix, we determine the expression of the angular frequency of the vibrational modes and show that it is positively defined.

Using the expression~\eqref{4.13} of the Coulomb potential between the nuclei, we deduce the zeroth, first and second order terms of the series expansion~\eqref{4.14}, i.e.
\begin{align}\label{4.15}
&V_{\mathcal{N}-\mathcal{N}\,(0)} = \frac{e^{2}}{8\pi\varepsilon_{0}}\,\sum_{\substack{\mu,\nu=1\\ \mu\neq\nu}}^{N}\,\frac{Z_{\mu}\,Z_{\nu}}{\Vert\Delta\,\boldsymbol{R}_{\mu\nu}^{(0)}\Vert}\,,\nonumber\\
&V_{\mathcal{N}-\mathcal{N}\,(\alpha)}=\frac{e^{2}}{8\pi\varepsilon_{0}}\,\sum_{\substack{\mu,\nu=1\\ \mu\neq\nu}}^{N}Z_{\mu}\,Z_{\nu}\,\left(\,-\,\frac{\Delta\,\boldsymbol{Y}_{\mu\nu\alpha}\cdot\Delta\,\boldsymbol{R}_{\mu\nu}^{(0)}}{\Vert\Delta\,\boldsymbol{R}_{\mu\nu}^{(0)}\Vert^{3}}\right)\,,\nonumber\\
&V_{\mathcal{N}-\mathcal{N}\,(\alpha\beta)}=\frac{e^{2}}{8\pi\varepsilon_{0}}\,\sum_{\substack{\mu,\nu=1\\ \mu\neq\nu}}^{N}Z_{\mu}\,Z_{\nu}\,
\left(\, -\,\frac{\Delta\,\boldsymbol{Y}_{\mu\nu\alpha}\cdot\Delta\,\boldsymbol{Y}_{\mu\nu\beta}}{\Vert\Delta\,\boldsymbol{R}_{\mu\nu}^{(0)}\Vert^{3}}\right.\nonumber
\\
&\left.+\,3\ \frac{\left(\Delta\,\boldsymbol{Y}_{\mu\nu\alpha}\cdot\Delta\,\boldsymbol{R}_{\mu\nu}^{(0)}\right)\left(\Delta\,\boldsymbol{R}_{\mu\nu}^{(0)}\cdot\Delta\,\boldsymbol{Y}_{\mu\nu\beta}\right)}{\Vert\Delta\,\boldsymbol{R}_{\mu\nu}^{(0)}\Vert^{5}}\,\right)\,,
\end{align}
where $\Delta\,\boldsymbol{R}^{(0)}_{\mu\nu}\equiv\boldsymbol{R}^{(0)}_{\mu}-\,\boldsymbol{R}^{(0)}_{\nu}$ and $\Delta\,\boldsymbol{Y}_{\mu\nu\alpha}\equiv\boldsymbol{Y}_{\mu\alpha}-\,\boldsymbol{Y}_{\nu\alpha}\,$. Similarly, using the expression~\eqref{4.13} of the Coulomb potential between the nuclei and the electrons, we deduce the expressions for the zeroth, first and second order terms of the series expansion~\eqref{4.23}, i.e.
\begin{align}\label{4.24}
&V_{\mathcal{N}-e\,(0)}\,(\boldsymbol{q}_{(\boldsymbol{.})}) = -\,\frac{e^{2}}{4\pi\varepsilon_{0}}\,\sum_{\mu=1}^{N}\sum_{\nu=1}^{n}\,\frac{Z_{\mu}}{\Vert\,\Delta\,\boldsymbol{Q}^{(0)}_{\mu\nu}\Vert}\,,\nonumber\\
&V_{\mathcal{N}-e\,(\alpha)}\,(\boldsymbol{q}_{(\boldsymbol{.})})=-\,\frac{e^{2}}{4\pi\varepsilon_{0}}\,\sum_{\mu=1}^{N}\sum_{\nu=1}^{n}\,Z_{\mu}\,
\left(-\,\frac{\boldsymbol{Y}_{\mu\alpha}\cdot\Delta\,\boldsymbol{Q}^{(0)}_{\mu\nu}}{\Vert\,\Delta\,\boldsymbol{Q}^{(0)}_{\mu\nu}\Vert^{3}}\right)\,,\nonumber\\
&V_{\mathcal{N}-e\,(\alpha\beta)}\,(\boldsymbol{q}_{(\boldsymbol{.})})=-\,\frac{e^{2}}{4\pi\varepsilon_{0}}\,\sum_{\mu=1}^{N}\sum_{\nu=1}^{n}\,Z_{\mu}\,\left(\, -\,\frac{\boldsymbol{Y}_{\mu\alpha}\cdot\boldsymbol{Y}_{\mu\beta}}{\Vert\,\Delta\,\boldsymbol{Q}^{(0)}_{\mu\nu}\Vert^{3}}\right.\nonumber
\\
&\left.\phantom{=}+\,3\ \frac{\left(\boldsymbol{Y}_{\mu\alpha}\cdot\Delta\,\boldsymbol{Q}^{(0)}_{\mu\nu}\right)\left(\Delta\,\boldsymbol{Q}^{(0)}_{\mu\nu}\cdot\boldsymbol{Y}_{\mu\beta}\right)}{\Vert\,\Delta\,\boldsymbol{Q}^{(0)}_{\mu\nu}\Vert^{5}}\,\right)\,.
\end{align}
where $\Delta\,\boldsymbol{Q}^{(0)}_{\mu\nu}\equiv\boldsymbol{R}^{(0)}_{\mu}\,\mathbb{1}-\,\boldsymbol{\bar{q}}_{(\nu)}\,$. We define three symmetric and trace-free tensors, i.e. 
\begin{equation}\label{5.7}
\begin{split}
&\mathsf{B}_{\mu\nu} = \frac{e^2\,Z_{\mu}\,Z_{\nu}}{4\pi\varepsilon_0}\,\left(-\,\frac{\mathbb{1}}{\Vert \Delta\,\boldsymbol{R}_{\mu\nu}^{(0)}\Vert^3} + 3\,\frac{\Delta\,\boldsymbol{R}_{\mu\nu}^{(0)}\,\Delta\,\boldsymbol{R}_{\mu\nu}^{(0)}}{\Vert \Delta\,\boldsymbol{R}_{\mu\nu}^{(0)} \Vert^5}\,\right)\,,\\
&\mathsf{A}_{\mu} = \frac{e^2\,Z_{\mu}}{4\pi\varepsilon_0}\,\sum_{\nu=1}^{N}\,Z_{\nu}\,\left(-\,\frac{\mathbb{1}}{\Vert \Delta\,\boldsymbol{R}_{\mu\nu}^{(0)}\Vert^3} + 3\,\frac{\Delta\,\boldsymbol{R}_{\mu\nu}^{(0)}\,\Delta\,\boldsymbol{R}_{\mu\nu}^{(0)}}{\Vert \Delta\,\boldsymbol{R}_{\mu\nu}^{(0)} \Vert^5}\,\right)\,,\\
&\mathsf{E}_{\mu} = -\,\frac{e^2\,Z_{\mu}}{4\pi\varepsilon_0}\Bigg\langle\sum_{\nu=1}^{n}\left(-\,\frac{\mathbb{1}}{\Vert \Delta\,\boldsymbol{Q}_{\mu\nu}^{(0)}\Vert^3} + 3\,\frac{\Delta\,\boldsymbol{Q}_{\mu\nu}^{(0)}\,\Delta\,\boldsymbol{Q}_{\mu\nu}^{(0)}}{\Vert \Delta\,\boldsymbol{Q}_{\mu\nu}^{(0)} \Vert^5}\,\right)\!\Bigg\rangle
\end{split}
\end{equation}
where 
\begin{equation}\label{5.8}
\begin{split}
&\mathsf{B}_{\mu\nu} = \mathsf{B}_{\nu\mu}\,,\\
&\mathsf{A}_{\mu} = \sum_{\substack{\nu=1\\ \mu\neq\nu}}^{N}\,\mathsf{B}_{\mu\nu}  = \sum_{\nu=1}^{N}\,\mathsf{B}_{\mu\nu}\left(1-\,\delta_{\mu\nu}\right)\,.
\end{split}
\end{equation}
Using the definitions~\eqref{4.13} and~\eqref{5.7}, and the properties~\eqref{5.8}, the second-order term of the Coulomb potential $V_{\mathcal{N}-\mathcal{N}\,(\alpha\beta)}$ is expressed as,
\begin{align}\label{5.9}
&V_{\mathcal{N}-\mathcal{N}\,(\alpha\beta)} = \frac{1}{2}\,\sum_{\mu,\nu=1}^{N}\,\Delta\,\boldsymbol{Y}_{\mu\nu\alpha}\cdot\mathsf{B}_{\mu\nu}\left(1-\,\delta_{\mu\nu}\right)\cdot\Delta\,\boldsymbol{Y}_{\mu\nu\beta}\nonumber\\
&= \sum_{\mu,\nu=1}^{N}\,\boldsymbol{Y}_{\mu\alpha}\cdot\mathsf{A}_{\mu}\,\delta_{\mu\nu}\cdot\boldsymbol{Y}_{\nu\beta}\\ &\phantom{=}-\,\sum_{\mu,\nu=1}^{N}\boldsymbol{Y}_{\mu\alpha}\cdot\mathsf{B}_{\mu\nu}\left(1-\,\delta_{\mu\nu}\right)\cdot\boldsymbol{Y}_{\nu\beta}\,.\nonumber
\end{align}
Similarly, using the definitions~\eqref{4.13} and~\eqref{5.7}, the second-order term of the Coulomb potential $V_{\mathcal{N}-e\,(\alpha\beta)}$ is expressed as,
\begin{equation}\label{5.10}
V_{\mathcal{N}-e\,(\alpha\beta)} = \sum_{\mu,\nu=1}^{N}\,\boldsymbol{Y}_{\mu\alpha}\cdot\mathsf{E}_{\mu}\,\delta_{\mu\nu}\cdot\boldsymbol{Y}_{\nu\beta}\,.
\end{equation}
According to the expressions~\eqref{5.9} and~\eqref{5.10} of the second-order Coulomb potentials, the condition~\eqref{5.6} is recast in terms of the tensors~\eqref{5.7} as,
\begin{equation}\label{5.11}
\sum_{\mu=1}^{N}\,\boldsymbol{Y}_{\mu\alpha}\cdot\sum_{\nu=1}^{N}\bigg(\left(\mathsf{A}_{\mu} + \mathsf{E}_{\mu}\right)\,\delta_{\mu\nu} -\,\mathsf{B}_{\mu\nu}\left(1-\,\delta_{\mu\nu}\right)\bigg)\cdot\boldsymbol{Y}_{\nu\beta} = 0
\end{equation}
To simplify the notation, we define a symmetric and trace-free tensor $\mathsf{D}_{\mu\nu}$ that is a linear combination of the tensors $\mathsf{A}_{\mu}$, $\mathsf{B}_{\mu\nu}$ and $\mathsf{E}_{\mu}$, i.e.
\begin{equation}\label{5.16}
\mathsf{D}_{\mu\nu} = \left(\mathsf{A}_{\mu} + \mathsf{E}_{\mu}\right)\delta_{\mu\nu} -\,\mathsf{B}_{\mu\nu}\left(1-\,\delta_{\mu\nu}\right)\,,
\end{equation}
which implies that the relation~\eqref{5.11} is recast as,
\begin{equation}\label{5.11.bis}
\sum_{\mu=1}^{N}\,\boldsymbol{Y}_{\mu\alpha}\cdot\left(\,\sum_{\nu=1}^{N}\,\mathsf{D}_{\mu\nu}\cdot\boldsymbol{Y}_{\nu\beta}\right) = 0\,,
\end{equation}
where $\alpha\neq\beta\,$. The conditions~\eqref{3.18},~\eqref{3.17} and~\eqref{3.17bis} imply that the condition~\eqref{5.11.bis} is satisfied provided all the vibration modes $\boldsymbol{Y}_{\nu\beta}$ satisfy the condition,
\begin{equation}\label{5.12}
\sum_{\nu=1}^{N}\,\mathsf{D}_{\mu\nu}\cdot\boldsymbol{Y}_{\nu\beta} = c_{\beta}\,\sqrt{M_{\mu}}\,\boldsymbol{X}_{\mu\beta} + M_{\mu}\,\boldsymbol{a}_{\beta} + M_{\mu}\left(\boldsymbol{b}_{\beta}\times\boldsymbol{R}_{\mu}^{(0)}\right)
\end{equation}
where $c_{\beta}$ is a scalar parameter, $\boldsymbol{a}_{\beta}$ and $\boldsymbol{b}_{\beta}$ are vectorial parameters. 

Using the definition~\eqref{5.16}, the conditions~\eqref{3.16},~\eqref{3.17} and~\eqref{5.8} and taking the sum over all the nuclei in the relation~\eqref{5.12} yields,
\begin{equation}\label{5.13}
\sum_{\mu,\nu=1}^{N}\,\mathsf{D}_{\mu\nu}\cdot\boldsymbol{Y}_{\nu\beta} = \sum_{\mu=1}^{N}\,\mathsf{E}_{\mu}\cdot\boldsymbol{Y}_{\mu\beta} = \sum_{\mu=1}^{N}\,M_{\mu}\,\boldsymbol{a}_{\beta}\,.
\end{equation}
The definition~\eqref{3.20pre} and the relation~\eqref{5.13} yield,
\begin{equation}\label{5.14}
\boldsymbol{a}_{\beta} = \frac{1}{M}\,\sum_{\nu=1}^{N}\,\mathsf{E}_{\nu}\cdot\boldsymbol{Y}_{\nu\beta}\,.
\end{equation}
Using the conditions~\eqref{3.16} and~\eqref{3.17} and taking the sum over all the nuclei in the relation~\eqref{5.12} after multiplying by $\boldsymbol{b}_{\beta}\times\boldsymbol{R}_{\mu}^{(0)}$ yields,
\begin{equation}\label{5.15}
\begin{split}
&\sum_{\mu,\nu=1}^{N}\,\left(\boldsymbol{b}_{\beta}\times\boldsymbol{R}_{\mu}^{(0)}\right)\cdot\mathsf{D}_{\mu\nu}\cdot\boldsymbol{Y}_{\nu\beta}\\
&= \sum_{\mu=1}^{N}\,M_{\mu}\,\left(\boldsymbol{b}_{\beta}\times\boldsymbol{R}_{\mu}^{(0)}\right)\cdot\left(\boldsymbol{b}_{\beta}\times\boldsymbol{R}_{\mu}^{(0)}\right)\,.
\end{split}
\end{equation}
The definitions~\eqref{3.32} implies that the relation~\eqref{5.15} reduces to,
\begin{equation}\label{5.17}
\sum_{\mu,\nu=1}^{N}\,\left(\boldsymbol{e}_{j}\times\boldsymbol{R}_{\mu}^{(0)}\right)\cdot\mathsf{D}_{\mu\nu}\cdot\boldsymbol{Y}_{\nu\beta} = \boldsymbol{e}_{j}\cdot\mathsf{I}_{0}\cdot\boldsymbol{b}_{\beta}\,,
\end{equation}
which is recast as,
\begin{equation}\label{5.18}
\boldsymbol{b}_{\beta} = \left(\mathsf{I}_{0}^{-1}\cdot\boldsymbol{e}^{j}\right)\sum_{\rho,\nu=1}^{N}\,\left(\boldsymbol{e}_{j}\times\boldsymbol{R}_{\rho}^{(0)}\right)\cdot\mathsf{D}_{\rho\nu}\cdot\boldsymbol{Y}_{\nu\beta}\,.
\end{equation}
Substituting the relations~\eqref{5.14},~\eqref{5.16} and~\eqref{5.18} into the condition~\eqref{5.12} yields the eigenvalue equation,
\begin{align}\label{5.19}
&\sum_{\nu=1}^{N}\left(\mathsf{D}_{\mu\nu} -\frac{M_{\mu}}{M}\,\mathsf{E}_{\nu} -M_{\mu}\bigg(\!\left(\boldsymbol{e}_{k}\times\boldsymbol{R}_{\mu}^{(0)}\right)\!\!\sum_{\rho,\nu=1}^{N}\!\left(\boldsymbol{e}^{k}\!\cdot\mathsf{I}_{0}^{-1}\!\cdot\boldsymbol{e}^{j}\right)\right.\nonumber\\
&\left.\cdot\left(\boldsymbol{e}_{j}\times\boldsymbol{R}_{\rho}^{(0)}\right)\cdot\mathsf{D}_{\rho\nu}\bigg)\vphantom{\frac{M_{\mu}}{M}}\right)\cdot\boldsymbol{Y}_{\nu\beta}= c_{\beta}\,\sqrt{M_{\mu}}\,\boldsymbol{X}_{\mu\beta}\,.
\end{align}
Using the conditions~\eqref{3.18},~\eqref{3.16} and~\eqref{3.17} and taking the sum over all the nuclei in the relation~\eqref{5.19} after multiplying by $\boldsymbol{Y}_{\mu\beta}$ yields,
\begin{equation}\label{5.20}
\sum_{\mu=1}^{N}\,\boldsymbol{Y}_{\mu\beta}\cdot\left(\,\sum_{\nu=1}^{N}\,\mathsf{D}_{\mu\nu}\cdot\boldsymbol{Y}_{\nu\beta}\right) = c_{\beta}\,.
\end{equation}
At equilibrium, the energy of the stable molecular system is minimal. Thus, in the neighbourhood of the equilibrium, the relation~\eqref{5.20} is a positive definite quadratic form. Identifying the relations~\eqref{5.6} and~\eqref{5.11.bis} and comparing them to the relation~\eqref{5.20}, we conclude that, i.e.
\begin{equation}
\omega^2_{\beta} = V_{\mathcal{N}-\mathcal{N}\,(\beta\beta)} + \Big\langle\,V_{\mathcal{N}-e\,(\beta\beta)}\,(\boldsymbol{q}_{(\boldsymbol{.})})\,\Big\rangle\,,
\end{equation}
where $\omega^2_{\beta} \equiv c_{\beta} > 0$ is identified physically as the square of the angular frequency of the vibration eigenmodes.


\bibliography{references}

\end{document}